\documentclass[12pt]{article}
\usepackage[margin=1in]{geometry}
\usepackage{amsmath, amssymb, amsthm}
\usepackage{authblk} 
\usepackage{graphicx}
\usepackage{booktabs}
\usepackage{hyperref}
\hypersetup{
    colorlinks = true,
    linkcolor  = blue!70!black,
    citecolor  = blue!70!black,
    urlcolor   = blue!70!black
}
\usepackage[authoryear,round]{natbib}
\bibpunct{(}{)}{;}{a}{}{,}
\usepackage{microtype}
\usepackage{enumitem}
\usepackage{subcaption}
\usepackage{xcolor}
\usepackage{times}
\usepackage[section]{placeins}

\usepackage{algorithm}
\usepackage{algpseudocode}
\usepackage{float}
\makeatletter
\newenvironment{breakablealgorithm}
  {\begin{center}
     \refstepcounter{algorithm}
     \renewcommand{\caption}[2][\relax]{
       \vspace{4pt}
       {\raggedright\textbf{Algorithm~\thealgorithm} ##2\par}
       \vspace{4pt}\hrule\vspace{4pt}
     }
  }
  {\vspace{4pt}\hrule\relax
   \end{center}}
\makeatother

\usepackage{nomencl}
\usepackage{multicol}
\makenomenclature

\renewcommand{\nompreamble}{\begin{multicols}{2}}
	\renewcommand{\nompostamble}{\end{multicols}}

\DeclareMathOperator*{\argmax}{arg\,max}

\usepackage{pgffor}
\usepackage{setspace}
\newcommand{\keywords}[1]{\par\noindent\textbf{Keywords:} #1}

\title{A Finite Mixture Failure-rate based Heterogeneous Step-stress Accelerated Life Testing ($h$-SSALT) Model}
\author[1]{Pranoy Palit}
\author[1]{Ayan Pal}
\author[2]{Kiran Prajapat\thanks{Corresponding author. Email: kiranprajapat92@gmail.com, kiran.prajapat@newcastle.ac.uk (Kiran Prajapat)}}

\affil[1]{\small Department of Statistics, The University of Burdwan, Burdwan, West Bengal, India}
\affil[2]{\small School of Mathematics, Statistics and Physics, Newcastle University, Newcastle upon Tyne, UK}

\date{}	

\begin{document}
	
	\maketitle
	
	\begin{abstract}

        Traditional step-stress accelerated life testing models assume that test units originate from a homogeneous population. Recently, \cite{lu2025} proposed a heterogeneous cumulative exposure based SSALT model to account for the inhomogeneous aging patterns among test units belonging to the same production batch. This paper introduces an alternative yet flexible failure-rate based heterogeneous simple SSALT ($h$-SSALT) model with Weibull-distributed Type-II censored failure times, allowing heterogeneity to emerge at the second stress level through a finite mixture of $m$ latent subgroups, each characterized by its own failure behavior. The expectation-maximization algorithm is developed for maximum likelihood estimation of the model parameters, exploiting the incomplete data structure arising from both unknown group membership and Type-II censoring. Interval estimation is performed using the missing information identity of \cite{louis1982finding} with transformation-based confidence intervals respecting parameter constraints. An extensive simulation study evaluates the finite-sample performance of the proposed estimators and demonstrates, through a quantile-based comparison, that ignoring population heterogeneity leads to systematic bias in lifetime predictions across the entire quantile range, with the most severe consequences at early failure quantiles of direct relevance to warranty period design. A special case comparison confirms that the proposed Weibull failure-rate based formulation reduces to the existing model of \cite{lu2025} when the shape parameter equals unity, validating the proposed framework as a proper generalization. The practical application of the model is further illustrated through simulated and real data analysis examples.
	\end{abstract}
	
	 \\[2pt]

% \tableofcontents

% \doublespacing
\newcommand{\bea}{\begin{eqnarray}}
\newcommand{\eea}{\end{eqnarray}}
\newcommand{\nn}{\nonumber}
\newcommand{\bee}{\begin{eqnarray*}}
\newcommand{\eee}{\end{eqnarray*}}
\newcommand{\lb}{\label}
\newcommand{\nii}{\noindent}
\newcommand{\ii}{\indent}

\newtheorem{theorem}{Theorem}[section]
\newtheorem{corollary}{Corollary}[theorem]
\newtheorem{proposition}{Proposition}[section]
\newtheorem{lemma}{Lemma}[section]
\newtheorem{remark}{Remark}[section]
    
\section{Introduction}
Evaluating the operational lifespan of highly reliable products under normal operating conditions (NOC) is often challenging due to the extensive time and cost required for life testing.In many modern engineering applications such as aerospace components, electronic devices, and safety-critical mechanical systems failures occur very rarely under NOC. As a result, collecting sufficient failure data under NOC becomes impractical. Accelerated life testing (ALT) addresses this challenge by subjecting test units to elevated stress levels, such as increased temperature, voltage, pressure, or load, to induce failures more rapidly and allow inference on product reliability at normal conditions. The theoretical foundations of ALT were first laid by \cite{nelson1978}, who developed optimal accelerated censored life tests for Weibull and extreme value distributions. A detailed account of ALT methodology and its applications can be found in \cite{nelson1990, meeker1998, bagdonavicius2001accelerated} and \cite{kundu2017}.

A significant advancement within ALT is step-stress ALT (SSALT), in which stress levels are increased sequentially during the experiment, allowing more efficient use of test units. In a general SSALT setup, $n$ units are initially tested at stress level $S_0$, with the stress increases to $s_{1},s_{2},\dots,s_{m}$ at preassigned time points $\tau_1<\tau_2<\dots<\tau_m$, respectively. If there are only two stress levels then the experiment is referred to as a simple step-stress life testing experiment. Optimal design and inference for step-stress accelerated life tests (SSALT) have received  plenty of attention in the literature. \cite{miller1983} developed optimal simple SSALT plans for exponentially distributed lifetimes without censoring while \cite{bai1989optimum} extended these results to censored observations. Optimal designs for Weibull lifetimes under Type-I censoring were later examined by \cite{bai1993optimum}. Statistical inference for SSALT has been widely studied in the literature, primarily under homogeneous population assumptions and common censoring schemes such as Type-I and Type-II censoring. In reliability experiments with Type-II censoring, all test units are placed on test simultaneously and the experiment is terminated once a predetermined number of failures say $r$ is observed. \cite{balakrishnan2007point} investigated point and interval estimation for a simple SSALT model assuming exponential lifetimes under Type-II censoring, and \cite{kateri2008} extended this framework to weibull lifetimes. \cite{watkins2001} emphasized that inference in step-stress models should preferably be based on the original model parameters rather than reparameterized forms. For a detailed insight see \cite{kundu2017} and the references cited therein.

There are two well established modeling approaches are available in the literature for analyzing data from step-stress accelerated life testing (SSALT) experiments namely  the cumulative exposure model (CEM) by \cite{sedyakin1966} and tampered failure-rate model (TFRM) introduced by \cite{bhattacharyya1989}. This concept was further developed by Khamis and Higgins (1996). All these and other related models have been discussed in detail in \cite{kundu2017}. In the present study, we adopt the flexible failure-rate based SSALT model recently  developed by \cite{kateri2015}. It states that the overall hazard function $h(t)$ under the step-stress scheme for $i=1,2$ is defined in a piecewise form as 
\[
h(t) = \begin{cases}
	h_{1}(t),&\text{if } 0<t\leq\tau\\
	h_{2}(t),&\text{if } \tau<t<\infty,
\end{cases}
\]
where $h_i(t)$ is the hazard function corresponding to the cumulative distribution function (CDF) $G_i(t),i=1,2.$ Inferential procedures and optimal design for SSALT within this framework have been investigated by several authors, including \cite{pal2021}, \cite{samanta2021}, \cite{cramer2025}.

Most existing SSALT models are developed under the assumption that all test items originate from a single homogeneous population, implying a common failure mechanism and lifetime distribution at each stress level. However,many ALT experiments may exhibit heterogeneous aging patterns, particularly under higher stresses. For example, \cite{meeker1995} observed distinct clustering of circuit-board failures in a humidity-based ALT experiment. A similar pattern is also seen in battery testing, where cells often appear nearly identical at the beginning of the experiment but begin to differ substantially as the number of charge–discharge cycles increases (see \cite{harris2017}). Increased variability and subgroup patterns have also been observed in metal-film resistors tested at elevated temperatures (see Example $21.1$ in \cite{meeker1998}). Such heterogeneity may arise not only from known factors, such as differences across production lines, but also among units from the same batch that behave similarly at early stages but start to differ at later stages due to small material or structural variations. If this type of heterogeneity is ignored, lifetime estimates may be biased, and extrapolation to normal operating conditions may become unreliable. Therefore, identifying and quantifying such latent subgroup structure is important not only for obtaining unbiased lifetime predictions but also for practical 
decision-making. For instance, knowing the proportion of units belonging to a weaker subgroup allows manufacturers to determine what fraction of a production batch should be screened before deployment, a process commonly known as burn-in testing, by reducing the risk of early field failures and unwarranted warranty claims. \cite{WilsonFarrow2021} also highlighted the importance of reliability demonstration testing and
appropriate sample-size determination for obtaining reliable statistical
evidence about product reliability in the presence of heterogeneity due to location effects. This kind of actionable inference is only possible when the heterogeneity in the population is explicitly modeled, which motivates the development of the proposed heterogeneous step-stress accelerated life testing ($h$-SSALT) model.

Several studies in the literature have addressed heterogeneity within the  accelerated life testing framework, considering situations in which group membership is either known in advance or remains unobserved. \cite{LIN2017} proposed a reliability assessment framework for heterogeneous populations in multiple-stress accelerated testing based on finite mixture distributions and EM-based estimation procedures, demonstrating that explicit modeling of latent subpopulations substantially improves reliability assessment accuracy. Finite mixture models have become an important and widely used tool for analyzing data generated from populations that exhibit unobserved heterogeneity. The central idea behind these models is that the overall population can be represented as a finite mixture of latent homogeneous subgroups, each associated with a distinct failure mechanism. Foundational developments and broad applications of finite mixture methodology are comprehensively documented in several classical monographs, including those by \cite{everitt1981, titterington1985, mclachlan1988, mclachlan2000}. \cite{al2006accelerated} introduced finite mixture models in ALT and derived maximum likelihood estimators for a two component Weibull mixture with unknown group membership, each component representing a distinct failure mechanism. Their approach was later extended to progressive stress ALT and step partially accelerated life testing models (see \cite{abdel2007progressive,abdel2008step}. When group membership is available, several Bayesian, regression-based and mixed-effects approaches have been proposed by \cite{leon2007, lv2015, Lin2017ress, Seo2016techno, WilsonFarrow2021, Zhuang2023ress}. These approaches have contributed significantly to modeling heterogeneous ALT data, but most of them assume either known group membership or focus on constant-stress ALT settings rather than SSALT.

In the SSALT setting, heterogeneity has received comparatively less attention. Most existing SSALT models that incorporate group effects generally assume that group membership is known in advance, such as the mixed-effects SSALT models of \cite{seo2017} and the nonlinear mixed-effect Weibull model proposed by \cite{wang2020}. These approaches are less suitable when heterogeneity emerges during testing and subgroup labels are unobserved. Beyond traditional lifetimes, finite mixtures have been adopted to complex censored data, as in \cite{lin2025} work on multivariate contaminated normal  distribution and censored regression models using the AECM algorithm for estimation and outlier detection. Although their work is not tailored to ALT or SSALT, it illustrates the flexibility of mixture-based approaches for censored data with latent structure. More recently, \cite{pal2023} and \cite{prajapati2024} developed finite mixture models for dependent competing risks under a progressive censoring  scheme, further demonstrating the usefulness of mixture models in a complex reliability framework.

A recent contribution by \cite{lu2025} proposed a simple heterogeneous SSALT ($h$-SSALT) model under Type-II censoring  based on exponential lifetimes and a cumulative exposure model formulation. Their model assumes that test units behave homogeneously at the initial stress level and split into latent subgroups after stress increases with parameter estimation carried out through an EM algorithm. Their simulation results showed that ignoring heterogeneity can lead to  overestimation of lifetime when extrapolating to NOC. Although the exponential model is convenient, it may be too restrictive in situations where the hazard rate is not constant. To address this limitation, we propose a failure-rate-based heterogeneous SSALT ($h$-SSALT) model with Weibull lifetimes under Type-II censoring. The proposed model assumes homogeneous behavior at the first stress level while allowing heterogeneity to emerge at the second stress level through a finite mixture of m latent subgroups, each characterized by its own failure behavior. This structure extends existing two group mixture models and provides greater flexibility for capturing diverse aging patterns that arise after stress elevation. Since subgroup membership is unobserved, parameter estimation is performed using a likelihood-based approach with an expectation-maximization (EM) algorithm, and interval estimation is carried out using the missing information identity of \cite{louis1982finding} with transformation-based confidence intervals.

The remainder of the paper is organised as follows. Section~\ref{sec:model_formulation} introduces the proposed $h$-SSALT model and the likelihood function under Type-II censoring. Section~\ref{sec:estimation} develops the maximum likelihood estimation framework, including a detailed derivation of the EM algorithm with closed-form M-step updates for all parameters, interval estimation via Louis' method, and transformation-based confidence intervals. Section~\ref{sec:simulation} presents an extensive Monte Carlo simulation study evaluating finite-sample performance through point estimation, a quantile-based comparison of the $h$-SSALT and homogeneous SSALT models across seven quantile levels, and a special case validation against the exponential model under CEM of \cite{lu2025}. Section~\ref{sec:data_analysis} illustrates the practical application of the proposed model through simulated and real data analysis examples.. Concluding remarks and directions for future work are given in Section~\ref{sec:conclusion}.

\section{Finite Mixture $h$-SSALT Model Formulation}
\label{sec:model_formulation}
Consider a simple SSALT experiment where $n$ test units are subjected simultaneously to a life testing experiment   at the initial 
stress level $s_1$,  and then the stress level is increased later to a higher stress condition $s_2$ at a pre-fixed stress changing time point $\tau$. Finally, the experiment is allowed to  terminate as soon as the $r$-th failure occurs (often referred to as Type-II censoring), where $r \leq n$ is fixed in advance. Let $T_{1:n} \leq T_{2:n} \leq \cdots \leq T_{N_1:n}<\tau \leq T_{N_1`+1:n} \leq \cdots T_{r:n}$ denote the ordered time-to-failure random variables, and let $N_1$ denotes the random  number of failures observed at the first stress level $s_1$ before the stress-changing  time point $\tau$. Furthermore, let the realizations of the random variables  $T_{i:n}$ and  $N_1$,
 be denoted by $t_{i:n}$ and  $n_1$  respectively. At time $\tau$, the stress is elevated to a higher level $s_2$, and the remaining units continue to be tested until the $r$-th failure is observed.

Under the Weibull lifetime model, the lifetime of a test unit follows a Weibull distribution with a common shape parameter $\alpha > 0$ and a stress-dependent scale parameter $\lambda > 0$. The probability density function (PDF) is $f(t;\alpha,\lambda) = \alpha\lambda\, 
t^{\alpha-1}\exp(-\lambda t^{\alpha})$, $t > 0$, where $\lambda = \lambda_1$ at stress $s_1$ and $\lambda = \lambda_{2j}$ for subgroup $j$ at stress $s_2$, as specified below.

While units may behave homogeneously at the initial stress level $s_1$, elevated stress $s_2$ on the second phase of the experiment can trigger diverse aging mechanisms, causing units from the same production batch to exhibit considerably different failure behavior. In order to capture such latent heterogeneity, the population at the second phase $s_2$ is modeled using a finite mixture of $m$ subgroups, each characterized by its own failure behavior, with mixing proportions $\pi_1, \pi_2, \ldots, \pi_{m-1}, \pi_m$, where $\pi_m = 1 - \sum\limits_{j=1}^{m-1} \pi_j$.

To describe the failure behavior across the  stress levels within a unified framework, we adopt the failure-rate based SSALT model (see  \cite{kateri2015}) which unlike the CE  model (see \citep{sedyakin1966, nelson1980})  considered by \cite{lu2025} specifies the overall failure-rate  function  in piecewise form. A mixture of hazard rates at the second stress level is a natural consequence of the finite mixture structure. Since each $j$-th cluster has a hazard $h_{2j}(t)$ with probability $\pi_j$, the total hazard at second phase $s_2$ is the mixture-weighted sum $\sum\limits_{j=1}^{m} \pi_j h_{2j}(t)$ \citep{mclachlan2000}. Under the failure-rate based $h$-SSALT model assumptions, the failure-rate function associated with the time-to-failure of the test units is given by
\[
h(t)=\begin{cases}
	h_{1}(t) &\text{if } 0<t\leq\tau,\\
	\displaystyle\sum_{j=1}^{m-1}\pi_j\,h_{2j}(t) + \left(1 - \displaystyle\sum_{j=1}^{m-1}\pi_j\right)\,h_{2m}(t) &\text{if } \tau<t<\infty,
\end{cases}
\]
where $h_{1}(\cdot)$ is the failure-rate function corresponding to the time-to-failure distribution at $s_1$ and $h_{2j}(\cdot)$ is the failure-rate function corresponding to the $j$-th cluster -specific time-to-failure distribution at $s_2$. 

We assume that the time-to-failure distribution of the test units corresponding to the stress level $s_1$ conforms to Weib($\alpha, \lambda$), while the time-to-failure distribution  of the test units for the $j-$  th cluster associated with the stress level $s_2$ conforms to Weib($\alpha, \lambda_{2j}$); $j = 1, \dots, m.$  The resulting failure-rate  function $h(t)$ then becomes:
\begin{equation}
	h(t)=\begin{cases}
		\alpha\lambda_{1}\,t^{\alpha-1} & \text{if }  0<t\leq\tau, \\
		\alpha\,t^{\alpha-1}\left\{\displaystyle\sum_{j=1}^{m-1}\pi_j\lambda_{2j} + \left(1 - \sum_{j=1}^{m-1}\pi_j\right)\lambda_{2m}\right\} &\text{if }  t>\tau.
	\end{cases}
\end{equation}

The associated cumulative hazard function (denoted by $H(t)$) can be obtained using the relation $H(t) = \displaystyle \int_{0}^{t} h(y)\,dy$  and is given by:
\begin{eqnarray}
    H(t) &=&
    \begin{cases}
        \lambda_{1} t^{\alpha} 
        & \text{if }  0 < t \leq \tau, \\
        \lambda_{1} \tau^{\alpha} +
        \left\{\displaystyle\sum_{j=1}^{m-1}
        \pi_j\lambda_{2j} + \left(1 - 
        \sum_{j=1}^{m-1}\pi_j\right)\lambda_{2m}
        \right\}(t^{\alpha} - \tau^{\alpha}) 
        &\text{if } t > \tau.
    \end{cases}
\end{eqnarray}
Hence, the compound CDF and the PDF associated with  the finite mixture failure-rate based $h$-SSALT model are respectively obtained as:
\begin{align}
\label{eq:cdf}
    G(t) = & 1 - \exp\!\left( - H(t) \right)& \nonumber \\
    = &  \begin{cases}
			1 - \exp\!\Big(-\lambda_1 t^{\alpha}\Big) & \text{if } 0 < t \le \tau, \\ 
			1 - \exp\!\Bigg(-\lambda_1 \tau^{\alpha} -\left\{\displaystyle\sum_{j=1}^{m-1}\pi_j\lambda_{2j} + \Big(1 - \sum_{j=1}^{m-1}\pi_j\Big)\lambda_{2m}\right\}(t^{\alpha}-\tau^{\alpha})\Bigg) & \text{if } t > \tau,
		\end{cases} &
\end{align}
and 
	\begin{eqnarray}
		g(t) = &
		\begin{cases}
			g_1(t) & \text{if } 0<t\le \tau,\\
			g_2(t) & \text{if } t>\tau,
		\end{cases} & 
        \label{Weib_PDF}
	\end{eqnarray}
     where
\begin{align*}
    g_1(t) = & \alpha \lambda_1 \, t^{\alpha - 1} \exp(-\lambda_1 t^{\alpha}), & \\
	g_2(t) = & \alpha t^{\alpha-1}\left\{\displaystyle\sum_{j=1}^{m-1}\pi_j\lambda_{2j} + \left(1 - \sum_{j=1}^{m-1}\pi_j\right)\lambda_{2m}\right\} & \\
    & \qquad\quad \times \exp\!\left(-\lambda_1 \tau^{\alpha}-\left\{\displaystyle\sum_{j=1}^{m-1}\pi_j\lambda_{2j} + \left(1 - \sum_{j=1}^{m-1}\pi_j\right)\lambda_{2m}\right\}(t^{\alpha}-\tau^{\alpha})\right). &
\end{align*}
% \textbf{Note that $S(t) = 1 - G(t) = \exp(-H(t))$ and is used in Section~\ref{sec:simulation} to define the $q$-th quantile $t_q$ as the solution to $1 - S(t_q) = q$.}

Let $n_i$  denote the observed number of  failures at the  stress level $s_i;\:i=1,2$, and  $\bar{n}_2 = n_1+n_2=r$  denote the total number of observed failures. Further, $\boldsymbol{\theta} = (\alpha, \lambda_1, \boldsymbol{\lambda_2}, 
\boldsymbol{\pi})$ defines the set of unknown model parameters, where 
$\boldsymbol{\lambda_2} = (\lambda_{21}, \lambda_{22}, \ldots, \lambda_{2m})$ 
and $\boldsymbol{\pi} = (\pi_1, \pi_2, \ldots, \pi_{m-1})$, with 
$\pi_m = 1 - \sum_{j=1}^{m-1}\pi_j$. 

%Further,  $\boldsymbol{\theta} = (\alpha, \lambda_1, \boldsymbol{\lambda_{2}}, \boldsymbol{\pi})$ defines the set of  unknown model parameters, where $\boldsymbol{\lambda_{2}}=(\lambda_{21},\lambda_{22},\ldots,\lambda_{2m}),\:\text{and}\:\boldsymbol{\pi}=(\pi_1,\pi_2,\ldots,\pi_{m-1})$. 
Now,  based on the observed Type-II censored data $\boldsymbol{\mathcal{D}} = (t_{1:n}, t_{2:n}, \ldots, t_{n_1:n}, t_{n_1+1:n}, \ldots,t_{r:n})$, the likelihood 
function is given by:
\begin{equation}
    L(\boldsymbol{\theta}\,\mid \boldsymbol{\mathcal{D}}) 
    \;\propto\;
    \left( \prod_{i=1}^{n_{1}} 
    g_{1}\!\left(t_{i:n}\right) \right)
    \left( \prod_{i=n_{1}+1}^{\bar{n}_{2}} 
    g_{2}\!\left(t_{i:n}\right) \right)
    \times \big[ S(t_{r:n}) \big]^{\,n-\bar{n}_{2}}.
\end{equation}
Substituting the Weibull-based densities $g_1$, 
$g_2$  from \eqref{Weib_PDF} and the survival function $S(.)=1-G(.),$ we have: 
\begin{eqnarray}
    %\begin{aligned}
    L(\boldsymbol{\theta}\mid\boldsymbol{\mathcal{D}})  
    &\propto & \left(\prod_{i=1}^{n_{1}}
    \alpha\lambda_{1}t^{\alpha-1}_{i:n}
    \exp\left(-\lambda_{1}t^{\alpha}_{i:n}
    \right)\right) \nonumber \\
    & &\times\prod_{i=n_{1}+1}^{\bar{n}_{2}} 
    \left[ \alpha t^{\alpha-1}_{i:n}
    \left\{ \sum_{j=1}^{m-1}\pi_j\lambda_{2j} 
    + \left( 1 - \sum_{j=1}^{m-1}\pi_j \right)
    \lambda_{2m} \right\} \right. \nonumber \\
    & & \left.\times\exp\left(-\lambda_{1}
    \tau^{\alpha} - \left\{
    \sum_{j=1}^{m-1}\pi_j\lambda_{2j} + 
    \left(1 - \sum_{j=1}^{m-1}\pi_j\right)
    \lambda_{2m}\right\}(t^{\alpha}_{i:n}-
    \tau^{\alpha})\right) \right] \nonumber \\
    & &\times\left[\exp\left(-\lambda_{1}\tau^{\alpha} 
    - \left\{\sum_{j=1}^{m-1}\pi_j\lambda_{2j} + 
    \left(1 - \sum_{j=1}^{m-1}\pi_j\right)
    \lambda_{2m}\right\}(t^{\alpha}_{r:n}-
    \tau^{\alpha})\right)\right]^{n-\bar{n}_{2}}.
   % \end{aligned}
    \label{eq:likelihood}
\end{eqnarray}

    \section{Parameter Estimation}
    \label{sec:estimation}
	\subsection{Maximum likelihood estimation}
	
	In this Section, we develop the maximum likelihood estimation (MLE) procedures for the proposed finite mixture failure-rate based $h$-SSALT model.  From the likelihood function obtained in equation \eqref{eq:likelihood} based on the observed Type-II censored   SSALT data $\boldsymbol{\mathcal{D}}$, the corresponding log-likelihood function (up to an additive constant) is obtained as follows:
	
	\begin{eqnarray}
\ell(\boldsymbol{\theta}\mid\boldsymbol{\mathcal{D}}) & = & n_{1}\ln\lambda_{1}+(\alpha-1)\sum_{i=1}^{n_{1}}\ln t_{i:n}-\lambda_{1}\left(\sum_{i=1}^{n_{1}}t^{\alpha}_{i:n}+(n-n_{1})\tau^{\alpha}\right)\nonumber \\
			& & + \:\bar{n}_{2}\ln\alpha+(\bar{n}_{2}-n_{1})\ln\left(\sum_{j=1}^{m-1}\pi_j\lambda_{2j} + \left(1 - \sum_{j=1}^{m-1}\pi_j\right)\lambda_{2m}\right)\nonumber \\
			& & + \:(\alpha-1)\sum_{i=n_{1}+1}^{\bar{n}_{2}}\ln t_{i:n}-\left\{\sum_{j=1}^{m-1}\pi_j\lambda_{2j} + \left(1 - \sum_{j=1}^{m-1}\pi_j\right)\lambda_{2m}\right\} \nonumber \\
			& & \times \left(\sum_{i=n_{1}+1}^{\bar{n}_{2}}t^{\alpha}_{i:n}+(n-\bar{n}_{2})t^{\alpha}_{r:n}-(n-n_{1})\tau^{\alpha}.\right)
            \label{org_log_lik}
		\end{eqnarray}
	
	It is to note here that  the log-likelihood function $\ell(\boldsymbol{\theta}\mid\boldsymbol{\mathcal{D}})$ can be decomposed into three components as follows:
	\[
	\ell(\boldsymbol{\theta} \mid \boldsymbol{\mathcal{D}}) = \ell^{(1)}(\alpha, \lambda_1 \mid \boldsymbol{\mathcal{D}}) + \ell^{(2)}(\alpha, \boldsymbol{\lambda_{2}}, \boldsymbol{\pi} \mid \boldsymbol{\mathcal{D}}) + \ell^{(3)}(\alpha \mid \boldsymbol{\mathcal{D}}),
	\]
	
	where
    \begin{eqnarray}
        \ell^{(1)}(\alpha,\lambda_{1}\mid\boldsymbol{\mathcal{D}}) &= & n_{1}\ln\lambda_{1}-\lambda_{1}\left(\sum_{i=1}^{n_{1}}t_{i:n}^{\alpha}+(n-n_{1})\tau^{\alpha}\right),  \nonumber \\
        \ell^{(2)}(\alpha,\boldsymbol{\lambda_{2}}, \boldsymbol{\pi} \mid\boldsymbol{\mathcal{D}}) & = & (\bar{n}_{2}-n_{1})\ln\left(\sum_{j=1}^{m-1}\pi_j\lambda_{2j} + \left(1 - \sum_{j=1}^{m-1}\pi_j\right)\lambda_{2m}\right) \nonumber \\
		& & -\left\{\sum_{j=1}^{m-1}\pi_j\lambda_{2j} + \left(1 - \sum_{j=1}^{m-1}\pi_j\right)\lambda_{2m}\right\} \left(\sum_{i=n_{1}+1}^{\bar{n}_{2}}t^{\alpha}_{i:n}+(n-\bar{n}_{2})t^{\alpha}_{r:n}-(n-n_{1})\tau^{\alpha}\right),  \nonumber  \\
	  \ell^{(3)}(\alpha\mid\boldsymbol{\mathcal{D}}) & = & \bar{n}_{2}\ln\alpha+(\alpha-1)\sum_{i=1}^{\bar{n}_{2}}\ln t_{i:n}. \nonumber  
	\end{eqnarray} 
	
	The MLEs of the unknown model parameters (denoted by $\widehat{\boldsymbol{\theta}})$
	can be   obtained by maximizing the  log-likelihood function given by \eqref{org_log_lik}. Differentiating partially the  log-likelihood function $\ell(\boldsymbol{\theta}\mid\boldsymbol{\mathcal{D}})$ with respect to  $\lambda_1$, one can  have the explicit solution of $\lambda_1$ conditional on $\alpha$  as follows: 
	\begin{eqnarray}
		\frac{\partial \ell}{\partial \lambda_1} 
		&=& \frac{n_1}{\lambda_1} 
		- \left(\sum_{i=1}^{n_1} t_{i:n}^{\alpha} + (n-n_1) \tau^{\alpha}\right) = 0 \nonumber \\
		\implies  \hat{\lambda}_1^{(\alpha)} 
		&= & \frac{n_1}{\sum_{i=1}^{n_1} t_{i:n}^{\alpha} + (n-n_1) \tau^{\alpha}}.
        \label{lambda_1}
	\end{eqnarray}

However, from the remaining  $2m$ likelihood equations, it is not possible to obtain explicit solutions for any one of the remaining model parameters. In particular, $\widehat{\alpha}$ must be obtained first, after which $\widehat{\lambda}_1$ follows directly as $\widehat{\lambda}_1^{
(\widehat{\alpha})}$ from equation~\eqref{lambda_1}. Thus,  $\widehat{\boldsymbol{\theta}}$ can only be obtained by solving a $2m$-dimensional maximization problem.  Additionally, the MLEs of any of the model parameters   cannot be
 obtained in closed form and the complexity of the maximization  problem grows as $m$ (the number of clusters) increases.

%Note that this expression is derived conditional on $\alpha$; 
%since $\alpha$ is unknown in practice, it is estimated jointly 
%with all other parameters using the EM algorithm described in 
%Section~\ref{sec:estimation}, where $\lambda_1^{(k+1)}$ is 
%updated using $\alpha^{(k+1)}$ at each iteration.

\begin{remark}
The hazard function at the second stress level depends on the mixture parameters $(\pi_j, \lambda_{2j})$ only through the 
weighted sum $\bar{\lambda}_2 = \sum_{j=1}^{m} \pi_j \lambda_{2j}$. Consequently, the observed-data likelihood is 
a function of $\bar{\lambda}_2$ alone, and the individual components $\pi_j$ and $\lambda_{2j}$ are not separately 
identifiable from the observed data without additional structure. Identifiability of the individual components is 
achieved through the complete data likelihood, which the EM algorithm exploits using the latent group membership indicators 
$W_{ij}$, introduced in Section~\ref{subsec:EM}. In practice, reliable separation of the components requires that 
the scale parameters $\lambda_{2j}$ be sufficiently distinct; in the simulation study of Section~\ref{sec:simulation}, the 
values $\lambda_{21} = 0.10$ and $\lambda_{22} = 1.0$ differ by a factor of $10$, ensuring practical identifiability. 
See \cite{mclachlan2000} for a general treatment of identifiability in finite mixture models.
\end{remark}

From the log-likelihood function in equation \eqref{eq:likelihood}, it is evident that the parameters $(\alpha, \lambda_{2j}, \pi_j)$ appear through the term $\ln\!\left(\sum_{j=1}^{m} \pi_j \lambda_{2j}\right)$, which arises due to the mixture structure at the second stress level. This logarithm of a sum prevents separation of the likelihood into simpler components and leads to highly nonlinear likelihood equations. Furthermore, the presence of unobserved subgroup memberships and Type-II censoring makes the data incomplete, adding additional complexity to the likelihood function. As a result, the normal equations do not admit closed-form solutions for $(\alpha, \lambda_{2j}, \pi_j)$. Mixture likelihoods are also known to exhibit multiple local maxima \citep{mclachlan2000}, making direct numerical maximisation unreliable in general. The EM algorithm \citep{dempster1977} is therefore the natural choice here, exploiting the incomplete data structure by treating the latent group membership indicators as missing data, for which the complete data log-likelihood is tractable while direct maximisation of the observed-data likelihood is not.

	\subsection{EM Algorithm}
    \label{subsec:EM}
	In this Section, we develop an expectation-maximization (EM) algorithm for estimating the unknown model  parameter vector $\boldsymbol{\theta}.$ %where the observed lifetimes are assumed to arise from a mixture of $m$ left-truncated Weibull distributions corresponding to latent subpopulations.
    In addition, the data are subject to Type-II censoring, where only a subset of failure times is fully observed and the remaining lifetimes are censored after a pre-specified number of failures. Due to these factors, unobserved component memberships and censored observations, the likelihood function becomes complex and does not admit closed form solutions for the maximum likelihood estimates. 
	The  EM algorithm efficiently addresses incomplete data by treating unknown component indicators as latent variables. It alternates between estimating the expected missing values (E-step) and maximizing the expected complete data 
	log-likelihood (M-step) until convergence, yielding effective estimates of the mixture probabilities and Weibull model parameters.

	Let $T_i$ be  the random variable denoting the time-to-failure  of the $i$-th unit  ($i = n_1+1, \dots, n$), and let the random censoring time be defined by $C = T_{r:n}$. The  observed time-to-failure   (or possibly censored) random variable  and the associated  censoring indicator random variable are then defined by 
\[
	T_{i:n}^* = \min(T_{i:n}, C) \text{ and }
	\Delta_i =
	\begin{cases}
		1, & \text{if } T_{i:n} \leq C ~(\text{time-to-failure is observed})\\
		0, &  \text{otherwise,}
	\end{cases},
\] respectively.   The binary indicator random variable $\Delta_i$ denotes whether the $i$-th lifetime random variable $T_{i:n}$ is observed or censored depending on whether its failure order exceeds a prefixed threshold $r$. The observed data can thus be represented as the vectors
	$\boldsymbol{t}^* = (t_{n_1+1}^*, \dots, t_n^*)$ and 
	$\boldsymbol{\delta} = (\delta_{n_1+1}, \dots, \delta_n)$. 
	Here, $t^*_{i:n}$ and $\delta_{i}$ are realizations of the random variables 
	$T^*_{i:n}$ and $\Delta_i$, respectively.
	Let us now introduce latent component indicator variables (denoted by  $W_{ij}$ ) as follows: 
	\[
	W_{ij} =
	\begin{cases}
		1, & \text{if } t_i^* \text{ belongs to subgroup } j\\
		0, & \text{otherwise,}
	\end{cases}, \quad j=1, 2,\ldots, m,
	\]
	such  that $\displaystyle\sum_{j=1}^m W_{ij} = 1$  for $i=n_1+1, \dots, n$. It is noteworthy that the complete dataset is then $(\boldsymbol{t}^*, \boldsymbol{\delta}, \boldsymbol{w})$, while the observed dataset  is $(\boldsymbol{t}^*, \boldsymbol{\delta})$. Here, $\boldsymbol{w}=(\boldsymbol{w}_{n_1+1},\boldsymbol{w}_{n_2+1},\ldots,\boldsymbol{w}_n),$ where,  $\boldsymbol{w}_i = (w_{i1}, \ldots, w_{im})$  denote the latent group membership indicators for the $i$-th observation $(t_i^*, \delta_i)$. Assuming that the censoring mechanism is independent of the group memberships, the contribution of the $i$-th observation to the complete data likelihood at the second stress level  is given by: 
	\begin{equation}
		L_i(\boldsymbol{\theta_2 }\mid t_i^*, \delta_i, w_i)
		\propto
		\prod_{j=1}^{m}
		\left[ \pi_j\, f_{X_j}^{\text{trunc}}(t_{i:n}\mid\,w_{i},\alpha, \lambda_{2j})^{\delta_i}\,S_{X_j}^{\text{trunc}}(t_{r:n}\mid\,w_{i},\alpha, \lambda_{2j})^{1-\delta_i} \right]^{w_{ij}},
	\end{equation}
	where $f_{X_j}^{\text{trunc}}(.)$ is the PDF associated with  the left-truncated Weib$(\alpha, \lambda_{2j})$ random variable  
	and $S_{X_j}^{\text{trunc}}(.)$ is the corresponding survival function, i.e.,
	\begin{align*}
		f_{X_j}^{\text{trunc}}(t_{i:n}\mid\alpha,\lambda_{2j}) 
		&= \alpha\lambda_{2j}\, t_{i:n}^{\alpha - 1}\, e^{-\lambda_{2j}(t_{i:n}^{\alpha}-\tau^{\alpha})},\\
		S_{X_j}^{\text{trunc}}(t_{r:n}\mid\alpha, \lambda_{2j})
		&= e^{-\lambda_{2j}(t_{r:n}^{\alpha}-\tau^{\alpha})},
	\end{align*}
	and $\boldsymbol{\theta}_2 = (\alpha, \boldsymbol{\lambda_2}, \boldsymbol{\pi})$ 
denotes the parameter sub-vector associated with the second stress level. The mixture proportions $\pi_j = P(W_{ij} = 1)$ for $j = 1, 2, \ldots, m$ represent the group membership probabilities, with $\displaystyle\sum_{j=1}^m \pi_j = 1$.

%		with $f_{X_j}^{\text{trunc}}$ being the PDF of left truncated weibull distribution and $S_{X_j}^{\text{trunc}}$ be the corresponding survival function i.e., $f_{X_j}^{\text{trunc}}(t_{i:n}\mid\alpha_,\lambda_{2j}) 
%		= \alpha\lambda_{2j} t^{\alpha - 1} e^{-\lambda_{2j}( t_i^{\alpha}-\tau^{\alpha})},S_{X_j}^{\text{trunc}}(t_{r:n}\mid\alpha, \lambda_{2j})
%		= e^{-\lambda_{2j} (t_r^{\alpha}-\tau^{\alpha})}\ and $ \boldsymbol{\theta}=$(\alpha,\lambda_{2j},\pi_j)$, the set of model parameters at second stress to be estimated. Also,the mixture proportions $\pi_j= P(W_{ij} = 1)$ for $j = 1, 2, \ldots, m$ represent the group membership probabilities, with $\displaystyle\sum_{j=1}^m \pi_j = 1$.

		Thus, the complete data log-likelihood over all observations $i=n_1+1, \dots, n$ i.e., in second stress level becomes 
		\begin{equation}
        \label{eq_10}
			\ell_c^{(2)}(\boldsymbol{\theta_2)} = \sum_{i=n_1+1}^{n} \sum_{j=1}^{m} W_{ij} \Big[ 
			\ln \pi_j + \delta_i \big( \ln \alpha + \ln \lambda_{2j} + (\alpha-1) \ln t_{i:n} 
			- \lambda_{2j} ( t_{i:n}^{\alpha} - \tau^{\alpha}) \big) 
			- (1-\delta_i)\, \lambda_{2j} (t_{r:n}^{\alpha} - \tau^{\alpha}) \Big].
		\end{equation}

\paragraph{\textbf{\textit{E-step}}:}

In the E-step, we compute the expectation of the complete log-likelihood function with respect to the distribution of the unobserved latent variables $W_{ij}$, given the current estimates of the parameters and the observed data $O= \{(t_i^{*}, \delta_i);~ i = n_1+1, \ldots, n\}$.
It is clear that the $W_{ij}$’s are Bernoulli random variables in the complete log-likelihood, and hence it is sufficient to evaluate
$\eta_{ij}^{(k)} = E(W_{ij} \mid \boldsymbol{\theta_2}^{(k)}, O), \quad i = n_1+1, \dots, n; j = 1, 2, \dots, m.$ Where, $\boldsymbol{\theta_2}^{(k)} = (\alpha^{(k)},\boldsymbol{\lambda_{2}}^{(k)}, \boldsymbol{\pi}^{(k)})$ denotes the vector of parameter estimates at the $k$-th iteration step, with $\displaystyle\sum_{j=1}^m \pi_j = 1$.
Thus,the posterior probability $\eta_{ij}^{(k)}$\ corresponding to the $i$-th censored failure observation belongs to the $j$-th group is as follows, 
\begin{align}
\label{eq_11}
	\eta_{ij}^{(k)} = E(W_{ij} \mid \boldsymbol{\theta_2}^{(k)}, O) 
	&= P(W_{ij}=1 \mid t_i^{*}, \delta_i, \boldsymbol{\theta_2}^{(k)}) \notag \\
	&= \left[
	\frac{
		\pi_j^{(k)} f_{X_j}\!\left(t_{i:n} \mid \alpha^{(k)}, \lambda_{2j}^{(k)}\right)
	}{
		\displaystyle\sum_{j=1}^{m} \pi_j^{(k)} f_{X_j}\!\left(t_{i:n} \mid \alpha^{(k)}, \lambda_{2j}^{(k)}\right)
	}
	\right]^{\delta_i}
	\left[
	\frac{
		\pi_j^{(k)} S_{X_j}\!\left(t_{r:n} \mid \alpha^{(k)}, \lambda_{2j}^{(k)}\right)
	}{
		\displaystyle\sum_{j=1}^{m} \pi_j^{(k)} S_{X_j}\!\left(t_{r:n} \mid \alpha^{(k)}, \lambda_{2j}^{(k)}\right)
	}
	\right]^{1-\delta_i}.
\end{align}
Thus, in the E-step, the missing indicators $W_{ij}$ in the complete log-likelihood 
function are replaced by their posterior $\eta_{ij}^{(k)}$ computed above. Let 
$Q(\boldsymbol{\theta_2}\mid \boldsymbol{\theta_2}^{(k)})$ denote the conditional expectation 
of the complete log-likelihood at the second stress level, i.e.,
\begin{equation}
	Q(\boldsymbol{\theta_2} \mid \boldsymbol{\theta_2}^{(k)}) = 
	\mathbb{E}\!\left[ \ell_{c}^{(2)}(\boldsymbol{\theta_2}) \,\middle|\, 
	\boldsymbol{t}^*, \boldsymbol{\delta}, \boldsymbol{\theta_2}^{(k)} \right].
\end{equation}
Hence,
\begin{align}
    \label{eq_13}
	Q(\boldsymbol{\theta_2}\mid \boldsymbol{\theta_2}^{(k)})=\sum_{i=n_1+1}^{n} \sum_{j=1}^{m} \eta_{ij}^{(k)} \Big[ & \ln \pi_j + \delta_i \big( \ln \alpha + \ln \lambda_{2j} + (\alpha - 1) \ln t_{i:n} - \lambda_{2j} (t_{i:n}^{\alpha} - \tau^{\alpha}) \big) & \nonumber \\
    & - (1-\delta_i) \lambda_{2j} (t_{r:n}^{\alpha} - \tau^{\alpha}) \Big].&
\end{align} 

% Now, let $\boldsymbol{\theta }= (\alpha, \lambda_1,{\boldsymbol\lambda_{2}},{\boldsymbol\pi })$ be the complete parameter vector to be estimated in the experiment. 
Since $\alpha$  is common to both stress levels, estimating it separately for each stress level would break the internal dependence structure of the step-stress model. Consequently,the conditional expectation of the complete data log-likelihood is expressed through a unified Q-function that incorporates information from both stresses,i.e.,
\[
Q(\boldsymbol{\theta} \mid \boldsymbol{\theta}^{(k)}) = \ell_c^{(1)}(\alpha, \lambda_1) + Q^{}(\boldsymbol{\theta_2} \mid \boldsymbol{\theta_2}^{(k)}),
\]
where $\ell_c^{(1)}(\alpha,\lambda_1)$ is the log-likelihood contribution of the first stress level, which depends only on $(\alpha,\lambda_1)$.

\paragraph{\textbf{\textit{M-step}}:}
In the M-step of the EM algorithm, the parameter $\boldsymbol{\theta}$ is updated by maximizing 
the expected complete log-likelihood function $Q(\boldsymbol{\theta} \mid \boldsymbol{\theta}^{(k)})$ with respect to $\boldsymbol{\theta}$, using the current parameter estimate $\boldsymbol{\theta}^{(k)}= (\alpha^{(k)}, \lambda_1^{(k)}, {\boldsymbol\lambda_{2}}^{(k)},{\boldsymbol\pi}^{(k)} )$. The updated estimate is given by
\[
\boldsymbol{\theta}^{(k+1)} = \argmax_{\boldsymbol{\Theta}}\, Q(\boldsymbol{\theta}\mid\boldsymbol{\theta}^{(k)}).
\]
where $\boldsymbol{\Theta} = \bigl\{\,\boldsymbol{\theta} \in \mathbb{R}^{2m} : 
\alpha > 0,\; \lambda_1 > 0,\; \lambda_{2j} > 0,\; \pi_j > 0 
\text{ for } j = 1,\ldots,m,\; 
{\textstyle\sum_{j=1}^{m}}\pi_j = 1\,\bigr\}$ representing the parameter space.

To simplify the formulation, the censored component in equation \eqref{eq_13} is not treated separately in the subsequent derivation but is incorporated through the notation \( t_{i:n}^{*} \). Specifically, \( t_{i:n}^{*} \) is defined as  
\[
t_{i:n}^{*} =
\begin{cases}
t_{i:n}, & \text{for } n_1 + 1 \le i \le r \\[6pt]
t_{r:n}, & \text{for } r + 1 \le i \le n
\end{cases},
\]
which implies that for censored data, the observation is replaced by the last recorded failure time \( t_{r:n} \). A Lagrange multiplier \( \psi \) is introduced to satisfy the constraint \(\displaystyle \sum_{j=1}^{m} \pi_j = 1 \). With this adjustment, the above equation can be reformulated as  
\begin{align}
\boldsymbol{\theta}^{(k+1)} &= \argmax_{\boldsymbol{\Theta}}\Bigg[ n_{1}(\ln\alpha+\ln\lambda_{1})+(\alpha-1)\sum_{i=1}^{n_{1}}\ln t_{i:n}-\lambda_{1}\left(\sum_{i=1}^{n_{1}}t^{\alpha}_{i:n}+(n-n_{1})\tau^{\alpha}\right)\nonumber\\
&\quad\quad
+\sum_{i=n_1+1}^{n} \sum_{j=1}^{m} \eta_{ij}^{(k)} \left\{
\ln \pi_j + \delta_i \big( \ln \alpha + \ln \lambda_{2j} + (\alpha - 1) \ln t_{i:n} \big) 
- \lambda_{2j} \big( (t_{i:n}^{*})^{\alpha} - \tau^{\alpha} \big)
\right\} \nonumber \\
&\quad\quad\quad\quad + \psi \left( \sum_{j=1}^{m} \pi_j - 1 \right)\Bigg].
\end{align}

At the $(k+1)$-th iteration of the EM algorithm, the parameter estimates 
$( \alpha^{(k+1)},\lambda_1^{(k+1)},{\boldsymbol\lambda_2}^{(k+1)} ,{\boldsymbol\pi}^{(k+1)})$ are obtained by maximizing the 
expected complete data log-likelihood $Q(\boldsymbol{\theta}; \boldsymbol{\theta}^{(k)})$ 
with respect to each parameter. The normal equations obtained by taking the first order derivative of $Q(\boldsymbol{\theta}; \boldsymbol{\theta}^{(k)})$ with respect to the model parameters and equating them to zero are given by
\begin{align}
	\frac{\partial Q(\boldsymbol{\theta}\mid \boldsymbol{\theta}^{(k)})}{\partial \alpha} 
	& =
	\left[\frac{n_1}{\alpha}+\sum_{i=1}^{n_1} \ln t_{i:n} 
	- \lambda_1^{(k)} \left( \sum_{i=1}^{n_1} t_{i:n}^{\alpha} \ln t_{i:n} 
	+ (n-n_1)\tau^{\alpha} \ln \tau \right) \right] \nonumber\\
	&\quad
	+\sum_{i=n_{1}+1}^{n} \sum_{j=1}^{m} \eta_{ij}^{(k)} 
	\left[ \frac{\delta_{i}}{\alpha} + \delta_{i} \ln t_{i:n} 
	- \lambda_{2j}^{(k)} \left( (t_{i:n}^{*})^{\alpha} \ln t_{i:n}^{*} 
	- \tau^{\alpha} \ln \tau \right) \right] = 0, \\
	\frac{\partial Q(\boldsymbol{\theta}\mid \boldsymbol{\theta}^{(k)})}{\partial \lambda_1} 
	&=
	\frac{n_1}{\lambda_1} 
	- \left(\sum_{i=1}^{n_1} t_{i:n}^{\alpha} + (n-n_1) \tau^{\alpha}\right) = 0, \\
    \frac{\partial Q(\boldsymbol{\theta}\mid \boldsymbol{\theta}^{(k)})}{\partial \pi_j} 
	&=
	\sum_{i=n_1+1}^{n} 
	\frac{\eta_{ij}^{(k)}}{\pi_j}
	+ \psi
	= 0, \quad j=1,2,\ldots,m,
\end{align} and
\begin{align}
	\frac{\partial Q(\boldsymbol{\theta}\mid \boldsymbol{\theta}^{(k)})}{\partial \lambda_{2j}} 
	&=
	\sum_{i=n_1+1}^{n} 
	\eta_{ij}^{(k)}
	\left[
	\frac{\delta_i}{\lambda_{2j}}
	-
	\left( (t_{i:n}^{*})^{\alpha} - \tau^{\alpha} \right)
	\right]
	= 0, \quad j=1,2,\ldots,m.
\end{align}

This corresponds to the M-step of the EM algorithm. Since the above equation $\frac{\partial Q(\boldsymbol{\theta}; \boldsymbol{\theta}^{(k)})}{\partial \alpha}=0$
is not in closed form, the $\alpha^{(k+1)}$ update is obtained by solving the above non linear equation by iterative procedure and the resulting closed form update equations for other parameter are given by  
\begin{align}
& \lambda_{1}^{(k+1)} = \frac{n_{1}}{\displaystyle\sum_{i=1}^{n_{1}} t_{i:n}^{\alpha^{(k+1)}} + (n - n_{1}) \tau^{\alpha^{(k+1)}}}, \\
& \pi_j^{(k+1)} = \frac{\displaystyle\sum_{i=n_1+1}^{n} \eta_{ij}^{(k)}}{n - n_1}, 
\quad j=1, 2, \dots, m, 
\end{align} and
\begin{align}
& \lambda_{2j}^{(k+1)} = \frac{\displaystyle\sum_{i=n_{1}+1}^{n} \eta_{ij}^{(k)} \delta_{i}}{\displaystyle\sum_{i=n_{1}+1}^{n} \eta_{ij}^{(k)} \left( (t_{i:n}^*)^{\alpha^{(k+1)}} - \tau^{\alpha^{(k+1)}} \right)}, 
\quad j=1, 2,\dots, m.
\end{align}

Here $t_{i:n}^{*}$ represents the observed or censored lifetime, and $\delta_i$ is the censoring indicator, with ($\delta_i = 1$) if the $i$-th lifetime is observed and $\delta_i = 0$ for censored observations.

These updates are applied iteratively until convergence. The resulting values of $\alpha$,$\lambda_{1}$,${\boldsymbol\lambda_2}$ and ${\boldsymbol\pi}$ at convergence are taken as the EM based maximum likelihood estimates (MLE's) of the model parameters. In 
practice, multiple starting values are recommended to guard against local maxima, following the strategy of \cite{lu2025}.
The complete iterative procedure is summarised in Algorithm~\ref{algo:1}.

\begin{breakablealgorithm}
\caption{: EM Algorithm   }
\begin{algorithmic}[1]
	%\Procedure{Roy}{$a,b$}       \Comment{This is a test}
	\State Set the initial guesses for the parameters as $\pi_1^{(0)}, \ldots, \pi_{m}^{(0)}, \alpha^{(0)}, \lambda_1^{(0)}, \lambda_{21}^{(0)}, \ldots, \lambda_{2m}^{(0)}$.
	\State For the $(k)$-th iteration, compute the posterior probability $\eta_{ij}^{(k)}$ that the $i$-th unit belongs to the $j$-th subgroup for $i = n_1+1, \dots, n$ and $j = 1, \dots, m$ as
	\bea
	\eta_{ij}^{(k)} = \frac{\pi_{j}^{(k)}f_{X_{j}}(t_{i:n}^*|\alpha^{(k)},\lambda_{2j}^{(k)})^{\delta_i} S_{X_{j}}(t_{i:n}^*|\alpha^{(k)},\lambda_{2j}^{(k)})^{1-\delta_i}}{\displaystyle\sum_{j=1}^{m} \left[\pi_{j}^{(k)}f_{X_{j}}(t_{i:n}^*|\alpha^{(k)},\lambda_{2j}^{(k)})^{\delta_i} S_{X_{j}}(t_{i:n}^*|\alpha^{(k)},\lambda_{2j}^{(k)})^{1-\delta_i}\right]},\nonumber
	\eea
	where $t_{i:n}^*$ represents the observed or censored times and $\delta_i $ is censoring indicator.
	\State  Obtain the improved estimates for mixing weights $\pi_{j}$ at the $(k+1)$-th iterate as
	\bea
	\pi_{j}^{(k+1)} = \frac{\displaystyle\sum_{i=n_{1}+1}^{n}\eta_{ij}^{(k)}}{n-n_{1}}, \quad j=1,2,\dots,m.\nonumber
	\eea
	\State Obtain the common shape parameter $\alpha^{(k+1)}$ by numerically solving the following non linear equation 
	\bea
	\begin{aligned}
		&\left[ \frac{n_1}{\alpha}+\sum_{i=1}^{n_{1}}\ln t_{i:n} - \lambda_{1}^{(k)}\left(\displaystyle\sum_{i=1}^{n_{1}}t_{i:n}^{\alpha}\ln t_{i:n} + (n-n_{1})\tau^{\alpha}\ln \tau\right)\right]\\
		&\quad\quad
		+\sum_{i=n_{1}+1}^{n}\sum_{j=1}^{m}\eta_{ij}^{(k)}\left[\frac{\delta_{i}}{\alpha} + \delta_{i}\ln t_{i:n} - \lambda_{2j}^{(k)}\left((t_{i:n}^{*})^{\alpha}\ln t_{i:n}^{*} - \tau^{\alpha}\ln \tau\right)\right] = 0.\nonumber
	\end{aligned}
	\eea 
	\State Using the newly obtained $\alpha^{(k+1)}$ compute, the updates for $\lambda_1$ and $\lambda_{2j}$ at the $(k+1)$-th iterate as follows:
	\begin{align*}
        \lambda_{1}^{(k+1)} & = \frac{n_{1}}{\displaystyle\sum_{i=1}^{n_{1}}t_{i:n}^{\alpha^{(k+1)}}+(n-n_{1})\tau^{\alpha^{(k+1)}}}, &\\
        \lambda_{2j}^{(k+1)} & = \frac{\sum_{i=n_{1}+1}^{n}\eta_{ij}^{(k)}\delta_{i}}{\displaystyle\sum_{i=n_{1}+1}^{n}\eta_{ij}^{(k)}\left((t_{i:n}^{*})^{\alpha^{(k+1)}}-\tau^{\alpha^{(k+1)}}\right)}, \quad j=1,2,\dots,m.&
    \end{align*}
	\State Check the convergence of $(\pi_{j}^{(k+1)}, \lambda_1^{(k+1)}, \lambda_{2j}^{(k+1)}, \alpha^{(k+1)})$. If the convergence is met, terminate  the iteration, otherwise go back to Step 2. 
\end{algorithmic}
\label{algo:1}
\end{breakablealgorithm} \vspace{1em}

To facilitate the simulation study presented in Section \ref{sec:simulation}, we describes the procedure for generating a heterogeneous SSALT dataset with Type-II censoring in Algorithm \ref{algo:2}, which is used to evaluate the performance of the proposed model.

\begin{breakablealgorithm}
\caption{: Generating heterogenous SSALT with Type-II censored dataset }
\begin{algorithmic}[1]
	%\Procedure{Roy}{$a,b$}       \Comment{This is a test}
	\State Given parameter values : $\pi_1,\:\pi_2,\ldots,\pi_{m},\:\alpha,\:\lambda_1,\:\lambda_{21},\ldots,\lambda_{2m}$ and  set $\pi_0=0.$
	\State Generate $n$ independent random variables from a Uniform distribution $U(0, 1)$ and Sort them to obtain the order statistics $U_{1:n}, U_{2:n}, \dots, U_{n:n}$.
	\State Find $n_1$ such that $U_{n_1:n}  \le 1 - \exp\{-\lambda_1 \tau^{{\alpha}} \}< U_{n_1+1:n}$.
	For $i = 1, \dots, n_1$, calculate the sorted failure times at the first stress level as follows :
	\begin{equation*}
		T_{i:n} = \left[ -\frac{1}{{\lambda}_1} \ln(1 - U_{i:n}) \right]^{\frac{1}{{\alpha}}}.
	\end{equation*}
	
	\State Generate the number of failures for each subgroup at the second stress level. Draw a random vector $(n_{21}, n_{22}, \dots, n_{2m})$ from a Multinomial distribution with sample size $n - n_1$ and probability parameters $({\pi}_1, {\pi}_2, \dots,{\pi}_m)$, such that$\displaystyle\sum_{j=1}^{m} n_{2j} = n - n_1$.
	\State For each subgroup $j = 1, \dots, m$ generate a data sample of size $n_{2j}$ from $U(0, 1)$ and sort them to obtain $U^{(j)}_{1}, \dots, U^{(j)}_{n_{2j}}$. Now calculate the sorted failure times for this subgroup using the inverse transformation of CDF for $t > \tau$ as
	\begin{equation*}
		T_{k}^{(j)} =
		\left(
		\tau^{\alpha} - \frac{\ln\!\left(1 - U_{k:n_{2j}}\right)}{\lambda_{2j}}
		\right)^{\frac{1}{\alpha}},
	\end{equation*}
	where $k = 1, \dots, n_{2j}$.
	\State Combine the generated failure times from all $m$ subgroups $\{T^{(1)}, \dots, T^{(m)}\}$ into a single set of size $n - n_1$ and sort these combined values.
	\State Select the first $r - n_1$ sorted values from the combined set. These constitute the observed order statistics for the second stress level, denoted as $T_{n_1+1:n}, \dots, T_{r:n}$.
	\State Combine these with the $n_1$ observations from step(3) to form the final Type-II censored sample dataset 
	\begin{equation*}
		\mathcal{D} = \{ t_{1:n}, \dots, t_{n_1:n}, t_{n_1+1:n}, \dots, t_{r:n} \}.
	\end{equation*}
\end{algorithmic}
\label{algo:2}
\end{breakablealgorithm} \vspace{1em}

\noindent The generated dataset $\mathcal{D}$ serves as input to Algorithm~\ref{algo:1} for parameter estimation. The simulation results are discussed in detail in the following section.

\subsection{Interval Estimation using the  Louis’ Method}

Direct evaluation of the observed Fisher information matrix for the proposed heterogeneous
step-stress accelerated life testing ($h$-SSALT) model is analytically intractable due to the
presence of latent subgroup memberships and Type-II censoring. To overcome this difficulty, we adopt the method proposed by \cite{louis1982finding}, which is based on the missing information principle introduced by \cite{OrchardWoodbury1972}. This approach has been widely used in accelerated life testing and lifetime data (see, for example, \cite{BalakrishnanMitra2011,BalakrishnanLing2013}).
 According to this principle, the observed  information matrix can be expressed as
\[
\text{observed information}
=
\text{complete information}
-
\text{missing information}.
\]
 Let ${t_{i:n}^{*}} = (t_{n_1+1:n}^*, \ldots, t_{n:n}^*)$ and 
\(\boldsymbol{\delta} = (\delta_{n_1+1}, \ldots, \delta_n)\) denote the observed data corresponding to the second stress level. The latent component indicators $\mathbf{W}$ are defined in Section \ref{subsec:EM}; consequently, the complete data corresponding to the second stress level can be represented by $(\mathbf{t}^*, \boldsymbol{\delta}, \mathbf{W})$.

The complete-data log-likelihood can then be expressed as

\[
\ell_c(\boldsymbol{\theta};
\mathbf t^{*},
\boldsymbol{\delta},
\mathbf W)
=
\ell_c^{(1)}(\alpha,\lambda_1)
+
\ell_c^{(2)}(\boldsymbol{\theta}_2;
\mathbf t^{*},
\boldsymbol{\delta},
\mathbf W),
\]

where \(\ell_c^{(1)}(\alpha,\lambda_1)\) represents the contribution from the fully observed first stress level and \(\ell_c^{(2)}(\boldsymbol{\theta}_2;\mathbf t^{*},\boldsymbol{\delta},\mathbf W)\) corresponds to the contribution from the second stress level involving the latent subgroup memberships. 

From the complete-data log-likelihood
$\ell_c(\boldsymbol{\theta};\mathbf{t}^{*},\boldsymbol{\delta},\mathbf{W})$,
the complete-data score vector and information matrix are respectively defined as
\[
S_c(\boldsymbol{\theta};
\mathbf t^{*},
\boldsymbol{\delta},
\mathbf W)
=
\frac{\partial \ell_c(\boldsymbol{\theta};\mathbf{t}^{*},\boldsymbol{\delta},\mathbf{W})}
{\partial \boldsymbol{\theta}},
\qquad
I_c(\boldsymbol{\theta};
\mathbf t^{*},
\boldsymbol{\delta},
\mathbf W)
=
-\frac{\partial^{2}\ell_c(\boldsymbol{\theta};\mathbf{t}^{*},\boldsymbol{\delta},\mathbf{W})}
{\partial\boldsymbol{\theta}\,\partial\boldsymbol{\theta}^{\top}}.
\]
In addition, \(S({\boldsymbol\theta}; \mathbf{t}^*, \boldsymbol{\delta})\) denote the  score vector of the observed-data log-likelihood. According to \cite{louis1982finding}, the observed information matrix can be expressed as
\begin{align}
I({\boldsymbol\theta}; \mathbf{t}^*, \boldsymbol{\delta}) = &\mathbb{E}\bigl[ I_c({\boldsymbol\theta}; \mathbf{t}^*, \boldsymbol{\delta}, \mathbf{W}) \mid \mathbf{t}^*, \boldsymbol{\delta} \bigr] - \mathbb{E}\bigl[ S_c({\boldsymbol\theta}; \mathbf{t}^*, \boldsymbol{\delta}, \mathbf{W}) S_c^\top({\boldsymbol\theta}; \mathbf{t}^*, \boldsymbol{\delta}, \mathbf{W}) \mid \mathbf{t}^*, \boldsymbol{\delta} \bigr] & \nonumber\\
& + S({\boldsymbol\theta}; \mathbf{t}^*, \boldsymbol{\delta}) S^\top({\boldsymbol\theta}; \mathbf{t}^*, \boldsymbol{\delta}),&
\end{align}
\noindent where the expectations are taken with respect to the conditional distribution of the latent variables $\mathbf{W}$ given the observed data. At the maximum likelihood estimate $\hat{{\boldsymbol\theta}}$, the observed-data score satisfies $S(\hat{{\boldsymbol\theta}}; \mathbf{t}^*, \boldsymbol{\delta}) = 0$. Consequently, the last term vanishes, yielding the simplified expression
\begin{equation}
I_{\text{obs}}(\hat{{\boldsymbol\theta}}) = I_{\text{com}}(\hat{{\boldsymbol\theta}}) - I_{\text{mis}}(\hat{{\boldsymbol\theta}}),
\end{equation}
where
\begin{align}
I_{\text{com}}(\hat{{\boldsymbol\theta}})
&=
\mathbb{E}\!\left[
I_c(\hat{{\boldsymbol\theta}}; \mathbf{t}^*, \boldsymbol{\delta}, \mathbf{W})
\;\middle|\;
\mathbf{t}^*, \boldsymbol{\delta}
\right],
\\[6pt]
I_{\text{mis}}(\hat{{\boldsymbol\theta}})
&=
\mathbb{E}\!\left[
S_c(\hat{{{\boldsymbol\theta}}}; \mathbf{t}^*, \boldsymbol{\delta}, \mathbf{W})
S_c^\top(\hat{{\boldsymbol\theta}}; \mathbf{t}^*, \boldsymbol{\delta}, \mathbf{W})
\;\middle|\;
\mathbf{t}^*, \boldsymbol{\delta}
\right].
\end{align}

The quantity \(I_{\text{com}}(\hat{{\boldsymbol\theta}})\) represents the conditional expectation of the complete data information given observed data, 
while \(I_{\text{mis}}(\hat{{\boldsymbol\theta}})\) corresponds to  the conditional covariance of the complete data score function given the observed data, evaluated at the MLEs, and hence quantifies the information loss due to missing data.

Since all observations at the first stress level are fully observed and no latent component is involved, $\ell_c^{(1)}(\alpha,\lambda_1)$ contributes only to the complete information matrix. In contrast, the second stress level involves unobserved subgroup memberships, leading to an incomplete data structure. The complete data log-likelihood, given in equation~\eqref{eq_10}, is therefore expressed through the latent indicators $W_{ij}$. To compute the required expectations in the information matrices, we consider the conditional distribution of the latent variables $\mathbf{W}$ given the observed data. In the E-step of the EM algorithm, the conditional expectation of $W_{ij}$ is
$
\eta_{ij} = \mathbb{E}(W_{ij} \mid \mathbf{t}^*, \boldsymbol{\delta}; \hat{{\boldsymbol\theta}}),
$
where $\eta_{ij}$, defined in equation~\eqref{eq_11}, represents the posterior 
probability that the $i$-th observation belongs to 
component $j$. These posterior weights are used to 
evaluate all required expectations by replacing the 
latent indicators $W_{ij}$ with their conditional 
expectations $\eta_{ij}$. The explicit expressions 
for all non-zero elements of 
$I_{\text{com}}(\hat{{\boldsymbol\theta}})$ and 
$I_{\text{mis}}(\hat{{\boldsymbol\theta}})$ are provided in 
Appendix~\ref{app:louis}, and the asymptotic 
confidence intervals are constructed using the 
diagonal elements of 
$I_{\text{obs}}^{-1}(\hat{{\boldsymbol\theta}})$.

\subsubsection{Transformed Confidence Interval }

Let $\boldsymbol{\theta} = (\alpha, \lambda_1, \lambda_{21}, \lambda_{22}, \pi)^\top$ denote the parameter vector, and let $\hat{\boldsymbol{\theta}}$ be the corresponding maximum likelihood estimator obtained via the EM algorithm. The variance--covariance matrix $\widehat{\mathrm{Var}}(\hat{\boldsymbol{\theta}})$ is obtained from the inverse of the observed information matrix, computed using the identity of \cite{louis1982finding}. Under standard regularity conditions for incomplete data likelihood models, the estimator $\hat{\boldsymbol{\theta}}$ is asymptotically normally distributed  (see \cite{dempster1977,wu1983}). However, despite this asymptotic normality, direct confidence intervals based on normal approximation may be inadequate due to the constrained nature of the parameters. In particular, $\alpha, \lambda_1, \lambda_{21}, \lambda_{22} > 0$ and $\pi \in (0,1)$, so conventional Wald-type intervals may produce negative lower bounds for positive parameters or values of $\pi$ outside the admissible range. This issue is particularly evident in mixture models, where estimator distributions are often right-skewed, particularly in small or moderate samples, which may result in poor coverage performance. To address these issues, transformation-based confidence intervals are employed following \cite{meeker1998} and \cite{Balakrishnan2024}, which provide a more reliable alternative by ensuring that the resulting intervals respect the underlying parameter constraints while improving finite sample accuracy.
We consider the transformations
\begin{equation*}
\phi_j(\theta_j) = \log(\theta_j), \quad j = 1,2,3,4, \quad \text{and} \quad
\phi_\pi(\pi) = \log\left(\frac{\pi}{1-\pi}\right).
\end{equation*}

Now, applying the delta method to the asymptotic distribution of the estimators, we obtain approximate normality for the transformed quantities. Inverting these transformations yields the corresponding confidence intervals on the original scale.

For the positive parameters $\theta_j \in \{\alpha, \lambda_1, \lambda_{21}, \lambda_{22}\}$, the $100(1-\gamma)\%$ confidence interval is given by
\begin{equation*}
\left(
\hat{\theta}_j \exp\big(-z_{\gamma/2}\,\mathrm{SE}(\log \hat{\theta}_j)\big),
\;
\hat{\theta}_j \exp\big(z_{\gamma/2}\,\mathrm{SE}(\log \hat{\theta}_j)\big)
\right),
\end{equation*}
where
\begin{equation*}
\mathrm{SE}(\log \hat{\theta}_j)
=
\sqrt{\frac{\widehat{\mathrm{Var}}(\hat{\theta}_j)}{\hat{\theta}_j^2}},
\quad j = 1,\ldots,4.
\end{equation*}

This transformation ensures positivity of the bounds and improves performance under right-skewness, particularly in small samples.

For the mixing proportion $\pi \in (0,1)$, we use the logit transformation $\text{logit}(\pi) = \log\{\pi/(1-\pi)\}$. Again using delta method, the standard error of $\text{logit}(\hat{\pi})$ is given by
\begin{equation*}
\mathrm{SE}(\text{logit}(\hat{\pi}))
=
\sqrt{\frac{\widehat{\mathrm{Var}}(\hat{\pi})}{\hat{\pi}^2(1-\hat{\pi})^2}}.
\end{equation*}
Let
\begin{equation*}
L = \text{logit}(\hat{\pi}) - z_{\gamma/2}\,\mathrm{SE}(\text{logit}(\hat{\pi})), 
\quad
U = \text{logit}(\hat{\pi}) + z_{\gamma/2}\,\mathrm{SE}(\text{logit}(\hat{\pi})).
\end{equation*}
The $100(1-\gamma)\%$ confidence interval for $\pi$ is then given by
\begin{equation*}
\left(
\frac{e^L}{1+e^L}, \;
\frac{e^U}{1+e^U}
\right),
\end{equation*}
which automatically respects the natural bounds of $\pi$ without requiring artificial truncation.

\section{Simulation Study}
\label{sec:simulation}

To evaluate the performance of the proposed $h$-SSALT model and to understand the practical consequences of ignoring population heterogeneity, we carry out an extensive Monte Carlo simulation study. Data are generated using Algorithm \ref{algo:2} with true parameter values $\alpha = 1.20$, $\lambda_1 = 0.20$, $\lambda_{21} = 0.10$, $\lambda_{22} = 1.0$, and $\pi = 0.40$, under a two-component mixture ($m = 2$) at the second stress level. The choice $\alpha = 1.20$ reflects an increasing failure-rate, which is characteristic of components subject to gradual wear-out degradation, such as mechanical parts or electronic devices operating past their early-life phase. Three values of the stress-change time, $\tau \in \{1.60, 1.70, 1.80\}$, are considered alongside sample sizes $n \in \{15, 25, 35, 50, 100\}$, with censoring numbers $r$ chosen to reflect both moderate Type-II censoring and complete data scenarios. Choices of the sample size represent that the proposed model is applicable to a general class of sample sizes, including small, moderate, and large samples. This range of sample sizes is intentional as accelerated life testing is inherently expensive, and in industrial practice, engineers often have access to only a limited number of test units. Understanding the behaviour of the proposed model and its application across small, moderate, and large samples is therefore of direct practical relevance. Each configuration is replicated $1000$ times.

The parameter setting $\lambda_{21} = 0.10$, $\lambda_{22} = 1.0$, and $\pi = 0.40$ imply that 40\% of units belong to a short-lived subgroup with a scale parameter ten times smaller than the long-lived majority group, representing a realistic scenario of manufacturing heterogeneity where a minority of units are prone to early failure. In physical terms, this represents a production batch in which a substantial minority of units carry latent weaknesses, perhaps arising from extreme material inconsistencies or variability in the manufacturing process, that make them significantly more vulnerable to failure once the stress level is elevated. Crucially, this heterogeneity is invisible at the first stress level and only manifests after the stress change at $\tau$. A practitioner unaware of this subgroup structure would routinely apply the homogeneous SSALT model, and therefore the resulting estimation errors are quantified here throughout the simulation results presented in this section.

\FloatBarrier
\subsection{Point Estimation}
\label{subsec:point_est}

Table \ref{table:point_estimates} reports the average estimates (AE) and mean squared errors (MSE) of all model parameters  $\alpha, ~\lambda_1, ~\lambda_{21},~ \lambda_{22}$, and $\pi$ under the proposed $h$-SSALT model. Several patterns of the parameter estimates are evident and satisfactory.

First, the AEs approach toward their true values as $n$ grows across all parameters, confirming consistency of the EM-based MLEs. This is reassuring from an engineering standpoint. It means that with a sufficiently large sample test, the proposed model will reliably recover the true failure behavior of each subgroup, even if the subgroup membership was originally completely unknown. Second, the MSEs decrease monotonically with increasing $n$, indicating improved precision with larger samples. Third, for fixed $n$, complete samples ($r = n$) yield smaller MSEs than censored samples, as expected. For a practitioner designing an accelerated life test, this suggests that when the goal is to characterize subgroup structure, reducing the censoring fraction, even at the cost of a longer test, yields improvements in the estimation precision.

Among all parameters, the shape parameter $\alpha$ is the most challenging to estimate at small $n$, demonstrating the slowest convergence and the largest relative bias and MSE for $n = 15$ and $n = 25$. This is a structural feature of the model rather than a limitation of the algorithm, as the shape parameter common to both stress levels, $\alpha$ must be identified jointly from failure times at two distinct stress environments, and with few observations this is a difficult inference problem. Engineers working with small test batches should therefore treat estimates of $\alpha$ at $n < 35$ with appropriate caution.

The mixture proportion $\hat{\pi}$ converges reliably to its true value $\pi = 0.40$ already at $n = 25$ under both censored and complete scenarios. This is practically significant because $\hat{\pi}$ directly quantifies the fraction of production units 
belonging to the weaker subgroup. In practice, this estimate helps manufacturers decide what proportion of units from a production batch should be screened out before deployment, a process commonly known as burn-in testing, by reducing the risk of 
early field failures reaching the customer. When precise characterization of the weaker subgroup is needed, for example, to set a warranty period that accounts for early failures, larger sample sizes and less aggressive censoring are advisable, as the results at $n = 100$ show substantially improved precision for all subgroup parameters.

\begin{table}[ht]
\scriptsize
\centering
\caption{ AE and MSE of the model parameters assuming Weibull distribution, with true values of the parameters as $\alpha=1.20,\:\lambda_1=0.20,\:\lambda_{21}=0.10,\:\lambda_{22}=1.0,\:\pi=0.40$.}
\begin{tabular}{ccccccccccccc}
	\toprule
	& & &
	\multicolumn{2}{c}{$\alpha$} &
	\multicolumn{2}{c}{$\lambda_1$} &
	\multicolumn{2}{c}{$\lambda_{21}$} &
	\multicolumn{2}{c}{$\lambda_{22}$} &
	\multicolumn{2}{c}{$\pi$} \\
	\cmidrule(lr){4-5}
	\cmidrule(lr){6-7}
	\cmidrule(lr){8-9}
	\cmidrule(lr){10-11}
	\cmidrule(lr){12-13}
	$n$ & $r$ & $\tau$
	& AE & MSE
	& AE & MSE
	& AE & MSE
	& AE & MSE
	& AE & MSE \\
	\midrule
	
	15 & 13 & 1.60&1.7612&1.1309&0.1894&0.0116&0.2687&0.1975&1.2829&2.0795&0.5683&0.0633\\
	&   & 1.70&1.7824&1.3392&0.1827&0.0121&0.3146&0.3065&1.2873&2.0712&0.5993&0.0726 \\
	&   & 1.80 &1.6488&0.8498&0.1857&0.0107&0.3426&0.3363&1.3871&2.3328&0.6073&0.0758\\
	
	\addlinespace
	& 15 & 1.60&1.7252&0.8576&0.1837&0.0108&0.1610&0.1179&1.2611&2.0407&0.4464&0.0520\\
	&    & 1.70&1.7111&0.7796&0.1824&0.0104&0.1258&0.0540&1.2409&2.0951&0.4554&0.0516\\
	&    & 1.80&1.6348&0.6783&0.1829&0.0111&0.1579&0.0977&1.2727&1.9498&0.4645&0.0558\\
	
	\midrule
	25 & 21 & 1.60&1.5490&0.5764&0.1825&0.0076&0.2434&0.1358&1.5242&2.3200&0.5713&0.0593\\
	&    & 1.70&1.5213&0.5022&0.1812&0.0068&0.2788&0.2015&1.4536&2.1730&0.5850&0.061\\
	&    & 1.80&1.4904&0.4035&0.1837&0.0069&0.2549&0.1233&1.4524&2.1286&0.5988&0.0679\\
	\addlinespace
	& 25 & 1.60&1.4813&0.3435&0.1910&0.0070&0.1209&0.0279&1.2991&1.8223&0.4103&0.0337\\
	&    & 1.70&1.4361&0.2682&0.1925&0.0066&0.1205&0.0270&1.2400&1.5176&0.4166&0.0376\\
	&    & 1.80&1.4519&0.2917&0.1884&0.0063&0.1189&0.0232&1.2498&1.4926&0.4194&0.0385\\
	
	\midrule
	35 & 30 & 1.60&1.4063&0.2483&0.1870&0.0047&0.1929&0.0744&1.4242&1.6026&0.5313&0.0434\\
	&    & 1.70&1.4092&0.2267&0.1880&0.0046&0.1956&0.0674&1.4063&1.6896&0.5395&0.0481\\
	&    & 1.80&1.3866&0.2024&0.1876&0.0048&0.1968&0.0668&1.5583&2.2056&0.5475&0.0499\\
	\addlinespace
	& 35 & 1.60&1.4098&0.2043&0.1876&0.0045&0.1091&0.0184&1.1180&0.9863&0.3989&0.0276\\
	&    & 1.70& 1.4022&0.2127&0.1877&0.0048&0.1105&0.0164&1.2151&1.2954&0.4105&0.0299\\
	&    & 1.80& 1.3759&0.1905&0.1880&0.0045&0.1137&0.0149&1.2251&1.2356&0.4122&0.0299\\
	
	\midrule
	50 & 43 & 1.60&1.3504&0.1509&0.1913&0.0033&0.1570&0.0365&1.3965&1.2675&0.4994&0.0347\\
	&    & 1.70&1.3396&0.1437&0.1900&0.0036&0.1542&0.0298&1.3859&1.2848&0.5032&0.0369 \\
	&    & 1.80& 1.3234&0.1293&0.1893&0.0034&0.1710&0.0451&1.4404&1.4714&0.5114&0.0385\\
	\addlinespace
	& 50 & 1.60&1.3283&0.1216&0.1943&0.0032&0.1102&0.0110&1.1420&0.7737&0.3981&0.0217\\
	&    & 1.70&1.3169&0.1130&0.1911&0.0032&0.1122&0.0117&1.2101&1.0926&0.4002&0.0225\\
	
	&    & 1.80&1.3198&0.1080&0.1931&0.0031&0.1088&0.0096&1.1792&0.9380&0.4008&0.0217
	\\
	
	\midrule
	100 & 86 & 1.60&1.2723&0.0577&0.1951&0.0018&0.1199&0.0088&1.2396&0.6476&0.4552&0.0201\\
	&     & 1.70&1.2555&0.0552&0.1936&0.0015&0.1286&0.0130&1.2416&0.6012&0.4555&0.0214\\
	&     & 1.80&1.2638&0.0520&0.1960&0.0016&0.1320&0.0141&1.3228&0.8888&0.4685&0.0260\\
	\addlinespace
	& 100 & 1.60&1.2524&0.0402&0.1970&0.0016&0.1049&0.0049&1.0634&0.2766&0.4039&0.0111\\
	&     & 1.70&1.2592&0.0467&0.1978&0.0016&0.1031&0.0043&1.0549&0.2968&0.3597&0.0105\\
	&     & 1.80&1.2530&0.0444&0.1974&0.0016&0.1065&0.0049&1.1139&0.4462&0.4043&0.0112\\
	
	\bottomrule
\end{tabular}
\label{table:point_estimates}
\end{table}
\FloatBarrier

\subsection{Confidence Intervals}
\label{subsec:CI}

Table~\ref{table:interval_est} reports the coverage probabilities (CP) and average lengths (AL) of the 95\% transformed confidence intervals for all model parameters, under the same simulation configurations, $\alpha=1.20$, $\lambda_1=0.20$, $\lambda_{21}=0.10$, $\lambda_{22}=1.0$, $\pi=0.40$ and $\tau\in\{1.60,1.70,1.80\}$, as in Section~\ref{subsec:point_est}. The CPs for $\lambda_1$ are consistently close to the nominal level of $0.95$ 
across all configurations, as expected given that this parameter is estimated entirely from the first stress level where no latent mixture 
structure is involved. For $\alpha$, the CPs are slightly below nominal at $n = 25$, ranging from approximately $0.877$ to $0.915$, but improve 
steadily to near $0.95$ at $n = 100$. This mild coverage at small $n$ is consistent with the slow convergence of $\hat{\alpha}$ discussed in Section~\ref{subsec:point_est}, and practitioners should treat the estimated CI for $\alpha$ at $n < 35$ as approximate.
\begin{table}[htbp]
\scriptsize
\centering
\caption{CP and AL of the proposed transformation-based confidence intervals for the model parameters, with true parameter values $\alpha=1.20,\:\lambda_1=0.20,\:\lambda_{21}=0.10,\:\lambda_{22}=1.0,\:\pi=0.40$.}
\begin{tabular}{ccccccccccccc}
\toprule
& & &
\multicolumn{2}{c}{$\alpha$} &
\multicolumn{2}{c}{$\lambda_1$} &
\multicolumn{2}{c}{$\lambda_{21}$} &
\multicolumn{2}{c}{$\lambda_{22}$} &
\multicolumn{2}{c}{$\pi$} \\
\cmidrule(lr){4-5}
\cmidrule(lr){6-7}
\cmidrule(lr){8-9}
\cmidrule(lr){10-11}
\cmidrule(lr){12-13}
$n$ & $r$ & $\tau$
& CP & AL
& CP & AL
& CP & AL
& CP & AL
& CP & AL \\
\midrule

25 & 21 & 1.60 & 0.8910&2.2224&0.9340&0.3342&0.9010&4.0416&0.9220&26.6511&0.9460&0.7177   \\
   &    & 1.70 &  0.8770&2.2412&0.9330&0.3257&0.9000&3.7097&0.9250&30.6835&0.9560&0.7263 \\
   &    & 1.80 &  0.9150&2.0174&0.9480&0.3330&0.8970&4.0567&0.9280&34.0028&0.9500&0.7440  \\

\addlinespace
   & 25 & 1.60&0.8790&1.9373&0.9440&0.3237&0.9450&1.3839&0.9380&9.1274&0.9690&0.6116  \\
   &    & 1.70 &0.8880&1.8776&0.9390&0.3189&0.9360&1.2880&0.9390&7.7780&0.9740&0.6146  \\
   &    & 1.80 &0.8860&1.8316&0.9410&0.3197&0.9400&1.2359&0.9300&8.9394&0.9680&0.6210  \\

\midrule

35 & 30 & 1.60 & 0.9120&1.7057&0.9470&0.2776&0.9100&2.5070&0.9320&18.1190&0.9350&0.6645\\
   &    & 1.70 &0.8950&1.7075&0.9490&0.2703&0.9140&2.1622&0.9300&17.9402&0.9410&0.6688 \\
   &    & 1.80 &0.9200&1.6038&0.9510&0.2715&0.9100&2.2406&0.9450&17.6747&0.9400&0.6716   \\

\addlinespace
   & 35 & 1.60 &  0.9050&1.5130&0.9460&0.2691&0.9290&0.8716&0.9160&6.4123&0.9520&0.5566 \\
   &    & 1.70 &0.8860&1.4776&0.9320&0.2690&0.9430&0.7938&0.9190&5.7763&0.9530&0.5558 \\
   &    & 1.80 & 0.9100&1.4293&0.9470&0.2680&0.9380&0.8667&0.9260&7.6035&0.9590&0.5720  \\

\midrule

50 & 43 & 1.60 &  0.9230&1.3667&0.9570&0.2278&0.9230&1.5131&0.9670&8.9705&0.9610&0.5992  \\
   &    & 1.70 & 0.9250&1.3078&0.9390&0.2261&0.9350&1.5807&0.9490&8.2754&0.9620&0.6110  \\
   &    & 1.80 & 0.9310&1.2563&0.9520&0.2255&0.9210&1.7099&0.9550&10.2593&0.9650&0.6183 \\

\addlinespace
   & 50 & 1.60 &  0.9040&1.1979&0.9400&0.2255&0.9450&0.5879&0.9320&4.1709&0.9500&0.4959 \\
   &    & 1.70 & 0.9150&1.1771&0.9530&0.2230&0.9490&0.5545&0.9370&3.8053&0.9520&0.4951\\
   &    & 1.80 & 0.9280&1.1329&0.9450&0.2242&0.9470&0.5801&0.9330&4.2167&0.9540&0.5132  \\

\midrule

100 & 86 & 1.60 & 0.9340&0.8969&0.9510&0.1587&0.9370&0.7030&0.9680&3.2593&0.9480&0.4989 \\
    &     & 1.70 &0.9340&0.8557&0.9640&0.1585&0.9160&0.7270&0.9420&3.6375&0.9570&0.5029   \\
    &     & 1.80 &  0.9450&0.8310&0.9420&0.1578&0.9300&0.8111&0.9540&3.6799&0.9550&0.5116  \\

\addlinespace
    & 100 & 1.60 & 0.9420&0.7957&0.9350&0.1577&0.9640&0.3238&0.9420&2.1259&0.9480&0.3864\\
    &     & 1.70 & 0.9350&0.7779&0.9440&0.1564&0.9430&0.3157&0.9400&2.2728&0.9350&0.3867 \\
    &     & 1.80 & 0.9520&0.7571&0.9580&0.1554&0.9328&0.3244&0.9380&2.3839&0.9440&0.3942\\

\bottomrule
\end{tabular}
\label{table:interval_est}
\end{table}
\begin{figure}[htbp]
\centering
{\small\textbf{Complete samples}}\\[-0.1em]
\includegraphics[width=0.95\linewidth]{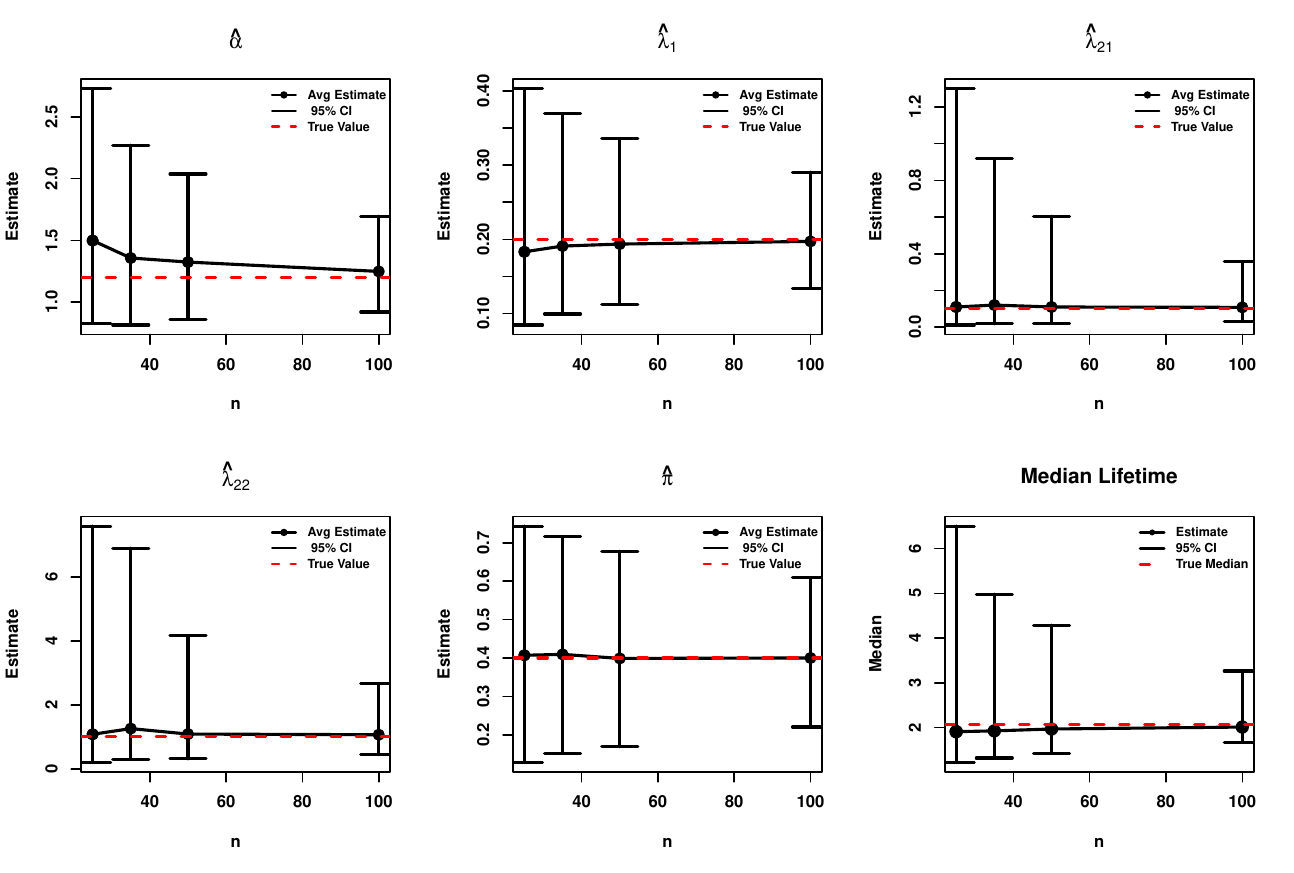}\\[-1em]
{\small\textbf{Type-II censored samples}}\\[-0.1em]
\includegraphics[width=0.95\linewidth]{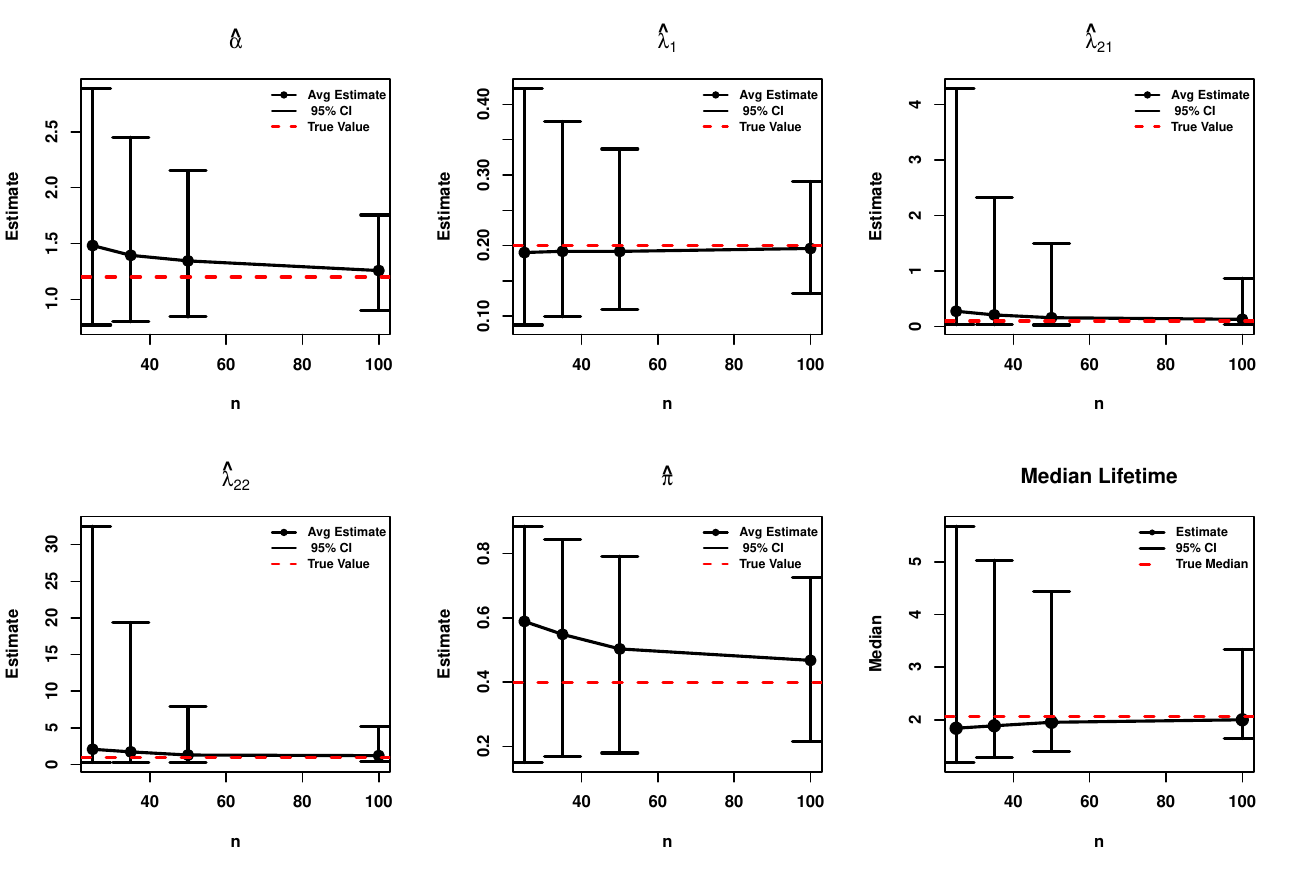}
\caption{Average estimates and 95\% transformed confidence intervals for $\alpha$, $\lambda_1$, $\lambda_{21}$, $\lambda_{22}$, $\pi$, and the 
median lifetime at $\tau = 1.70$, for complete samples with $n\in\{25,35,50,100\}$ (top) and censored samples with $(n,r)\in\{(25,21),(35,30),(50,43),(100,86)\}$ (bottom).}
\label{fig:CI}
\end{figure}
% \begin{figure}[htbp]
% \centering
%     \includegraphics[width=\linewidth]{Censored_data_CI_plot.pdf}
%     \caption{Average estimates and 95\% transformed confidence intervals for 
% $\alpha$, $\lambda_1$, $\lambda_{21}$, $\lambda_{22}$, $\pi$, 
% and the median lifetime under censoring with $(n,r)\in\{(25,21),(35,30),(50,43),(100,86)\}$.}
% \label{fig:CI_censored}
% \end{figure}
The mixture parameters $\lambda_{21}$ and $\lambda_{22}$ show the most interesting confidence interval behaviour. The CPs for $\lambda_{21}$ are 
broadly near the nominal level throughout, but the ALs are considerably wider under censoring than for complete samples at the same $n$, reaching approximately $4.0$ at $(n, r) = (25, 21)$ compared to $1.3$ at $(n, r) = (25, 25)$, narrowing to below $0.35$ for complete samples at $n = 100$. The behaviour of $\lambda_{22}$ is more striking: while the CPs remain above $0.92$ throughout, the ALs are extremely wide at small censored samples, reaching $34.0$ at $(n, r) = (25, 21)$. This follows directly from the high variability of $\hat{\lambda}_{22}$ at small $n$ seen in Table~\ref{table:point_estimates}: the long-lived subgroup spreads failure times over a much wider range, and under censoring many of these are unobserved. The intervals are wide enough to maintain coverage but offer limited precision until $n$ reaches $100$, where the ALs fall to roughly $2$ to $4$. The mixing proportion $\pi$ is well-covered throughout, with CPs at or above $0.93$ even at $n = 25$, and ALs decreasing from $0.74$ to $0.39$ as $n$ grows from $25$ to $100$, in line with the reliable estimation of $\hat{\pi}$ seen earlier in the point estimation. 

Across all parameters, ALs decrease monotonically with $n$, and complete samples consistently yield narrower intervals than censored ones, confirming the asymptotic efficiency of 
the EM-based estimators. Figure \ref{fig:CI} displays the average estimates and 95\% transformation based intervals for all parameters and for the median lifetime $t_{0.50}$, defined as the solution to $G(t_{0.50}) = 0.50$, for complete and censored configurations at $\tau = 1.70$. Both figures 
confirm that the true values are consistently covered and that interval widths narrow visibly with $n$, with the narrowing most pronounced for 
$\lambda_{21}$ and $\lambda_{22}$.
\FloatBarrier
\subsection{$h$-SSALT and SSALT: Quantile-Based Comparison}
\label{subsec:hssalt_vs_sslat}

\cite{lu2025} compared the SSALT and $h$-SSALT models based on the mean lifetime under the second stress level, which is a natural summary quantity in the exponential case. In our Weibull setting, however, the mean lifetime alone is insufficient to characterize the lifetime distribution, as it depends jointly on the shape parameter and the scale parameters. A single scalar life summary about life that captures the combined effect of all parameters is therefore needed. We propose using the $q$-th quantile $t_q$, defined as the solution to $G(t_q) = q$ for $q \in (0,1)$, where $G(t)$ is the $h$-SSALT CDF given in equation \eqref{eq:cdf} as the primary quantity of interest for this comparison. This choice is practically motivated. The $q$-th quantile, $t_q$ corresponds directly to the life used in reliability engineering, representing the time by which a proportion $q$ of units are expected to fail, and is therefore of direct relevance in warranty period design and safety-critical component qualification. Throughout this comparison, the true quantile $t_q$ is computed from the $h$-SSALT survival function $G(t)$ in equation \eqref{eq:cdf} evaluated at the true parameter values, since the data are generated from this model; the SSALT estimates are obtained by fitting a homogeneous model to the same data, ignoring the mixture structure. We consider $q \in \{0.25, 0.50, 0.75, 0.99\}$ for the main comparison, and additionally $q \in \{0.01, 0.05, 0.10\}$ for the early failure quantiles. For each configuration $(n, r, \tau)$, we compare the mean estimate and RMSE of $\hat{t}_q$ under both the $h$-SSALT and the homogeneous SSALT models, based on $1000$ Monte Carlo replications.

% \textcolor{blue}{Tables~\ref{table:comp_ssalt_q_010_050} and~\ref{table:comp_ssalt_q_075_099} compare the $h$-SSALT and homogeneous SSALT models in terms of the mean estimates and RMSEs of the $q$-th quantiles at $q \in \{0.10, 0.25, 0.50, 0.75, 0.99\}$, and Table \ref{table:early_quantiles} additionally reports results for the early failure quantiles $q \in \{0.01, 0.05\}$, corresponding to the $t_{0.01}$ and $t_{0.05}$ lives widely used in reliability engineering practice.}

\begin{table}[ht]
\scriptsize
\centering
\caption{Mean estimates and RMSE of the estimated quantiles $q_{0.25}$ and $q_{0.50}$ under the $h$-SSALT and SSALT schemes for different $(n,r,\tau)$ combinations with $\:\alpha=1.20,\:\lambda_1=0.20,\:\lambda_{21}=0.10,\:\lambda_{22}=1.0,\:\pi=0.40$.}
\setlength{\tabcolsep}{3pt}
\begin{tabular}{ccccccccccccc}

\toprule
$n$ & $r$ & $\tau$ &
\multicolumn{5}{c}{$q=0.25$} &
\multicolumn{5}{c}{$q=0.50$} \\

\cmidrule(lr){4-8}
\cmidrule(lr){9-13}

& & &
\textit{$t_{0.25}$} &
\multicolumn{2}{c}{$h$-SSALT} &
\multicolumn{2}{c}{SSALT} &
\textit{$t_{0.50}$} &
\multicolumn{2}{c}{$h$-SSALT} &
\multicolumn{2}{c}{SSALT} \\

\cmidrule(lr){5-6}
\cmidrule(lr){7-8}
\cmidrule(lr){10-11}
\cmidrule(lr){12-13}

& & &
& Mean & RMSE & Mean & RMSE
& & Mean & RMSE & Mean & RMSE \\

\midrule

15 & 13 & 1.60 & 1.3538&1.3660&0.3657&1.3842&0.4573&2.0082&2.0145&0.5000&2.6685&1.1060 \\
   &    & 1.70 & 1.3538&1.3822&0.3847&1.3842&0.4583&2.0731&2.0577&0.4109&2.6697&1.0394 \\
   &    & 1.80 & 1.3538&1.4044&0.4005&1.4013&0.4747&2.1384&2.1284&0.4301&2.7249&1.0470 \\

& 15 & 1.60 & 1.3538&1.3735&0.3474&1.3596&0.4800&2.0082&2.0209&0.4080&2.7722&1.1134 \\
   &    & 1.70 & 1.3538&1.4130&0.3826&1.3828&0.5139&2.0731&2.1023&0.4036&2.8605&1.1792 \\
   &    & 1.80 & 1.3538&1.4117&0.3989&1.3462&0.5159&2.1384&2.1335&0.3878&2.8420&1.0816 \\

\midrule

25 & 21 & 1.60 & 1.3538&1.3536&0.2964&1.3487&0.3735&2.0082&1.9907&0.2827&2.6181&0.9046 \\
   &    & 1.70 & 1.3538&1.3842&0.3186&1.3662&0.3839&2.0731&2.0283&0.2428&2.5867&0.8195 \\
   &    & 1.80 & 1.3538&1.4232&0.3366&1.3923&0.3979&2.1384&2.1179&0.2477&2.6608&0.8341 \\

& 25 & 1.60 & 1.3538&1.3630&0.2920&1.3081&0.4141&2.0082&2.0123&0.2865&2.7764&0.9753 \\
   &    & 1.70 & 1.3538&1.3941&0.3158&1.3193&0.4398&2.0731&2.0703&0.2629&2.7798&0.9409 \\
   &    & 1.80 & 1.3538&1.4083&0.3391&1.3045&0.4577&2.1384&2.1261&0.2563&2.8193&0.9460 \\

\midrule

35 & 30 & 1.60 & 1.3538&1.3712&0.2622&1.3536&0.3511&2.0082&1.9959&0.2100&2.6513&0.8530 \\
   &    & 1.70 & 1.3538&1.3850&0.2919&1.3466&0.3657&2.0731&2.0474&0.2042&2.6419&0.7875 \\
   &    & 1.80 & 1.3538&1.4125&0.3032&1.3594&0.3657&2.1384&2.1201&0.2029&2.6899&0.7744 \\

& 35 & 1.60 & 1.3538&1.3758&0.2607&1.3276&0.3830&2.0082&2.0211&0.2229&2.7677&0.9142 \\
   &    & 1.70 & 1.3538&1.3880&0.2885&1.2985&0.4129&2.0731&2.0660&0.2075&2.7610&0.8429 \\
   &    & 1.80 & 1.3538&1.3922&0.2941&1.2671&0.4120&2.1384&2.1204&0.2051&2.7853&0.8095 \\

\midrule

50 & 43 & 1.60 & 1.3538&1.3768&0.2372&1.3493&0.3137&2.0082&1.9970&0.1733&2.6367&0.7699 \\
   &    & 1.70 & 1.3538&1.3980&0.2564&1.3499&0.3270&2.0731&2.0638&0.1697&2.6777&0.7566 \\
   &    & 1.80 & 1.3538&1.3963&0.2690&1.3211&0.3297&2.1384&2.1265&0.1674&2.6944&0.7138 \\

& 50 & 1.60 & 1.3538&1.3774&0.2318&1.3233&0.3443&2.0082&2.0113&0.1704&2.7392&0.8375 \\
   &    & 1.70 & 1.3538&1.3745&0.2498&1.2758&0.3652&2.0731&2.0747&0.1726&2.7647&0.7967 \\
   &    & 1.80 & 1.3538&1.3821&0.2593&1.2489&0.3687&2.1384&2.1376&0.1778&2.8035&0.7679 \\

\midrule

100 & 86 & 1.60 & 1.3538&1.3651&0.1790&1.3185&0.2515&2.0082&2.0042&0.1143&2.6354&0.6981 \\
    &    & 1.70 & 1.3538&1.3774&0.1910&1.3037&0.2576&2.0731&2.0731&0.1150&2.6861&0.6901 \\
    &    & 1.80 & 1.3538&1.3803&0.1982&1.2799&0.2573&2.1384&2.1304&0.1118&2.7048&0.6499 \\

& 100 & 1.60 & 1.3538&1.3677&0.1861&1.3011&0.2862&2.0082&2.0093&0.1208&2.7236&0.7667 \\
    &    & 1.70 & 1.3538&1.3876&0.1908&1.2815&0.2909&2.0731&2.0771&0.1142&2.7906&0.7692 \\
    &    & 1.80 & 1.3538&1.3818&0.1976&1.2324&0.3054&2.1384&2.1385&0.1118&2.8115&0.7261 \\

\bottomrule
\end{tabular}
\label{table:comp_ssalt_q_025_050}
\end{table}

\begin{table}[ht]
\scriptsize
\centering
\caption{{Mean estimates and RMSE of the estimated quantiles $q_{0.75}$ and $q_{0.99}$ under the $h$-SSALT and SSALT schemes for different $(n,r,\tau)$ combinations  with $\:\alpha=1.20,\:\lambda_1=0.20,\:\lambda_{21}=0.10,\:\lambda_{22}=1.0,\:\pi=0.40$.} }
\setlength{\tabcolsep}{3pt}
\begin{tabular}{ccccccccccccc}
	
	\toprule
	$n$ & $r$ & $\tau$ &
	\multicolumn{5}{c}{$q=0.75$} &
	\multicolumn{5}{c}{$q=0.99$} \\
	
	\cmidrule(lr){4-8}
	\cmidrule(lr){9-13}
	
	& & &
	\textit{$t_{0.75}$}&
	\multicolumn{2}{c}{$h$-SSALT} &
	\multicolumn{2}{c}{SSALT} &
	\textit{$t_{0.99}$}&
	\multicolumn{2}{c}{$h$-SSALT} &
	\multicolumn{2}{c}{SSALT} \\
	
	\cmidrule(lr){5-6}
	\cmidrule(lr){7-8}
	\cmidrule(lr){10-11}
	\cmidrule(lr){12-13}
	
	& & &
	& Mean & RMSE & Mean & RMSE
	& & Mean & RMSE & Mean & RMSE \\
	
	\midrule
	
	15 & 13 & 1.60 & 2.7908 &2.8022&1.1923&5.1512&3.4640&6.0190&6.0280&4.6733&22.1837&26.3379 \\
	&    & 1.70 & 2.8516 &2.8539&1.0647&5.1241&3.3961&6.0712&6.2980&5.7145&20.7920&22.7231 \\
	&    & 1.80 & 2.9129 &2.9571&1.0778&5.1964&3.3874&6.1240&6.3120&4.5178&20.5033&22.6335  \\
	
	& 15 & 1.60 & 2.7908 & 2.8284&0.9760&5.6602&3.5983&6.0190&6.0665&4.0465&25.7583&24.3268 \\
	&    & 1.70 & 2.8516 & 2.9130&1.0176&5.7906&3.7579&6.0712&6.1666&4.5975&25.5150&24.0784 \\
	&    & 1.80 & 2.9129 &2.9221&0.9351&5.7828&3.6377&6.1240&6.0018&3.8651&24.7198&23.2517  \\
	
	\midrule
	
	25 & 21 & 1.60 &2.7908&2.7373&0.6394&4.9903&3.0049&6.0190&5.7461&2.6732&20.6571&21.3797 \\
	&    & 1.70 &2.8516& 2.7356&0.5714&4.7390&2.6899&6.0712&5.5344&2.3840&17.8957&18.1196 \\
	&    & 1.80 &2.9129&2.8420&0.5589&4.8599&2.7444&6.1240&5.7389&2.3206&17.9138&17.5621 \\
	
	& 25 & 1.60 &  2.7908&2.7899&0.6353&5.5611&3.2119&6.0190&5.8127&2.4237&25.7775&22.5225\\
	&    & 1.70 &2.8516&2.8319&0.5630&5.5837&3.2127&6.0712&5.7671&2.1903&24.6806&21.3489 \\
	&    & 1.80 & 2.9129&2.9115&0.5777&5.7192&3.3361&6.1240&5.9944&2.5942&24.9691&21.9146\\
	
	\midrule
	
	35 & 30 & 1.60 & 2.7908&2.7316  & 0.4529& 5.0940 & 2.8833& 6.0190 & 5.6701 & 1.9148& 21.4099& 20.2303 \\
	&    & 1.70 &2.8516&2.7755&0.4397&4.9853&2.7240&6.0712&5.6810&1.7815&20.0035&18.8892 \\
	&    & 1.80 &2.9129&2.8423&0.4367&4.9967&2.6610&6.1240&5.7363&1.7256&19.0474&17.2185\\
	
	& 35 & 1.60 & 2.7908 & 2.8073 & 0.4816 & 5.6251 & 3.1770&6.0190 &5.9094 & 1.9024& 26.3298 & 22.5805\\
	&    & 1.70 & 2.8516 & 2.8344 & 0.4366& 5.5555& 3.0313&6.0712 & 5.8507& 1.6606 & 25.1161& 21.2228\\
	&    & 1.80 & 2.9129 & 2.8835 & 0.4292 & 5.5959 & 3.0267& 6.1240 &5.8666& 1.6285 & 24.3424& 20.2554\\
	
	\midrule
	
	50 & 43 & 1.60 &2.7908&2.7406&0.3717&5.0539&2.6722&6.0190&5.7420&1.4718&22.1856&16.5261 \\
	&    & 1.70 &2.8516&2.8053&0.3701&5.0553&2.6107&6.0713&5.8039&1.4759&20.2739&17.4234 \\
	&    & 1.80 & 2.9129&2.8688&0.3647&5.0274&2.5169&6.1240&5.8487&1.4251&21.2311&16.2222\\
	
	& 50 & 1.60 & 2.7908&2.7776&0.3826&5.4983&2.9465&6.0190&5.8553&1.4903&25.8262&21.2570 \\
	&    & 1.70 & 2.8516&2.8623&0.3686&5.5500&2.9271&6.0712& 6.0224&1.3290&25.1804&20.4789 \\
	&    & 1.80 & 2.9129&2.9239&0.4044&5.6165&2.9211&6.1240&6.1065&1.5876&24.8460&20.1211 \\
	
	\midrule
	
	100 & 86 & 1.60 &2.7908&2.7793&0.2529&5.0571&2.4740 & 6.0190& 5.9377&1.0186&21.1506&16.9423\\
	&     & 1.70 & 2.8516&2.8478&0.2522&5.0978&2.4594&6.0712&6.0029&0.9885&20.4250&16.1513\\
	&     & 1.80 & 2.9129&2.8902&0.2447&5.0593&2.3690&6.1240&5.9806&0.9616&19.2847&14.8692\\
	
	& 100 & 1.60 & 2.7908&2.7882&0.2548&5.4745&2.8065&6.0190&5.9604&0.9466&25.8450&20.6092 \\
	&     & 1.70 &2.8516&2.8508&0.2518&5.5777&2.8496&6.0712&5.9926&0.9457&25.6453&20.3837\\
	&     & 1.80 & 2.9129&2.9127&0.2377&5.5763&2.7753&6.1240&6.0611&0.8757&24.7029&19.2642\\
	
	\bottomrule
\end{tabular}
\label{table:comp_ssalt_q_075_099}
\end{table}

Tables \ref{table:comp_ssalt_q_025_050} and \ref{table:comp_ssalt_q_075_099} report the mean estimates and RMSEs of the estimated quantiles at $q \in \{0.25, 0.50, 0.75, 0.99\}$. Several clear patterns emerge. The $h$-SSALT model consistently achieves substantially lower RMSEs than the homogeneous SSALT model across all quantile levels, sample sizes, and stress-change times. The SSALT model, by treating all units as arising from a single homogeneous population in the second phase, fails to account for the 40\% short-lived subgroup present in the data and consequently overestimates the upper quantiles. This overestimation bias is particularly severe at $q = 0.75$ and $q = 0.99$, where the SSALT model systematically overpredicts the true quantile, producing mean estimates substantially above the true value across all $(n, r, \tau)$ configurations. In contrast, the $h$-SSALT model produces mean estimates that are well-centered around the true quantile values even at moderate sample sizes.

For the median ($q = 0.50$), the $h$-SSALT model achieves RMSEs approximately two to four times smaller than those of the SSALT model across all configurations. For example, at $(n, r) = (35, 30)$ and $\tau = 1.70$, the RMSE of the median estimate under SSALT is $1.0394$, compared to $0.4109$ under $h$-SSALT, a reduction of more than 60\%. This finding confirms and extends the result of \cite{lu2025}, who showed that ignoring heterogeneity leads to biased lifetime estimation in the exponential case; our results demonstrate that this conclusion holds more broadly for the Weibull failure-rate-based formulation. As $n$ increases from $15$ to $100$, the RMSEs of both models decrease, but the rate of improvement is markedly faster for $h$-SSALT, indicating that the $h$-SSALT model extracts information about the subgroup structure more efficiently as the sample size grows.

The improvement of $h$-SSALT over SSALT is most pronounced at the higher quantiles $q = 0.75$ and $q = 0.99$. At $q = 0.99$, the SSALT model produces estimates with RMSEs an order of magnitude larger than $h$-SSALT at small sample sizes, reflecting the extreme sensitivity of the tail quantile to model misspecification. At $(n, r) = (15, 13)$ and $\tau = 1.60$, the RMSE of $\hat{t}_{0.99}$ under SSALT is $26.3379$, compared to $4.6733$ under $h$-SSALT. Even at $n = 100$, the SSALT model continues to exhibit noticeably larger RMSEs than $h$-SSALT at the upper tail, confirming that the misspecification penalty does not vanish at large samples but persists as a systematic bias. At the median and upper quantiles, the difference between complete and censored samples is less pronounced, with comparable RMSEs observed across both schemes at moderate sample sizes.

\begin{figure}[ht]
\centering

\noindent\hspace{0.5em}\textit{$q = 0.50$}

\vspace{3.5pt}

\includegraphics[width=0.32\textwidth]{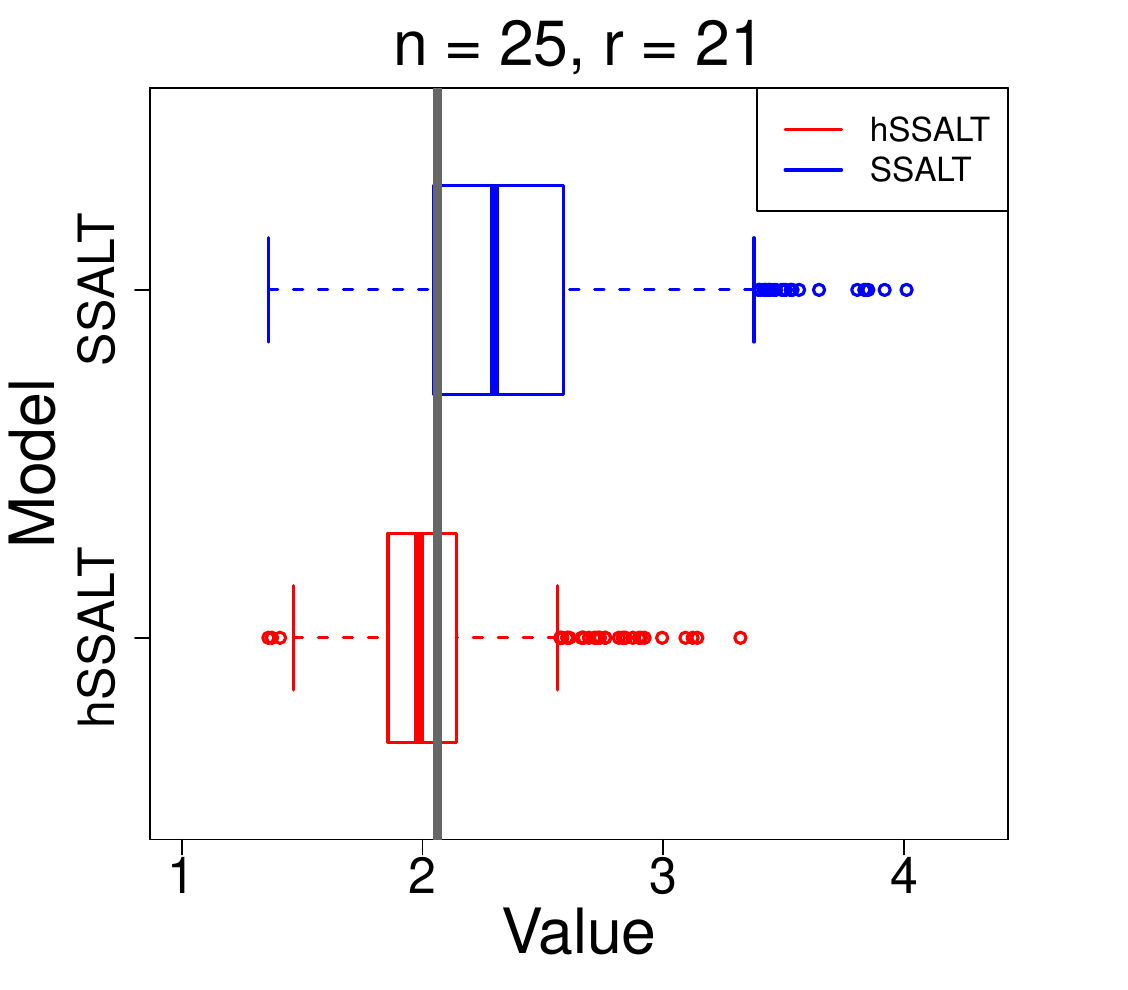}\hfill
\includegraphics[width=0.32\textwidth]{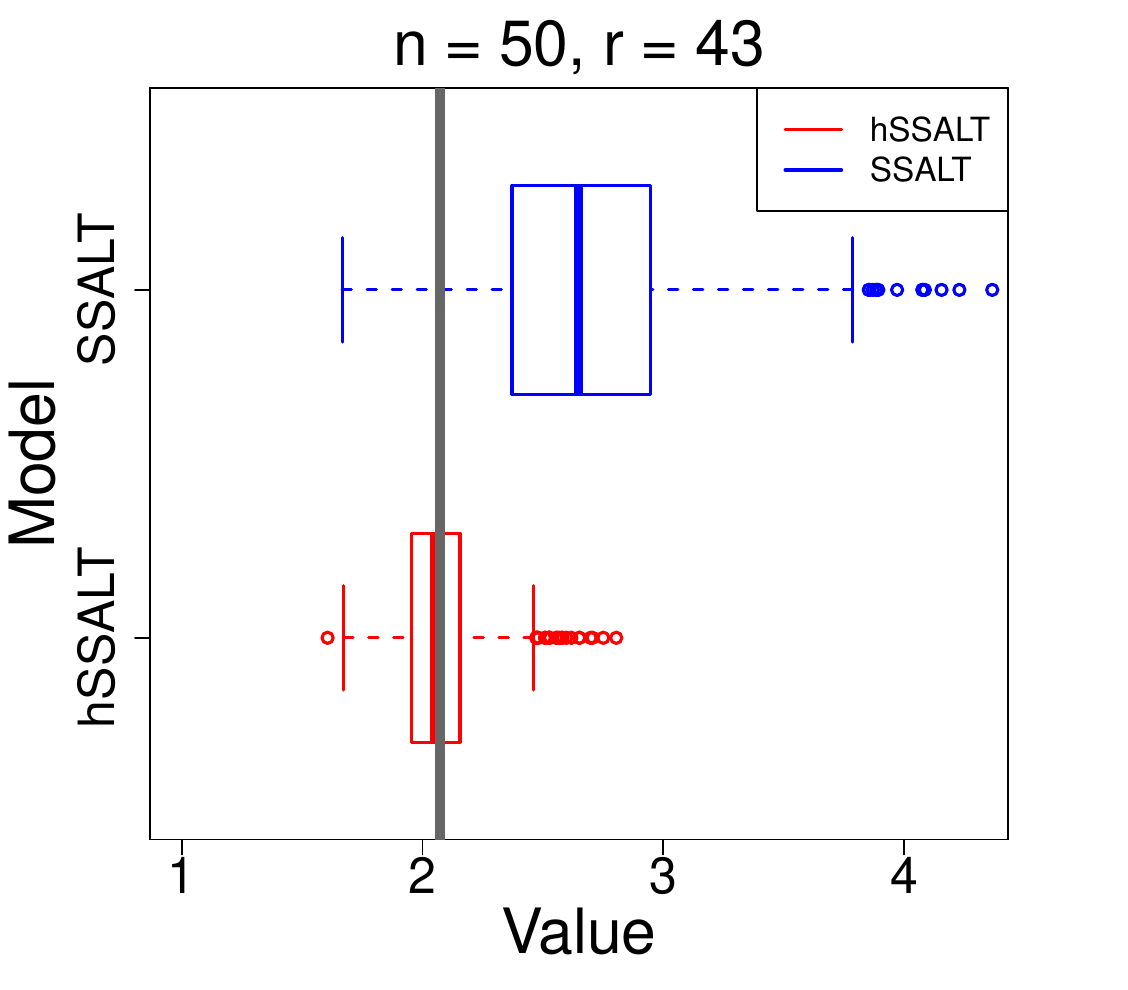}\hfill
\includegraphics[width=0.32\textwidth]{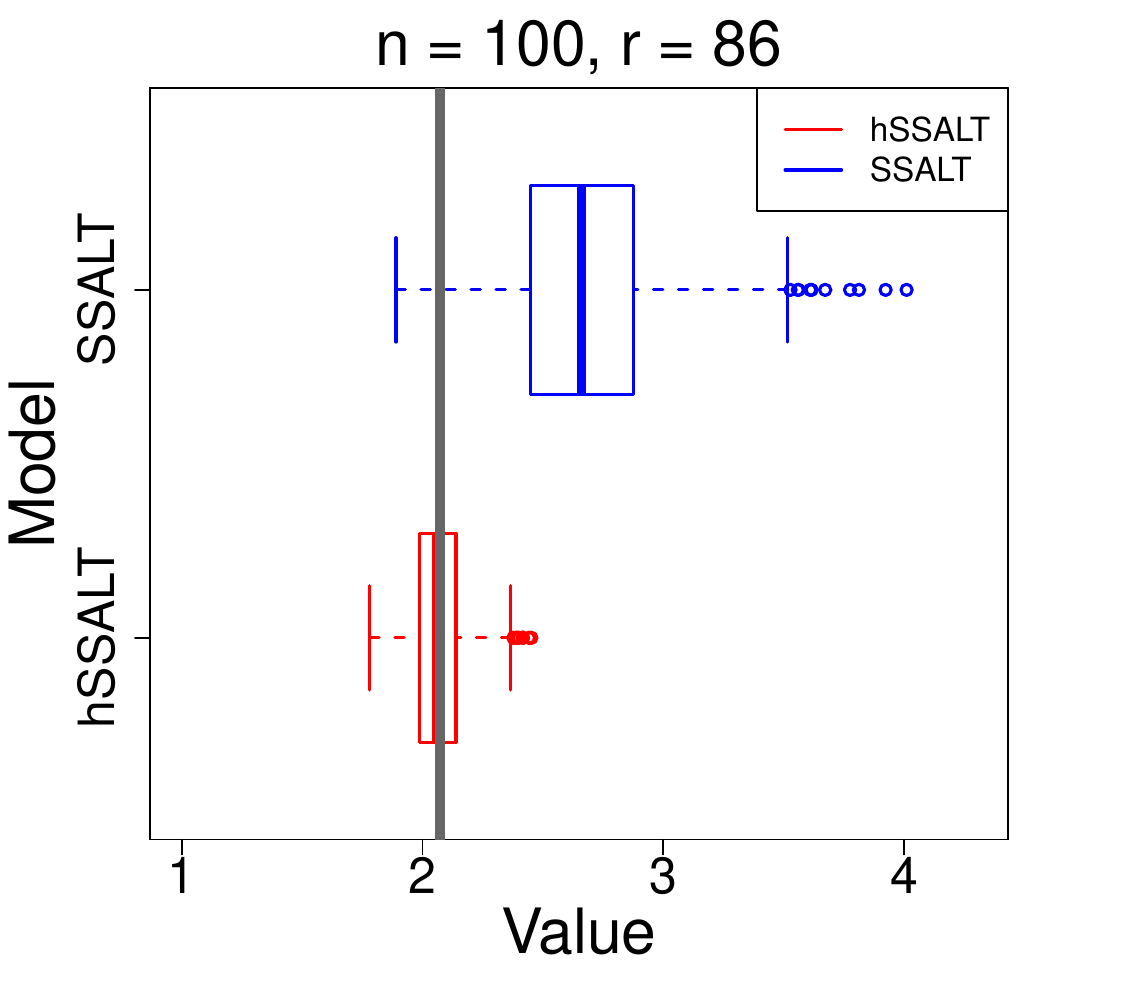}

\vspace{2pt}

\includegraphics[width=0.32\textwidth]{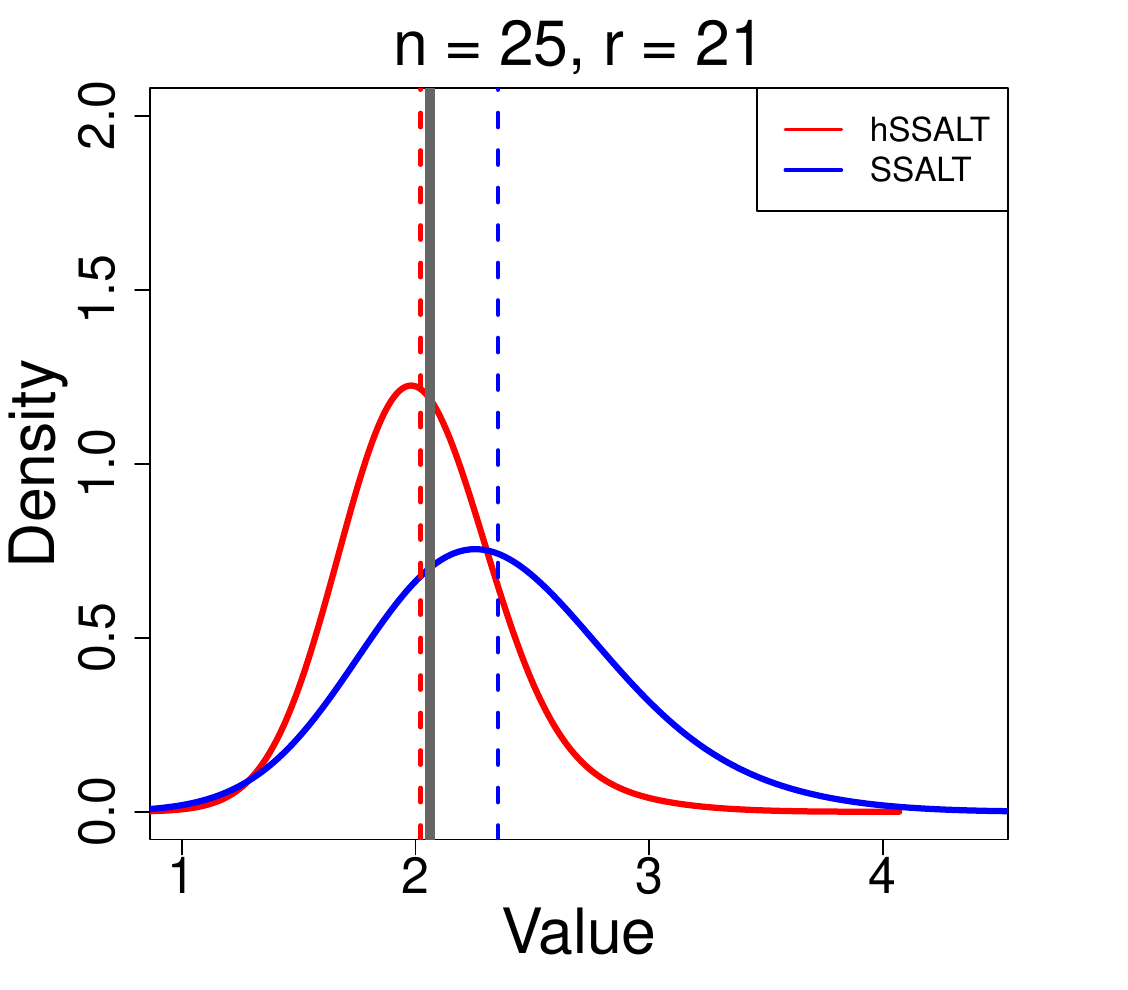}\hfill
\includegraphics[width=0.32\textwidth]{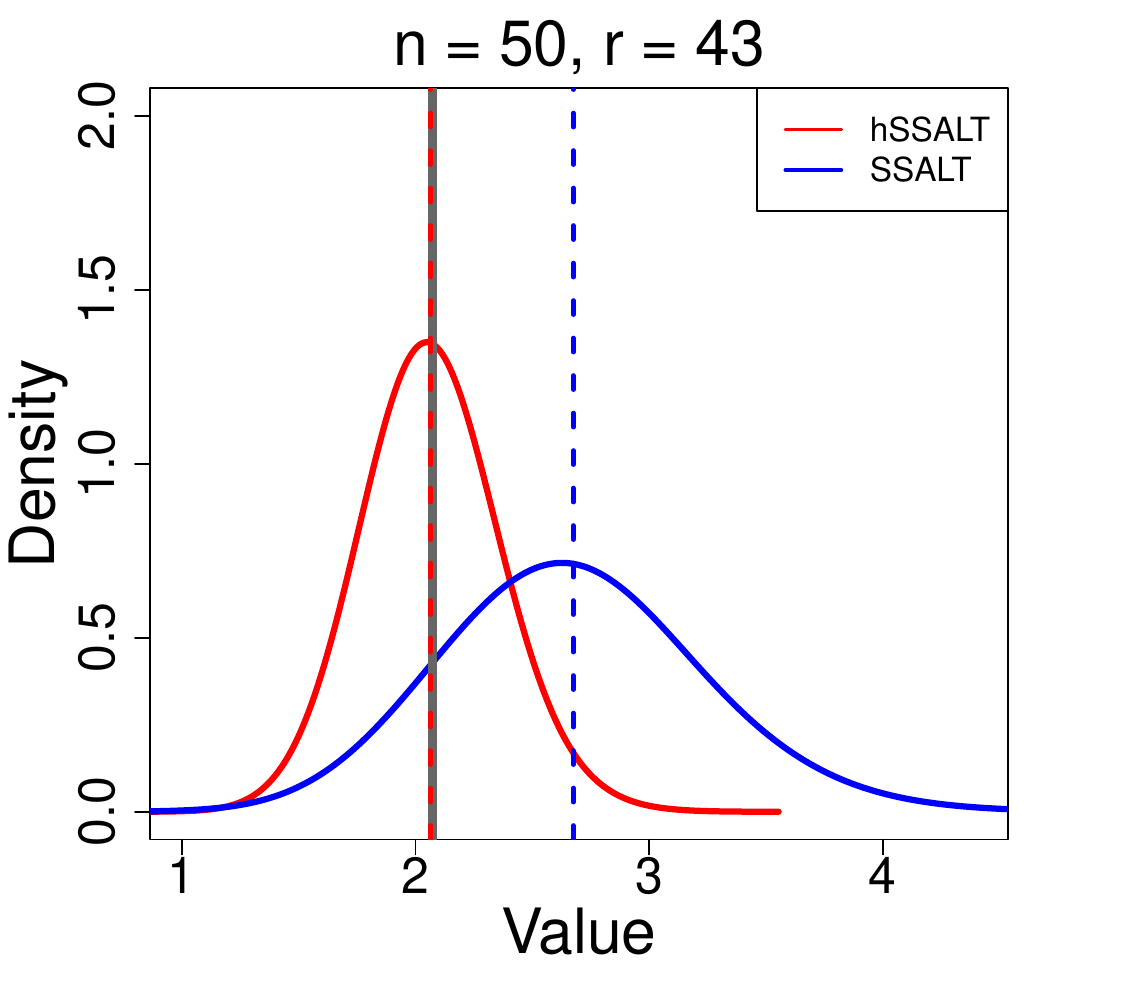}\hfill
\includegraphics[width=0.32\textwidth]{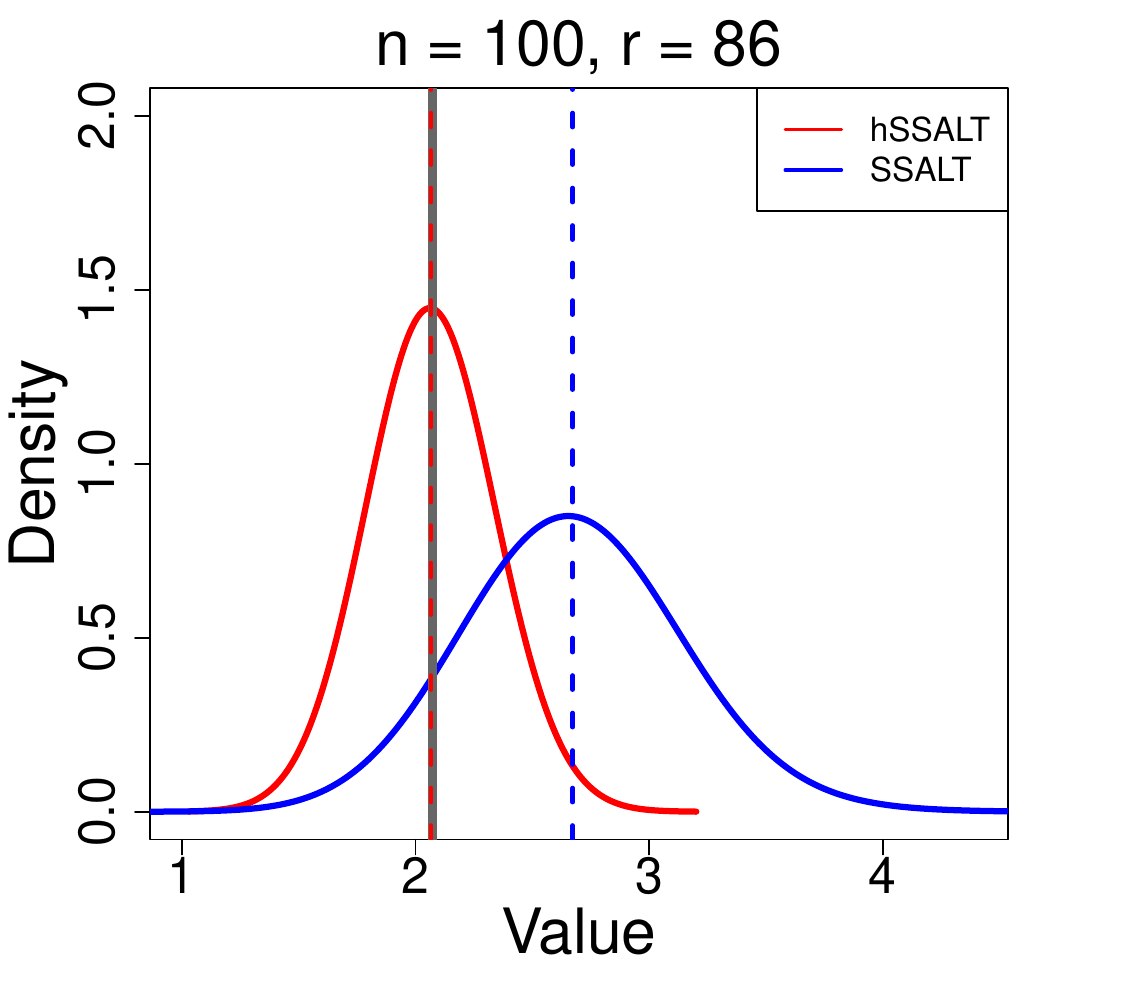}

\caption{Boxplots (top) and kernel density plots (bottom) of the estimated median quantile ($q = 0.50$) under the $h$-SSALT (red) and SSALT (blue) models, based on $1000$ simulations, with columns corresponding to $(n,r) \in \{(25,21),(50,43),(100,86)\}$. Dashed lines indicate mean estimates and the grey solid line represents the true quantile value.}
\label{fig:comp_ssalt_hSSALT_q050}
\end{figure}

\noindent Figure~\ref{fig:comp_ssalt_hSSALT_q050} provides a visualization of the median quantile (i.e., $q = 0.50$) comparison through boxplots and kernel density estimates, based on $1000$ replications for $(n, r) \in \{(25, 21), (50, 43), (100, 86)\}$. The $h$-SSALT estimates (red) are tightly concentrated around the true quantile value (grey line), while the SSALT estimates (blue) exhibit a pronounced rightward shift and substantially larger spread, confirming the bias introduced by ignoring heterogeneity. The kernel density plots confirm that the SSALT sampling distribution is right-skewed and displaced from the truth, whereas the $h$-SSALT distribution is approximately symmetric and centered at the true value. As $n$ increases, both distributions narrow, but the $h$-SSALT model converges to the truth considerably faster. 

\begin{table}[htbp]
\scriptsize
\centering
\caption{Mean estimates and RMSE of the $t_{0.01}$, $t_{0.05}$, and $t_{0.10}$ lives under the $h$-SSALT and SSALT models for selected $(n,r,\tau)$ combinations. These early failure quantiles are of direct relevance to warranty period design and safety-critical component qualification.}
\setlength{\tabcolsep}{3pt}
\begin{tabular}{ccccccccccccccccccc}
\toprule
$n$ & $r$ & $\tau$ &
\multicolumn{5}{c}{$q=0.01$ ($t_{0.01}$ life)} &
\multicolumn{5}{c}{$q=0.05$ ($t_{0.05}$ life)} &
\multicolumn{5}{c}{$q=0.10$ ($t_{0.10}$ life)} \\
\cmidrule(lr){4-8}
\cmidrule(lr){9-13}
\cmidrule(lr){14-18}
& & &
\textit{$t_{0.01}$} &
\multicolumn{2}{c}{$h$-SSALT} &
\multicolumn{2}{c}{SSALT} &
\textit{$t_{0.05}$} &
\multicolumn{2}{c}{$h$-SSALT} &
\multicolumn{2}{c}{SSALT} &
\textit{$t_{0.10}$} &
\multicolumn{2}{c}{$h$-SSALT} &
\multicolumn{2}{c}{SSALT} \\
\cmidrule(lr){5-6}
\cmidrule(lr){7-8}
\cmidrule(lr){10-11}
\cmidrule(lr){12-13}
\cmidrule(lr){15-16}
\cmidrule(lr){17-18}
& & & & Mean & RMSE & Mean & RMSE
      & & Mean & RMSE & Mean & RMSE
      & & Mean & RMSE & Mean & RMSE \\
\midrule
35 & 30 & 1.60 & 0.0827&0.1363&0.1260&0.0570&0.0571&0.3218&0.4005&0.2130&0.2508&0.1497&0.5862&0.6796&0.2702&0.4266&0.2617 \\
   &    & 1.70 & 0.0827&0.1367&0.1204&0.0625&0.0597&0.3218&0.4052&0.2105&0.2625&0.1438&0.5862&0.6774&0.2606&0.4297&0.2449 \\
   &    & 1.80 & 0.0827&0.1278&0.1132&0.0655&0.0604&0.3218&0.3924&0.2016&0.2698&0.1408&0.5862&0.6695&0.2567&0.4314&0.2309 \\
\addlinespace
   & 35 & 1.60 & 0.0827&0.1288&0.1074&0.0239&0.0633&0.3218&0.3971&0.1991&0.1644&0.1856&0.5862&0.6786&0.2528&0.3283&0.3137 \\
   &    & 1.70 & 0.0827&0.1255&0.1051&0.0253&0.0616&0.3218&0.3917&0.1999&0.1671&0.1777&0.5862&0.6774&0.2475&0.3189&0.3112 \\
   &    & 1.80 & 0.0827&0.1265&0.1005&0.0280&0.0599&0.3218&0.3889&0.1842&0.1687&0.1737&0.5862&0.6665&0.2413&0.3130&0.3095 \\
\midrule
50 & 43 & 1.60 & 0.0827&0.1175&0.0980&0.0461&0.0522&0.3218&0.3715&0.1711&0.2254&0.1396&0.5862&0.6513&0.2074&0.3801&0.2488 \\
   &    & 1.70 & 0.0827&0.1186&0.0905&0.0534&0.0524&0.3218&0.3772&0.1682&0.2426&0.1324&0.5862&0.6506&0.2081&0.3854&0.2445 \\
   &    & 1.80 & 0.0827&0.1128&0.0812&0.0530&0.0500&0.3218&0.3705&0.1618&0.2430&0.1304&0.5862&0.6500&0.2158&0.3943&0.2365 \\
\addlinespace
   & 50 & 1.60 & 0.0827&0.1189&0.0855&0.0205&0.0640&0.3218&0.3808&0.1624&0.1519&0.1811&0.5862&0.6535&0.2087&0.2980&0.3202 \\
   &    & 1.70 & 0.0827&0.1184&0.0833&0.0217&0.0632&0.3218&0.3804&0.1596&0.1526&0.1811&0.5862&0.6324&0.1836&0.2882&0.3167 \\
   &    & 1.80 & 0.0827&0.1178&0.0844&0.0245&0.0611&0.3218&0.3779&0.1603&0.1593&0.1758&0.5862&0.6299&0.1867&0.2877&0.3123 \\
\midrule
100 & 86 & 1.60 & 0.0827&0.1020&0.0558&0.0386&0.0503&0.3218&0.3547&0.1089&0.2096&0.1311&0.5862&0.6147&0.1316&0.3360&0.2631 \\
    &    & 1.70 & 0.0827&0.0972&0.0514&0.0407&0.0489&0.3218&0.3450&0.1067&0.2141&0.1279&0.5862&0.6187&0.1353&0.3476&0.2529 \\
    &    & 1.80 & 0.0827&0.0960&0.0508&0.0441&0.0464&0.3218&0.3422&0.1045&0.2216&0.1218&0.5862&0.6196&0.1364&0.3607&0.2413 \\
\addlinespace
    & 100 & 1.60 & 0.0827&0.1016&0.0502&0.0171&0.0663&0.3218&0.3547&0.1031&0.1382&0.1880&0.5862&0.6146&0.1333&0.2702&0.3240 \\
    &     & 1.70 & 0.0827&0.1006&0.0486&0.0184&0.0651&0.3218&0.3536&0.1012&0.1414&0.1852&0.5862&0.6220&0.1290&0.2716&0.3214 \\
    &     & 1.80 & 0.0827&0.0989&0.0505&0.0193&0.0642&0.3218&0.3489&0.1050&0.1434&0.1834&0.5862&0.6172&0.1295&0.2729&0.3197 \\
\bottomrule
\end{tabular}
\label{table:early_quantiles}
\end{table}

\begin{figure}[htbp]
\centering

% q = 0.01 block
\noindent\hspace{0.5em} $q = 0.01$

\vspace{3.5pt}

\includegraphics[width=0.32\textwidth]{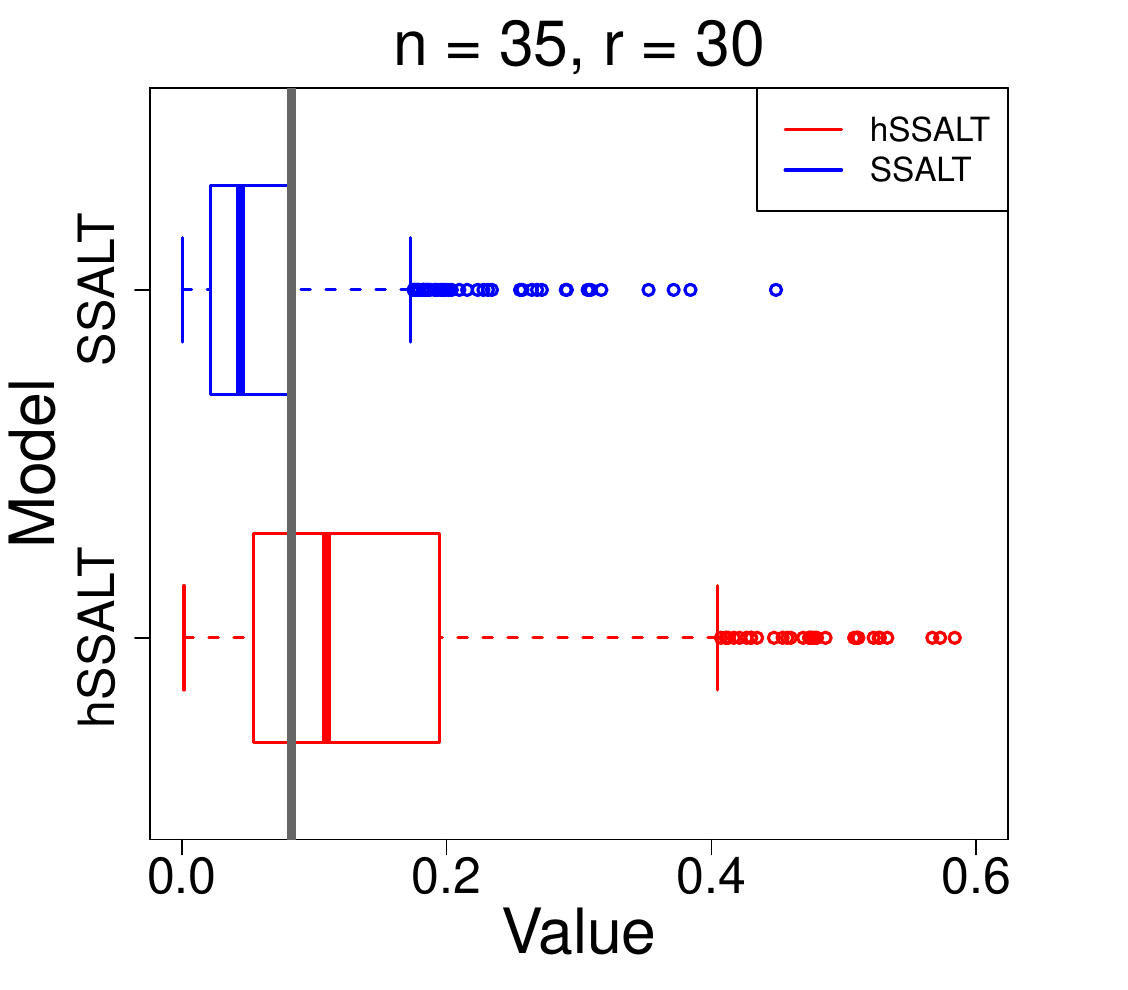}\hfill
\includegraphics[width=0.32\textwidth]{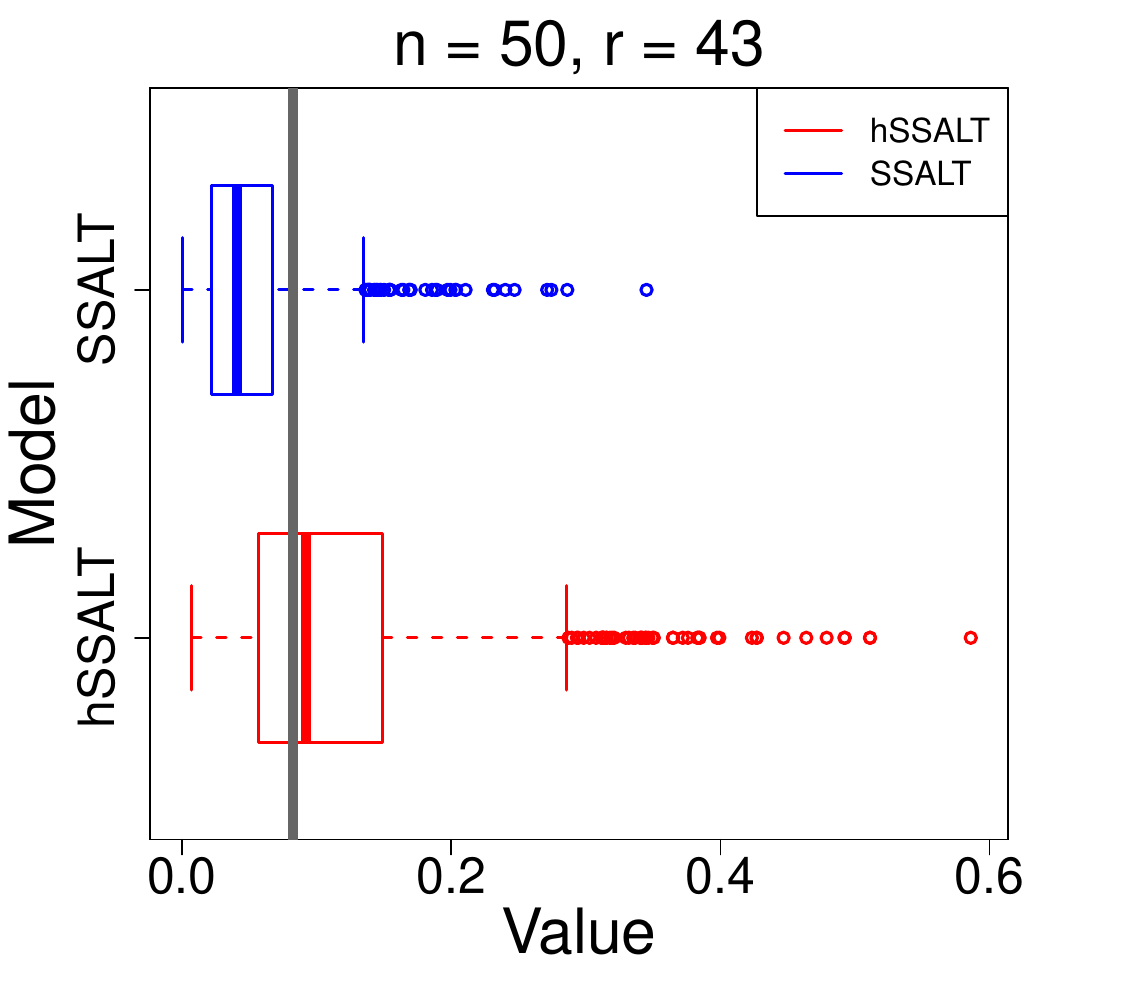}\hfill
\includegraphics[width=0.32\textwidth]{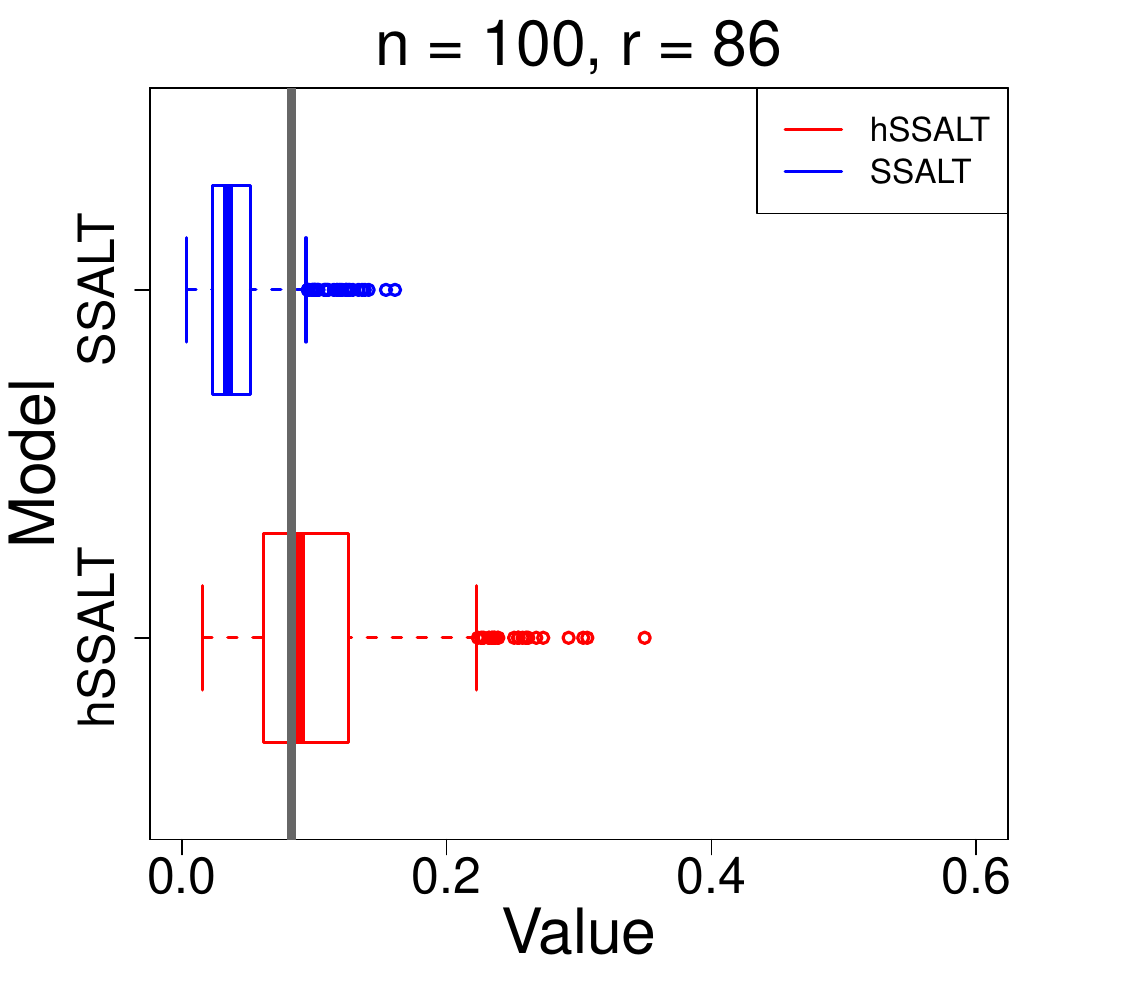}

\vspace{2pt}

\includegraphics[width=0.32\textwidth]{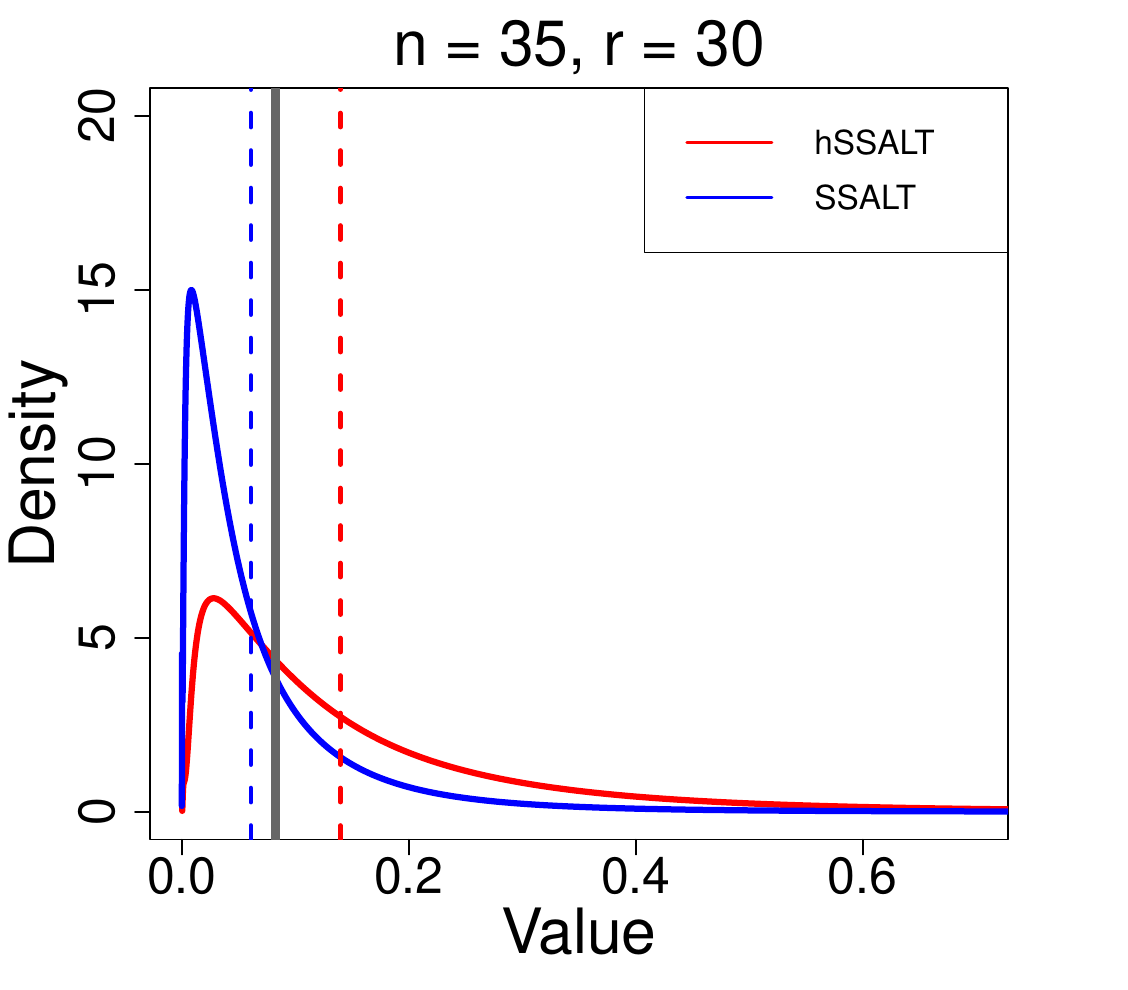}
\hfill
\includegraphics[width=0.32\textwidth]{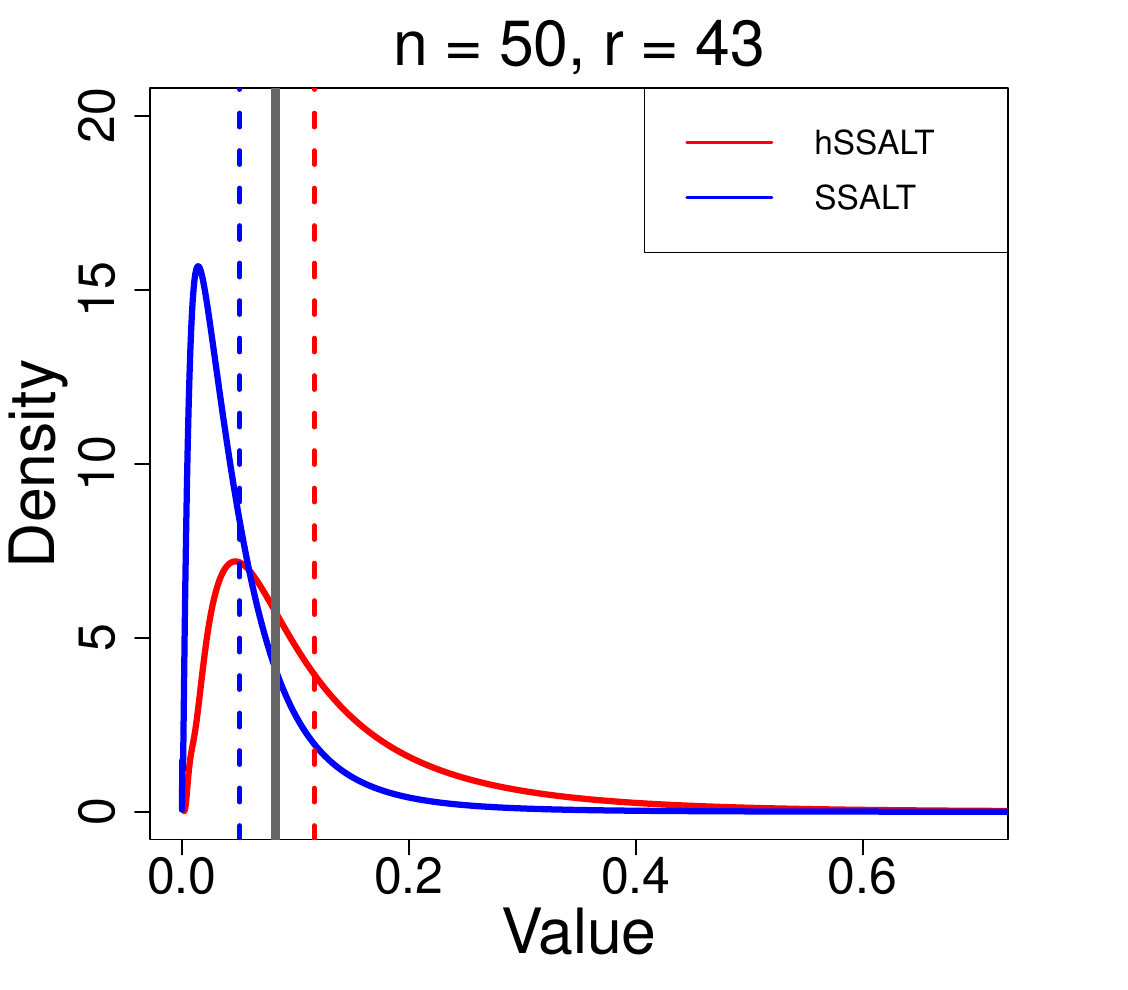}
\hfill
\includegraphics[width=0.32\textwidth]{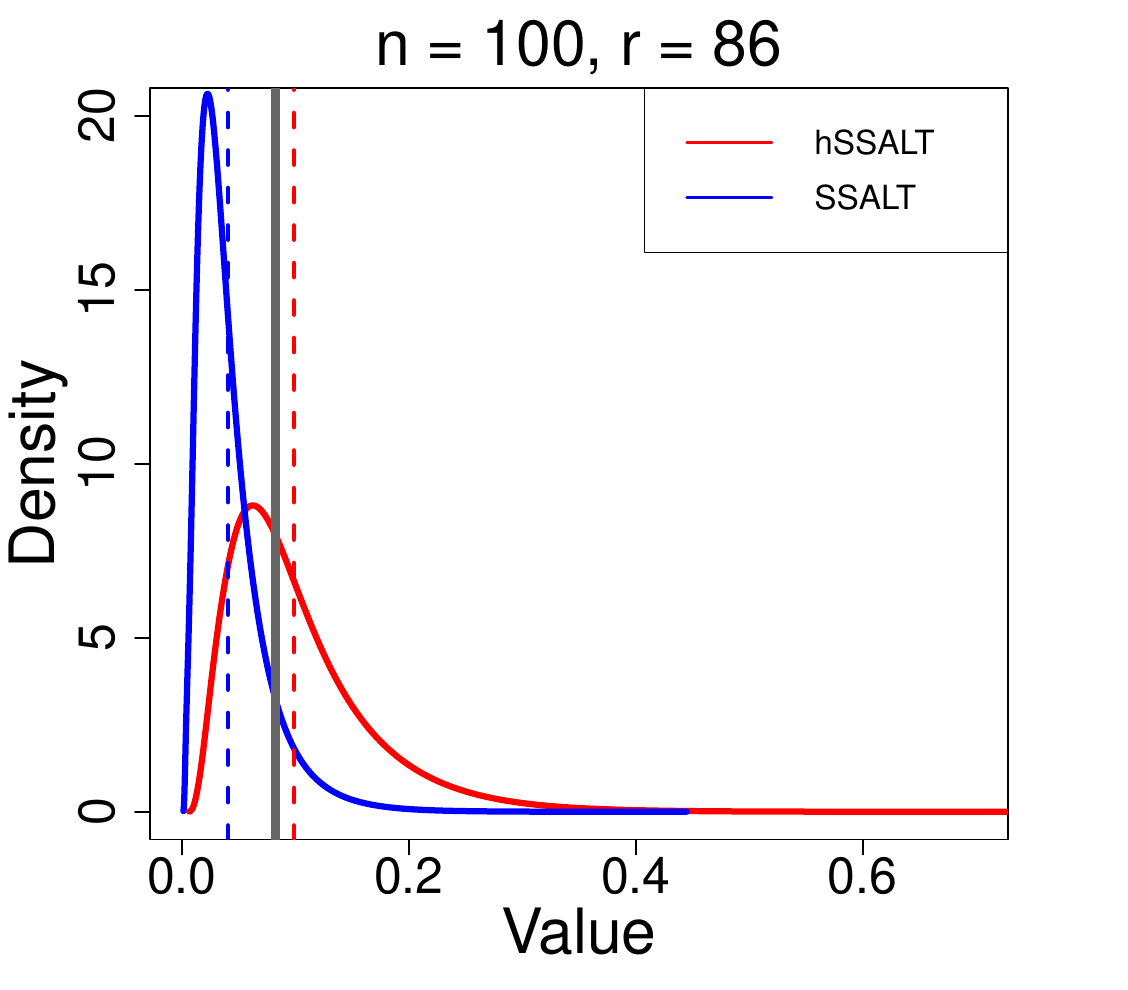}

\vspace{10pt}

% q = 0.05 block
\noindent\hspace{0.5em} $q = 0.05$

\vspace{3.5pt}

\includegraphics[width=0.32\textwidth]{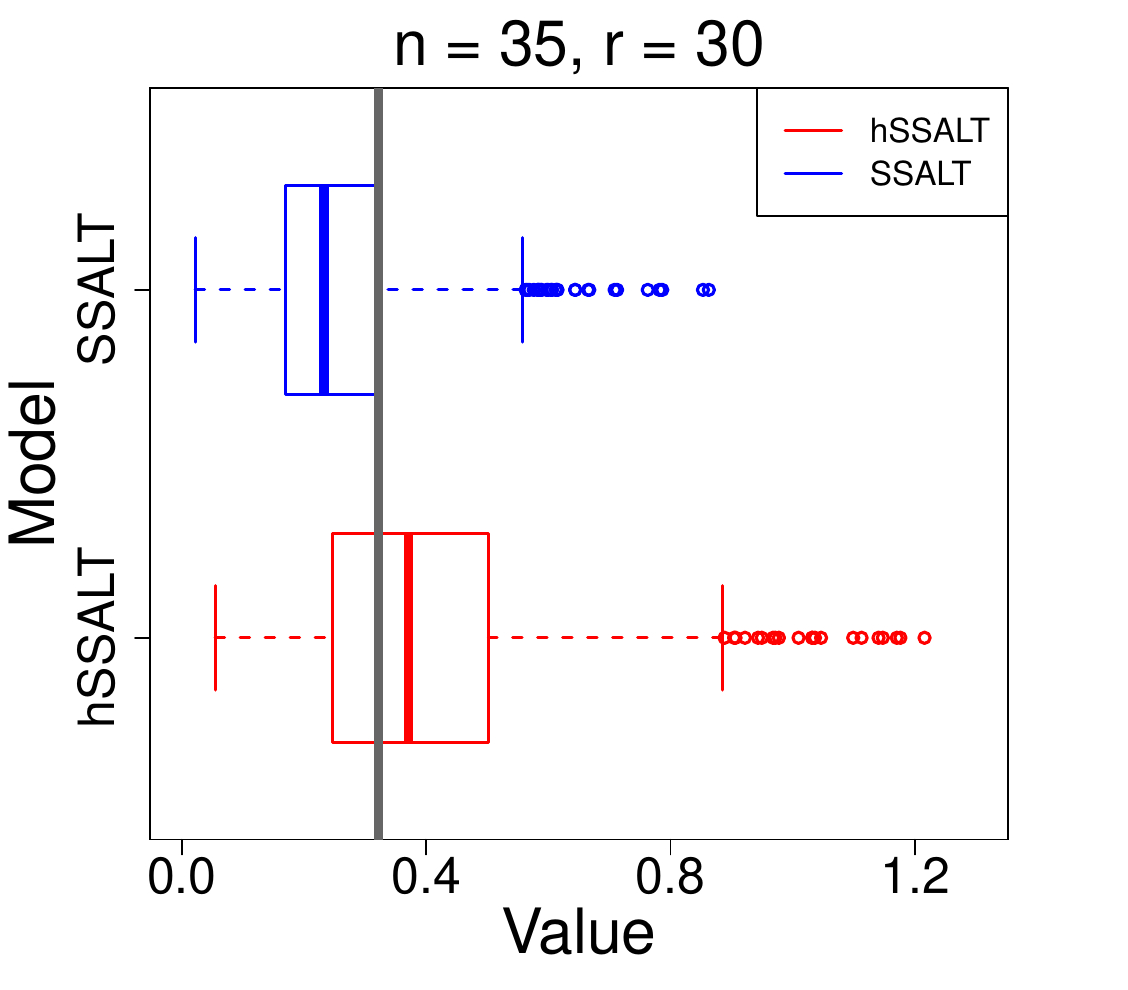}\hfill
\includegraphics[width=0.32\textwidth]{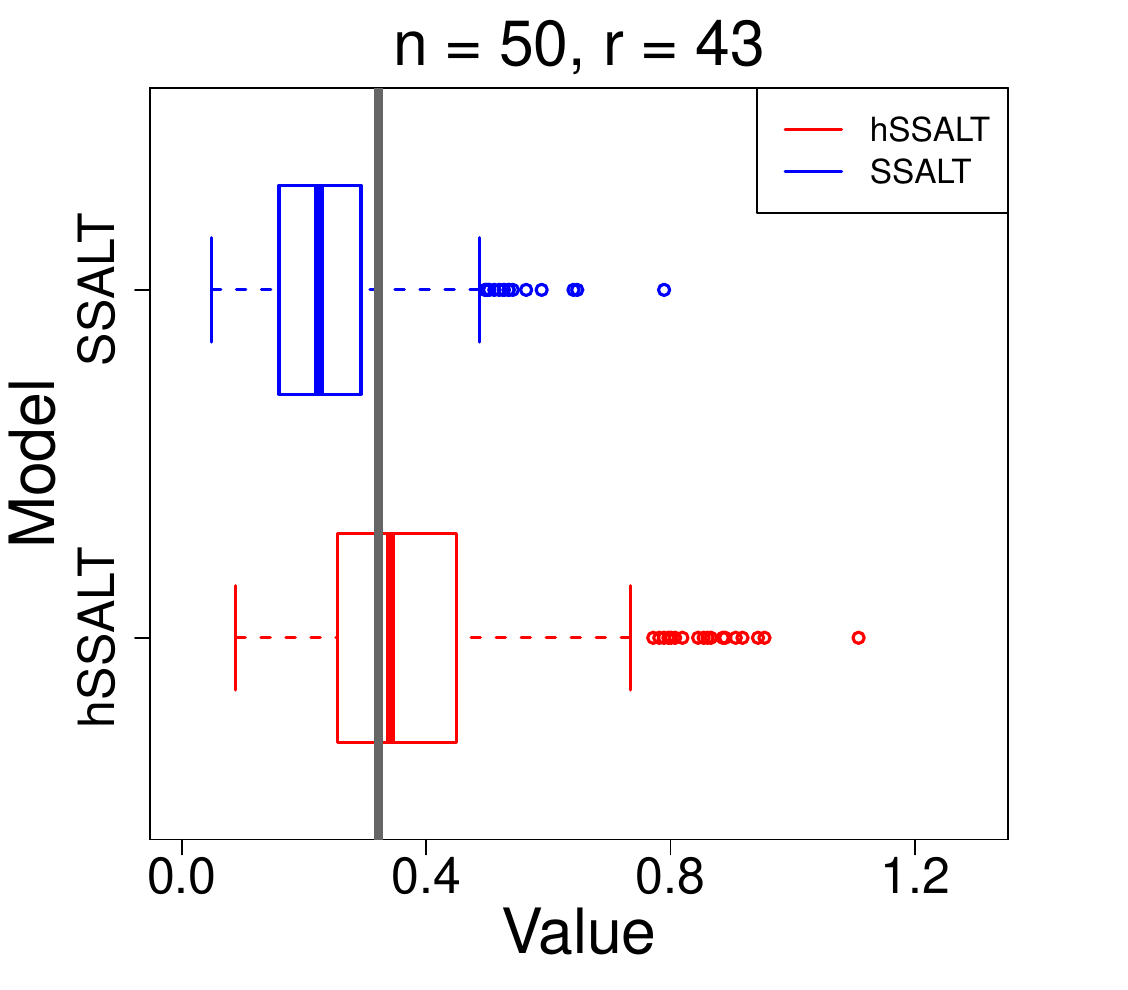}\hfill
\includegraphics[width=0.32\textwidth]{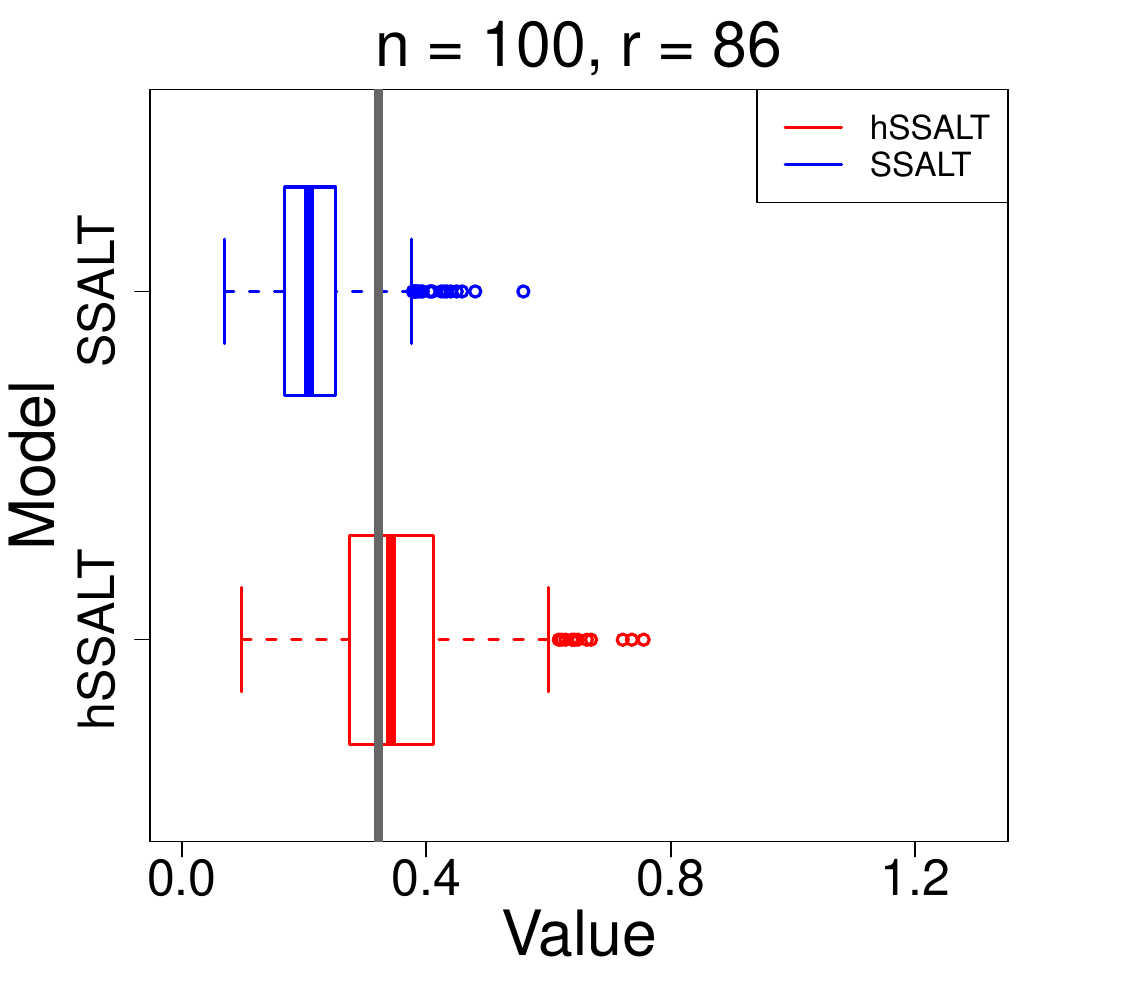}

\vspace{2pt}

\includegraphics[width=0.32\textwidth]{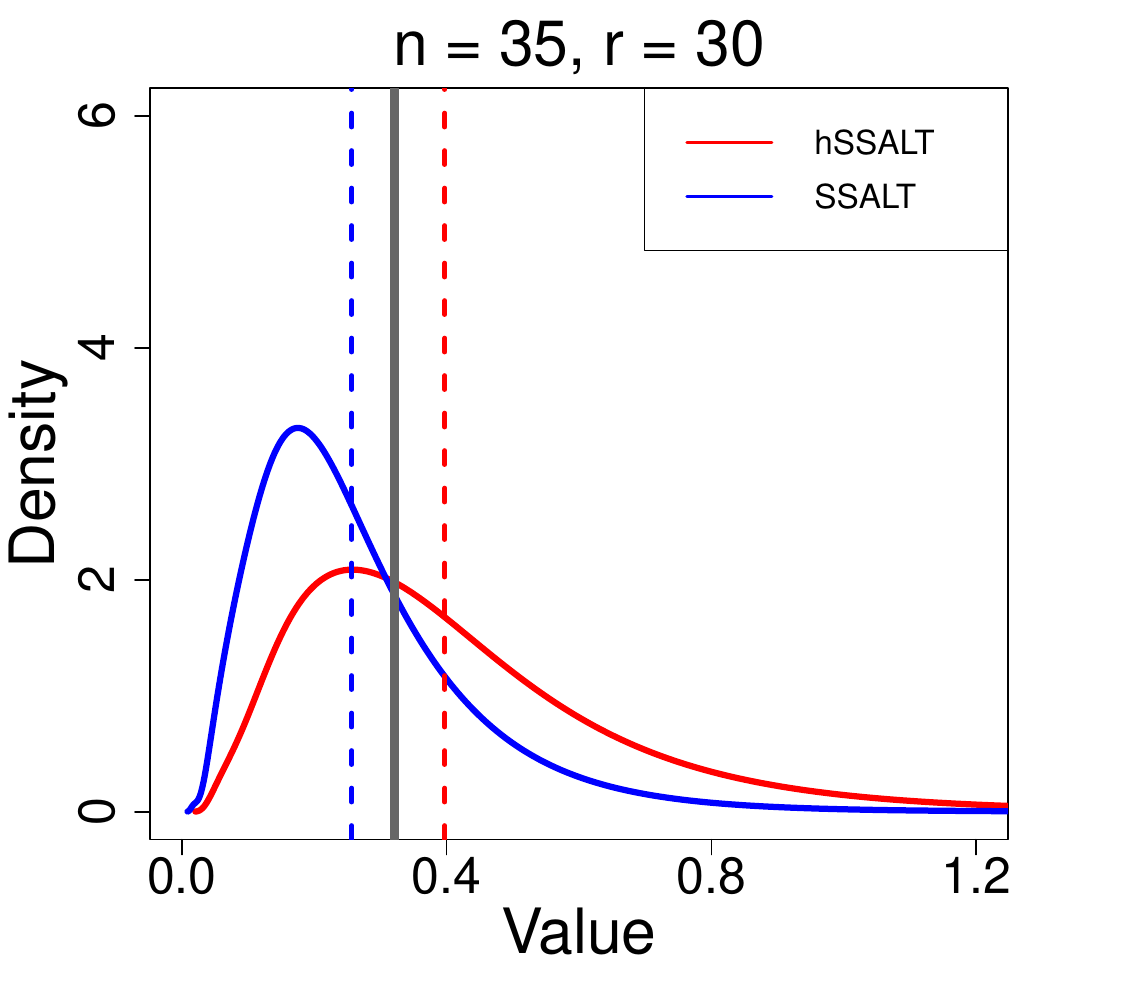}\hfill
\includegraphics[width=0.32\textwidth]{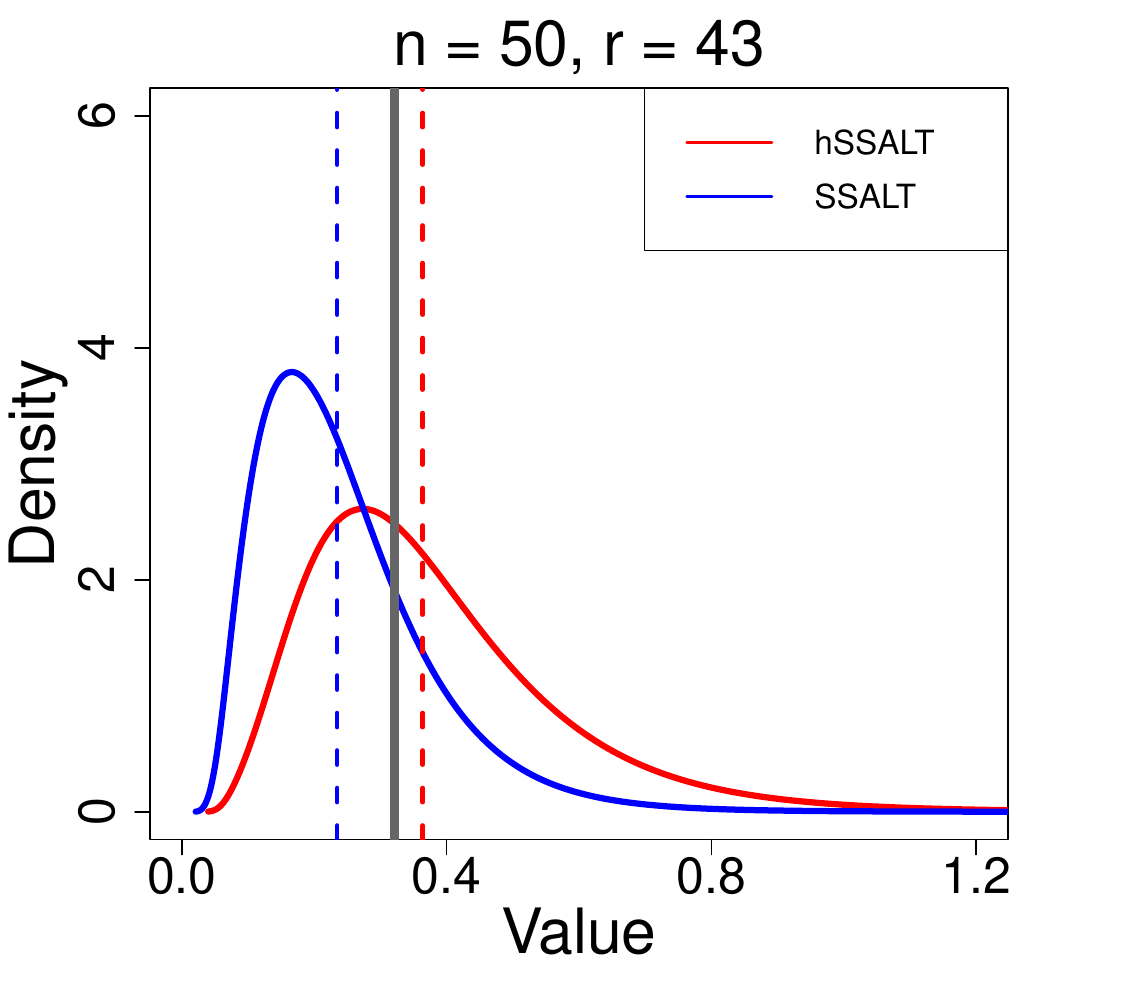}\hfill
\includegraphics[width=0.32\textwidth]{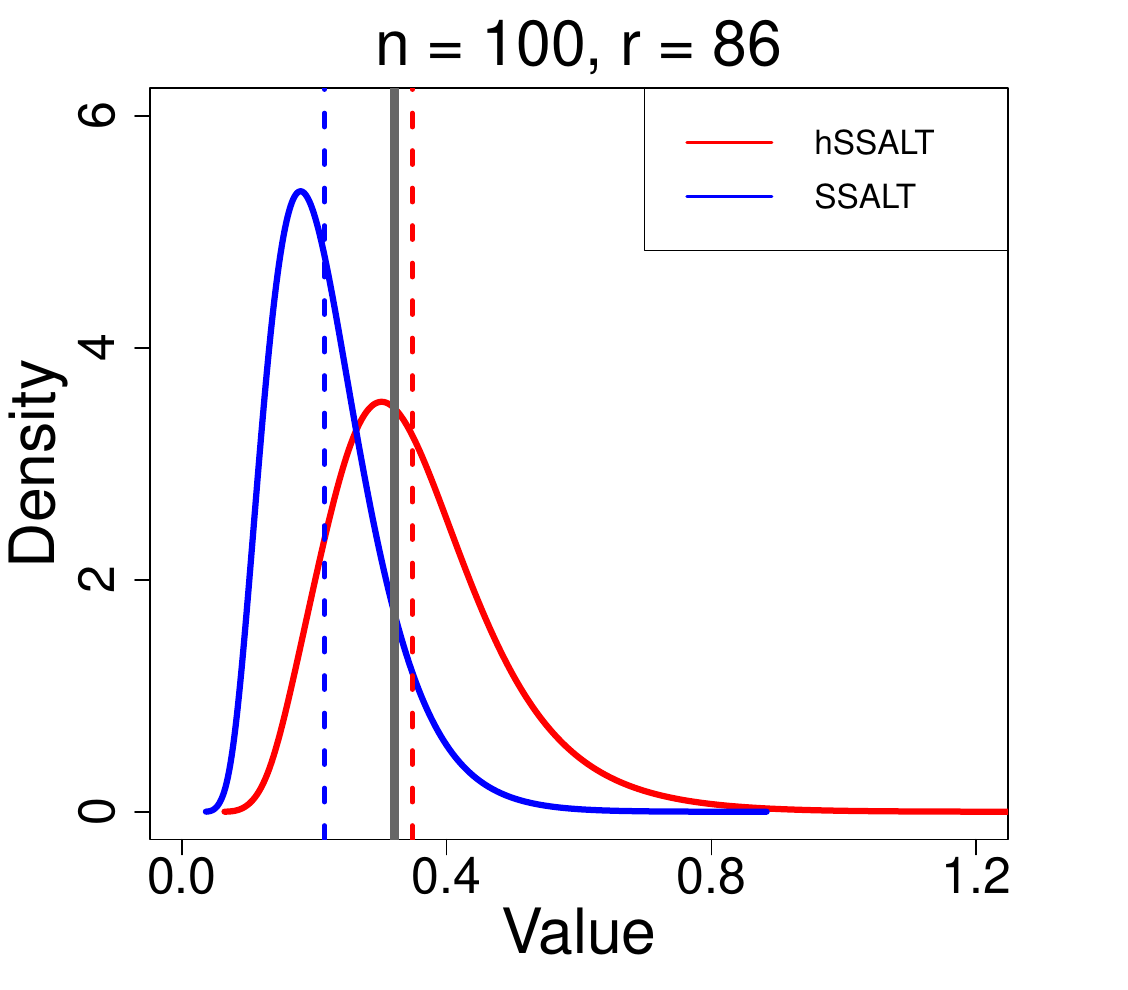}
\caption{Boxplots and kernel density plots of the estimated early failure quantiles ($q \in \{0.01, 0.05\}$), as in Figure~\ref{fig:comp_ssalt_hSSALT_q050} but with columns corresponding to $(n,r) \in \{(35,30),(50,43),(100,86)\}$.}
\label{fig:comp_ssalt_hSSALT_q001_005}
\end{figure}

The practical consequences of model misspecification are most severe at the early failure quantiles shown in Table \ref{table:early_quantiles}. The $t_{0.01}$ life at $q = 0.01$, the $t_{0.05}$ life at $q = 0.05$, and the $t_{0.10}$ life at $q = 0.10$ are the standard metrics used by engineers for warranty period specification and safety-critical component qualification, where even a small proportion of early failures can have serious consequences \citep{meeker1998}. The homogeneous SSALT model, by ignoring the mixture structure, consistently underestimates these early lives across all sample sizes and stress-change times considered. This underestimation is not a small-sample artifact: at $n = 100$ and $r = 86$, the SSALT mean estimate of $\hat{t}_{0.01}$ is $0.0386$ to $0.0441$, compared to the true value of $0.0827$, a systematic underestimation of approximately 50\%. For $q = 0.10$, the SSALT mean estimates range from $0.3360$ to $0.3607$ at $n = 100$, compared to the true value of $0.5862$, again less than two-thirds of the true value. This bias does not diminish as $n$ increases to $100$, confirming that it is a structural consequence of model misspecification rather than finite-sample noise, indicating serious consistency issues with the early quantile estimates under the SSALT model. In practice, a manufacturer relying on the SSALT model would underestimate how quickly early failures occur, leading to an overly optimistic assessment of product reliability at early times. In contrast, the $h$-SSALT model produces mean estimates that are well-centred around the true quantile lives across all sample sizes and configurations, with estimates slightly above the true quantile value in most configurations, a modest and diminishing bias confirming consistency. For complete samples ($r = n$), the $h$-SSALT model yields consistently lower RMSEs than censored samples of the same size at these early quantiles, as the full observation of failure times provides more information for identifying the short-lived subgroup. The SSALT model exhibits the opposite pattern: complete samples yield higher RMSEs than censored samples at early quantiles, since the additional observations at the second stress level expose the homogeneous misspecification more severely. 

\begin{figure}[htbp]
\centering

\noindent\hspace{0.5em}\textit{$q = 0.10$}

\vspace{3.5pt}

\includegraphics[width=0.32\textwidth]{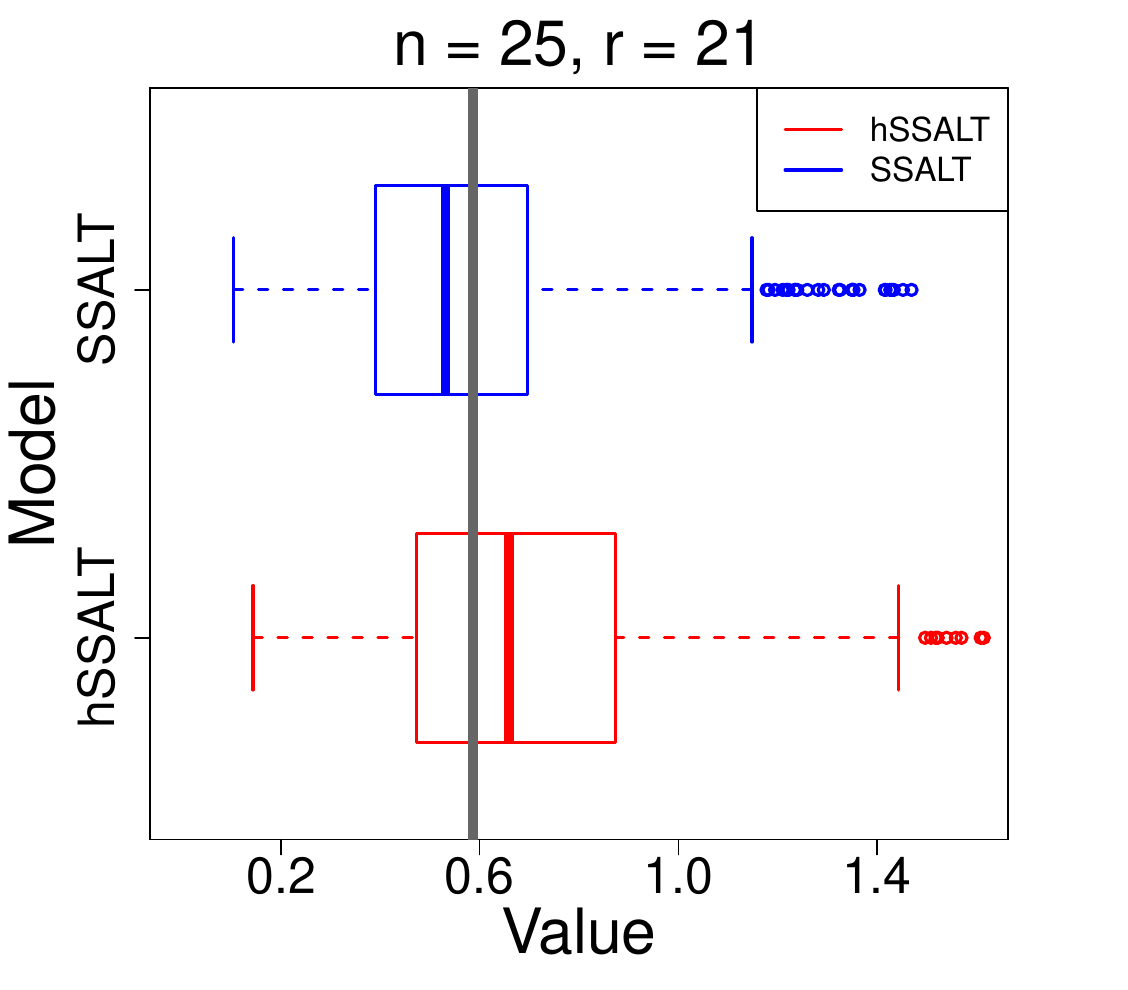}\hfill
\includegraphics[width=0.32\textwidth]{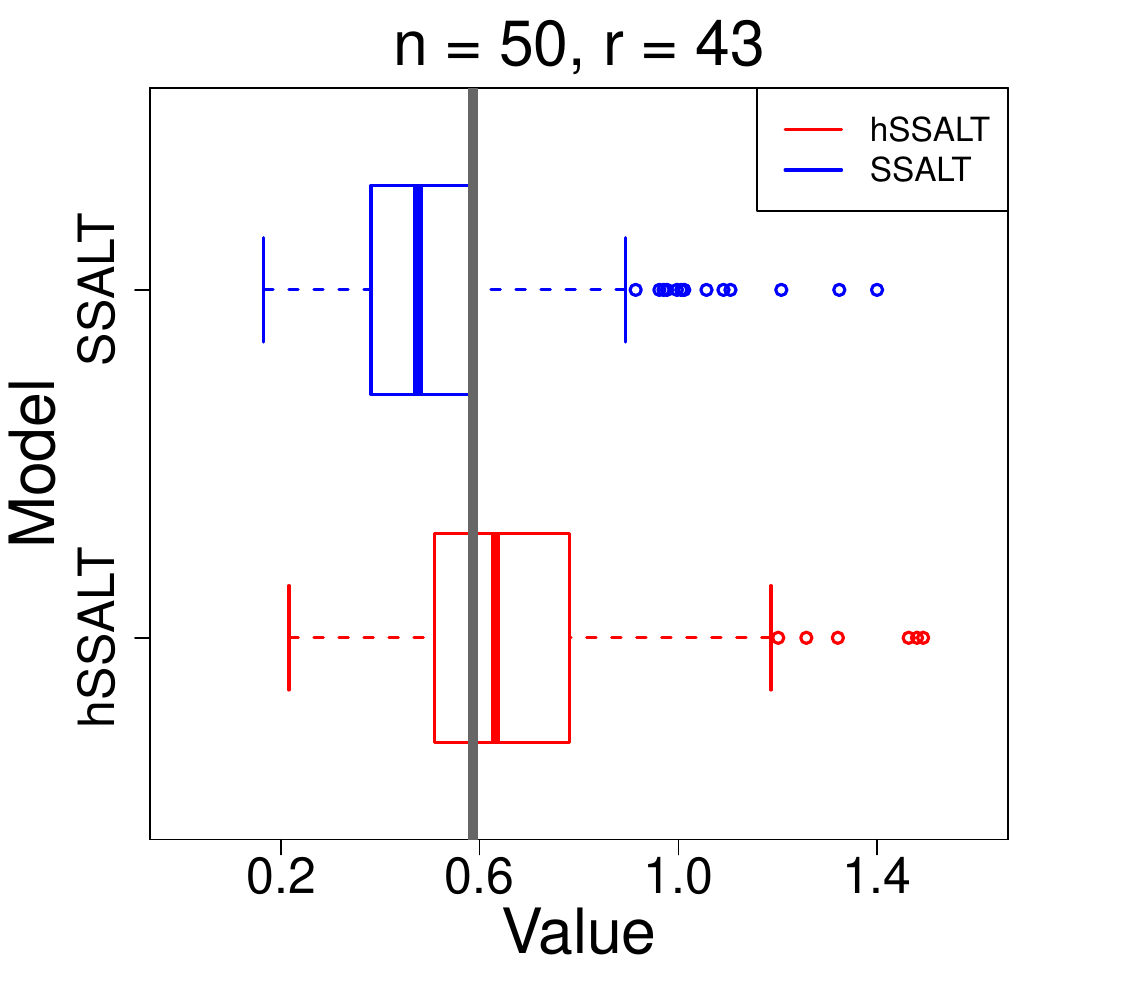}\hfill
\includegraphics[width=0.32\textwidth]{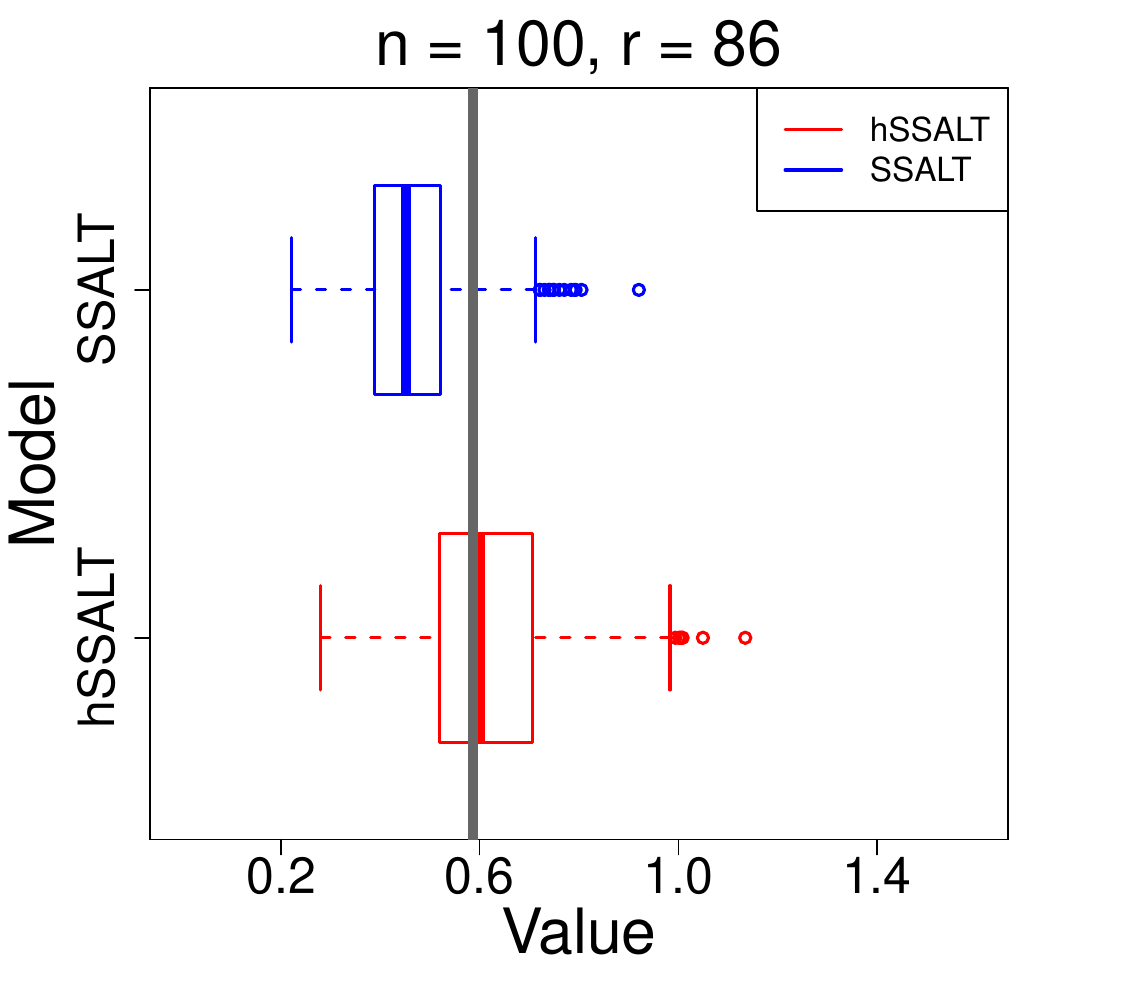}

\vspace{2pt}

\includegraphics[width=0.32\textwidth]{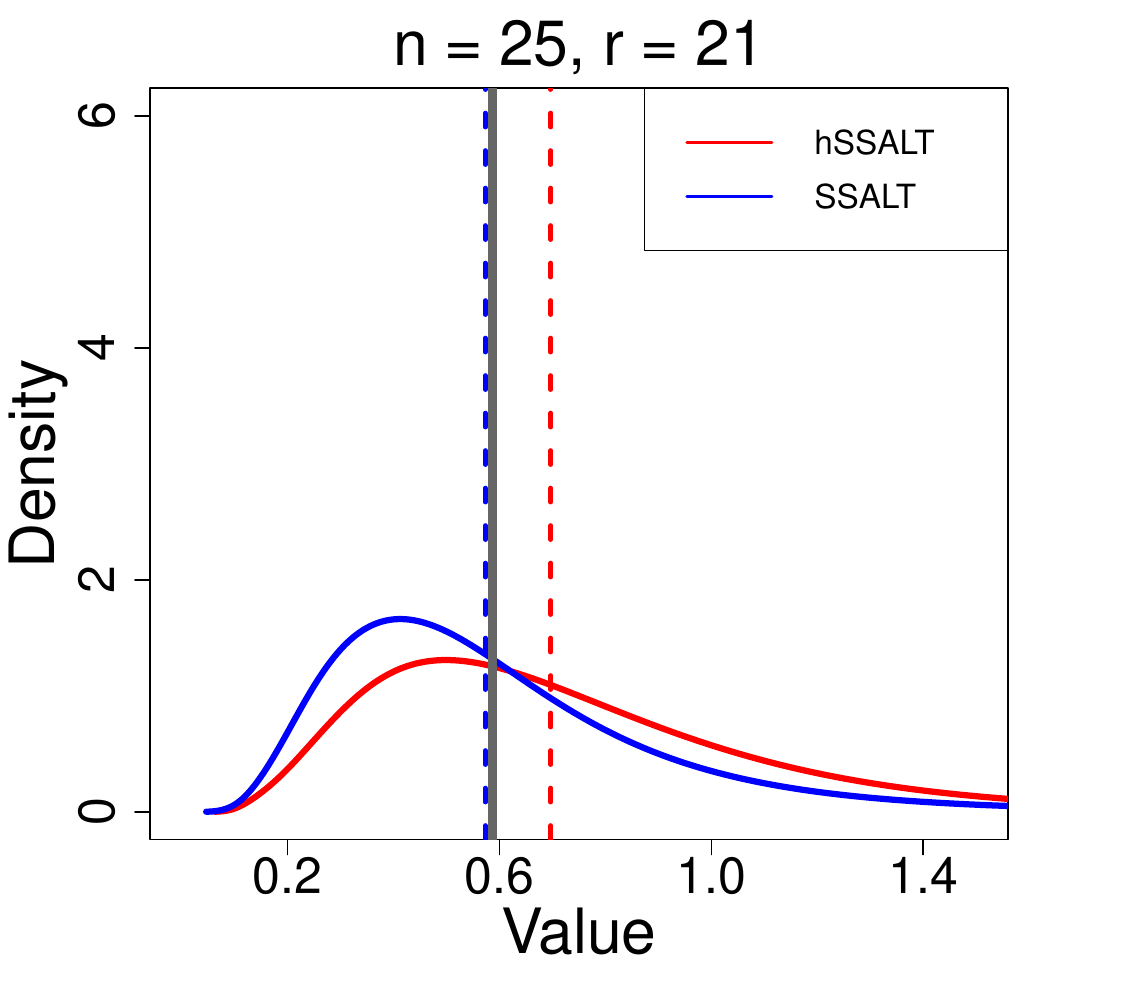}\hfill
\includegraphics[width=0.32\textwidth]{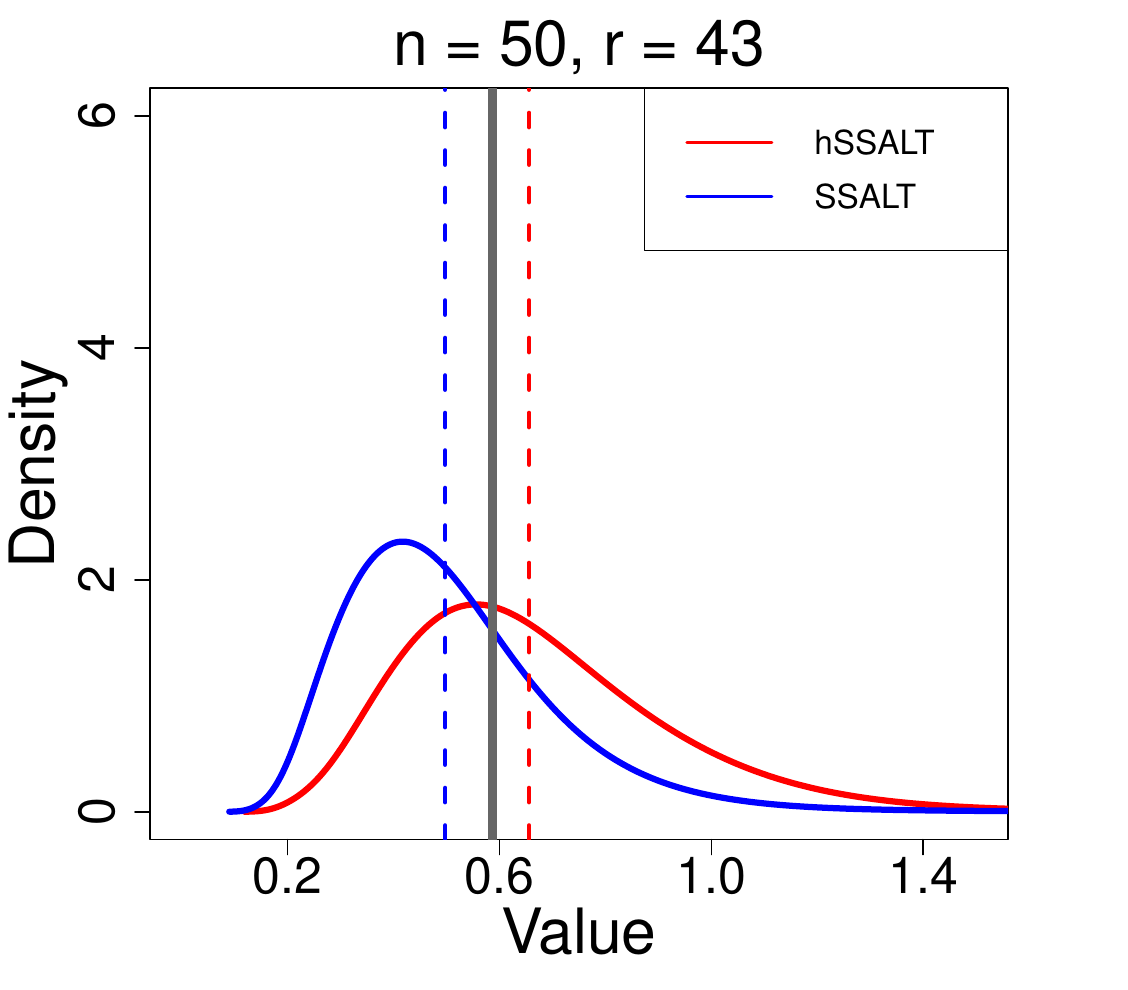}\hfill
\includegraphics[width=0.32\textwidth]{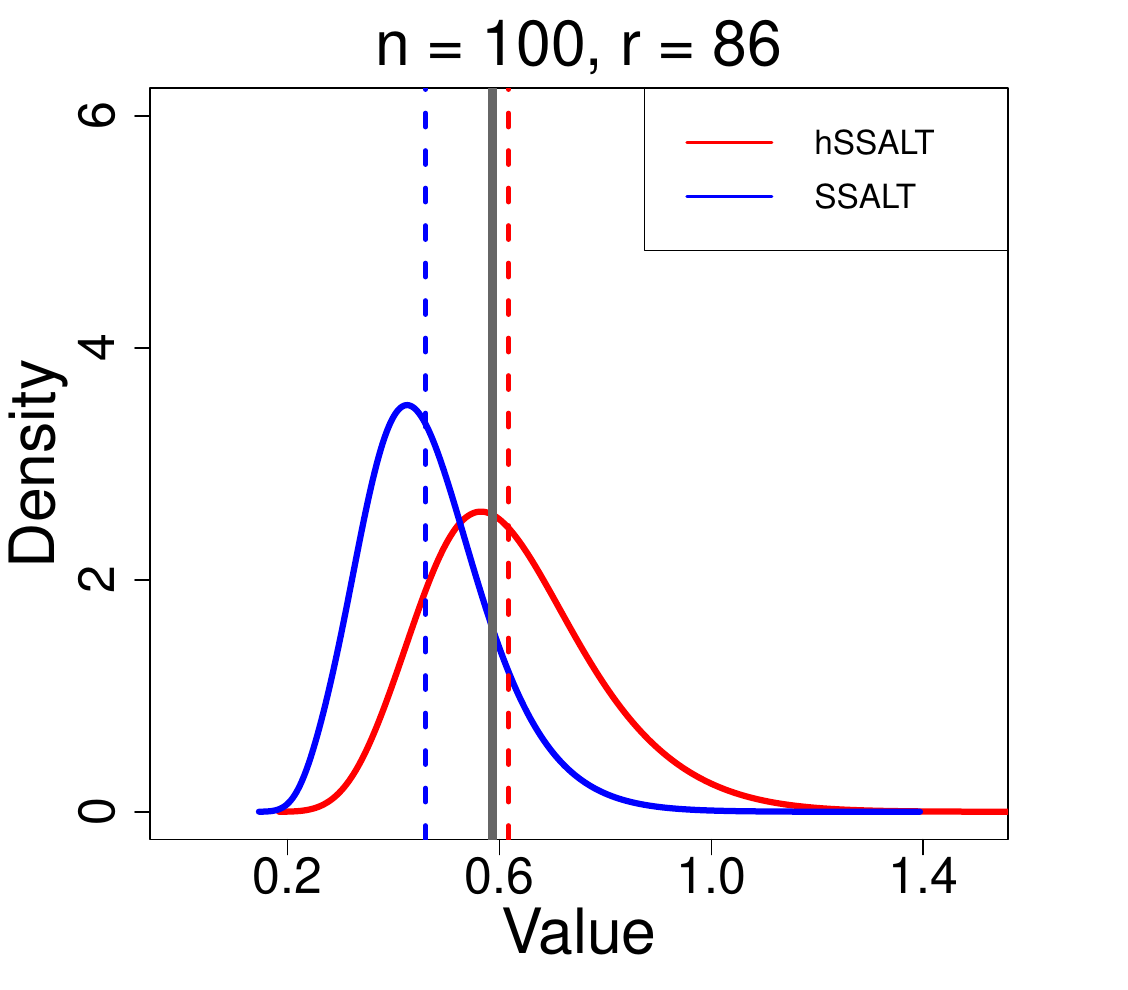}
\caption{Boxplots and kernel density plots of the estimated quantiles, as in Figure~\ref{fig:comp_ssalt_hSSALT_q050} 
but for $q = 0.10$, with columns corresponding to 
$(n,r) \in \{(25,21),(50,43),(100,86)\}$.}
\label{fig:comp_ssalt_hSSALT_q010}
\end{figure}

% \FloatBarrier
Figures~\ref{fig:comp_ssalt_hSSALT_q001_005} and~\ref{fig:comp_ssalt_hSSALT_q010} show the corresponding boxplots and kernel density plots for $q \in \{0.01, 0.05\}$ and $q = 0.10$ respectively, with columns corresponding to $(n,r) \in \{(35,30),(50,43),(100,86)\}$ and $(n,r) \in \{(25,21),(50,43),(100,86)\}$. In all three early failure quantiles, the SSALT distribution is shifted leftward relative to the true value and the $h$-SSALT estimates, visually confirming the consistent underestimation of the homogeneous model. This illustrates a classical bias-variance tradeoff: the SSALT model achieves lower variance at the cost of systematic bias, while $h$-SSALT sacrifices some variance to eliminate the bias, a tradeoff that is particularly visible at $q = 0.01$ for smaller sample sizes.

Across $\tau \in \{1.60, 1.70, 1.80\}$, the conclusions of the comparison remain qualitatively unchanged. For $h$-SSALT, 
the RMSEs at the median and lower quantiles increase slightly with $\tau$, reflecting fewer failures at the second stress level and 
consequently less information for estimating the mixture parameters. For the SSALT model, the overestimation bias at upper 
quantiles decreases modestly with $\tau$ as the true quantile value rises. Overall, the $h$-SSALT model outperforms SSALT 
consistently across all three stress-change times.

\FloatBarrier
\subsection{CEM and FRM: Exponential Special Case}
\label{subsec:CEM_vs_FRM}
To facilitate a direct comparison between the CEM of \cite{lu2025} and the FRM adopted in this paper, we consider the special case of 
Weibull lifetimes with shape parameter $\alpha = 1$, which reduces to the exponential distribution. Under this reduction, the Weibull 
scale parameters coincide with the exponential mean lifetimes using $\lambda_1 = 1/\theta_1$, $\lambda_{21} = 1/\theta_{21}$, and 
$\lambda_{22} = 1/\theta_{22}$. We fix the parameter values at $\theta_1 = e^{3.5}$, $\theta_{21} = e^{-0.2}$, $\theta_{22} = 
e^{2.0}$, and $\pi = 0.4$, matching exactly with the simulation settings of \cite{lu2025}. The corresponding Weibull scale parameters are 
$\lambda_1 = e^{-3.5} \approx 0.03$, $\lambda_{21} = e^{0.2} \approx 1.22$, and $\lambda_{22} = e^{-2.0} \approx 0.14$, obtained 
using the relation $\lambda = 1/\theta$. Under this exponential special case, the CEM and FRM formulations coincide in terms of the first stress level but differ in how the hazard function is specified after the stress change. The point estimates for the exponential case under the CEM, are therefore directly comparable to those obtained from our FRM-based $h$-SSALT model under the same parameter settings, allowing us to isolate the effect of the model assumption (CEM vs FRM) while keeping everything else fixed.

\begin{table}[ht]
\scriptsize
\centering
\caption{Comparison of FRM and CEM estimators with $\alpha = 1, \lambda_1 = 0.03, \lambda_{21} = 1.22, \lambda_{22} = 0.14, \pi = 0.4, \tau = 8$. Each entry is reported as FRM (CEM).}
\label{tab:FRM_CRM simulation_results}
\vspace{2mm}
\resizebox{\textwidth}{!}{%
\begin{tabular}{@{}lcccccccccc@{}}
\toprule
$(n, r)$ & \multicolumn{2}{c}{$\alpha$} & \multicolumn{2}{c}{$\lambda_1$} & \multicolumn{2}{c}{$\lambda_{21}$} & \multicolumn{2}{c}{$\lambda_{22}$} & \multicolumn{2}{c}{$\pi$} \\
\cmidrule(lr){2-3} \cmidrule(lr){4-5} \cmidrule(lr){6-7} \cmidrule(lr){8-9} \cmidrule(lr){10-11}
& AE & MSE & AE & MSE & AE & MSE & AE & MSE & AE & MSE \\
\toprule
(35, 30)   & 1.1262 & 0.1308 & 0.0311 & 0.0005 & 1.5415 & 2.0900 & 0.1804 & 0.0561 & 0.3968 & 0.0282 \\
           &  &  & (0.0298) & (0.0001) & (1.6222) & (1.3568) & (0.1531) & (0.0069) & (0.4169) & (0.0246) \\
\addlinespace[1.5mm]  
(35, 35)   & 1.1463 & 0.1240 & 0.0284 & 0.0004 & 1.3803 & 1.6913 & 0.1539 & 0.0361 & 0.4324 & 0.0319 \\
           &  &  & (0.0307) & (0.0001) & (1.4442) & (0.9948) & (0.1384) & (0.0029) & (0.4556) & (0.0327) \\
\addlinespace[1.5mm]
(100, 86)  & 1.0516 & 0.0482 & 0.0299 & 0.0002 & 1.4820 & 1.1234 & 0.1528 & 0.0138 & 0.4045 & 0.0165 \\
           &  &  & (0.0302) & (0.0000) & (1.4866) & (0.5648) & (0.1425) & (0.0021) & (0.3943)&(0.0140)\\ 
\addlinespace[1.5mm]          
(100, 100) & 1.0702 & 0.0511 & 0.0288 & 0.0002 & 1.3492 & 0.8934 & 0.1474 & 0.0137 & 0.4233 & 0.0159 \\
           &  &  & (0.0306) & (0.0000) & (1.3885) & (0.4559) & (0.1353) & (0.0008) & (0.4107) & (0.0143) \\
\bottomrule
\end{tabular}%
}
\end{table}

Table \ref{tab:FRM_CRM simulation_results} confirms that both models recover the true parameter values reliably, with the expected 
differences arising from their distinct structural assumptions. The CEM fixes $\alpha = 1$ by assumption rather than estimating it from data, so the corresponding AE and MSE entries for $\alpha$ are omitted as they are not estimated quantities; this reflects a model restriction rather than a genuine estimation advantage. The FRM, by estimating $\alpha$ from data, incurs additional variance for the scale parameters $\lambda_{21}$ and $\lambda_{22}$ at small sample sizes, which is expected for a more flexible model with one additional unknown parameter to estimate. This  penalty  on variance decreases as it $n$ rises, emphasizing being consistent with asymptotic efficiency. The mixture proportion $\pi$ is estimated with comparable accuracy by both models across all configurations, confirming that the mixture structure is identifiable regardless of the hazard specification used.

\section{Data Analysis}
\label{sec:data_analysis}
In this section, we analyze the performance of the proposed heterogeneous SSALT model, based on both the simulated and real data sets originally reported by \cite{basak2018} and \cite{Lee2013}, respectively.
\FloatBarrier
\subsection{Illustration under Simulated Data}
\label{subsec:analysis_simulated}
The original data is correspond to a two-stress-level experiment with sample size $n = 40$ and stress change time $\tau = 15$. In the present study, we adopt a Weibull framework. To maintain comparability with the original exponential setting, we consider the special case $\alpha = 1$, under which the Weibull distribution reduces to the exponential distribution. In this case, the Weibull parameters are related to the exponential parameters through the relation $\lambda = 1/\theta$.

The true parameter values are fixed at $\theta_1 = e^{3.0}$, $\theta_{21} = e^{-0.2}$, $\theta_{22} = e^{2.0}$, and $\pi = 0.4$, and are subsequently expressed in terms of the Weibull scale parameters $\lambda_1$, $\lambda_{21}$, and $\lambda_{22}$. To incorporate heterogeneity at the second stress level, the observations corresponding to $s_2$ are generated from a mixture of two Weibull distributions. A proportion 0.6 of  observations at $s_2$ is randomly selected and retained from the original data, while the remaining  proportion $\pi$=0.4 is replaced by values generated from a Weib$(\alpha,\lambda_{21})$ distribution shifted at $\tau$. This results in a two-component mixture structure at the second stress level and the resulting heterogeneous dataset is presented in Table \ref{tab:simulated}.

\begin{table}[htbp]
\scriptsize
\centering
\caption{Simulated $h$-SSALT dataset based on \cite{basak2018}.}
\label{tab:simulated}
\begin{tabular}{|c|c|p{8cm}|}
\toprule
\textbf{Stress Level} & \textbf{Parameter} & \textbf{Failure Times} \\
\toprule
$s_1$ & $\lambda_1 = e^{-3.0}$ &
0.22, 1.16, 1.45, 1.58, 2.92, 3.70, 4.30, 6.20, 7.23, 8.79, 9.35, 9.68, 9.89, 10.95, 11.55, 12.48, 13.56 \\
\midrule
$s_2$ & $\lambda_{21} = e^{0.2}$ &
15.05, 15.31, 15.32, 15.42, 15.45, 15.73, 15.74, 15.98, 17.06 \\
      & $\lambda_{22} = e^{-2.0}$ &
15.27, 15.37, 15.61, 16.38, 18.60, 19.42, 21.00, 22.29, 24.42, 24.82, 25.54, 28.92, 29.94, 40.19 \\
\bottomrule
\end{tabular}
\end{table}

% \begin{table}[!ht]
% \centering
% \caption{Simulated $h$-SSALT dataset based on \cite{basak2018}.}
% \label{tab:simulated}
% \begin{tabular}{lll}
% \toprule
% \textbf{Stress level} & \textbf{Parameter} & \textbf{Failure times} \\
% \midrule
% $s_1$ & $\lambda_1 = e^{-3.0}$ &
% 0.22 \; 1.16 \; 1.45 \; 1.58 \; 2.92 \; 3.70 \; 4.30 \; 6.20 \; 7.23 \; 8.79 \\
%       &                  &
% 9.35 \; 9.68 \; 9.89 \;10.95 \;11.55\;12.48\; 13.56 \\

% \midrule
% $s_2$ & $\lambda_{21}=e^{0.2}$ &
% 15.05 \; 15.31 \; 15.32 \; 15.42 \; 15.45 \; 15.73 \; 15.74 \; 15.98 \; 17.06\\
%       & $\lambda_{22} = e^{-2.0}$ &
% 15.27 \; 15.37 \; 15.61 \; 16.38 \; 18.60 \; 19.42 \; 21.00 \; 22.29 \;  24.42\\
%       &                         &
% 24.82 \; 25.54 \; 28.92 \; 29.94 \; 40.19 \\
% \bottomrule
% \end{tabular}
% \end{table}

\noindent To study the effect of censoring, we also consider Type-II censored data. In this case, only the first $r$ ordered failure times are observed, while the remaining observations are censored. The censored sample is obtained from the simulated dataset given in Table \ref{tab:simulated} and is presented in Table \ref{tab:censored}.
% \begin{table}[!ht]
% \centering
% \caption{Type-II censored sample from the simulated data  (Table \ref{tab:simulated} ) }
% \label{tab:censored}
% \begin{tabular}{lll}
% \toprule
% \textbf{Stress level} & \textbf{Parameter} & \textbf{Failure times} \\
% \midrule
% $s_1$ & $\lambda_1 = e^{-3.0}$ &
% 0.22 \; 1.16 \; 1.45 \; 1.58 \; 2.92 \; 3.70 \; 4.30 \; 6.20 \; 7.23 \; 8.79  \\
%       &                  &
% 9.35 \; 9.68 \; 9.89 \;10.95\;11.55 \;12.48 \; 13.56\\
% \midrule
% $s_2$ & $\lambda_{21}=e^{0.2}$ &
% 15.05 \; 15.31 \; 15.32 \; 15.42 \; 15.45 \; 15.73 \; 15.74 \; 15.98 \; 17.06 \\
%       & $\lambda_{22} = e^{-2.0}$ &
% 15.27 \; 15.37 \; 15.61 \; 16.38 \; 18.60 \; 19.42 \; 21.00 \; 22.29 \; 24.42\\
% \bottomrule
% \end{tabular}
% \end{table}

\begin{table}[htbp]
\scriptsize
\centering
\caption{Type-II censored sample from the simulated data (Table~\ref{tab:simulated}).}
\label{tab:censored}
\begin{tabular}{|c|c|p{8cm}|}
\toprule
\textbf{Stress Level} & \textbf{Parameter} & \textbf{Failure Times} \\
\toprule
$s_1$ & $\lambda_1 = e^{-3.0}$ &
0.22, 1.16, 1.45, 1.58, 2.92, 3.70, 4.30, 6.20, 7.23, 8.79, 9.35, 9.68, 9.89, 10.95, 11.55, 12.48, 13.56 \\
\midrule
$s_2$ & $\lambda_{21} = e^{0.2}$ &
15.05, 15.31, 15.32, 15.42, 15.45, 15.73, 15.74, 15.98, 17.06 \\
      & $\lambda_{22} = e^{-2.0}$ &
15.27, 15.37, 15.61, 16.38, 18.60, 19.42, 21.00, 22.29, 24.42 \\
\bottomrule
\end{tabular}
\end{table}

The model parameters $\alpha$, $\lambda_1$, $\lambda_{21}$, $\lambda_{22}$, and the mixing proportion $\pi$ are estimated using the EM algorithm as described in Section \ref{subsec:EM}. To examine the convergence of the EM algorithm, the estimation is carried out using different initial values for the parameters. In all cases, the algorithm converges to the same solution up to four decimal places, indicating stable convergence. The resulting estimates for both simulated (complete) and Type-II censored data are reported in Table \ref{tab:combined}. In the complete data case, the estimated parameters are very close to the corresponding true values used in the data generation, demonstrating the accuracy of the estimation procedure. For the Type-II censored sample, some deviation from the true values is observed, which is expected due to the partial loss of information under the censoring. In particular, the mixture parameter $\lambda_{21}$  exhibits comparatively larger variability under censoring, whereas the estimates of $\alpha$, $\lambda_1$, $\lambda_{22}$ and $\pi$ remain stable. 

The confidence intervals, obtained using a parametric bootstrap procedure, are also provided in \ref{tab:combined}. The interval for $\lambda_{21}$ is relatively wide in both cases, indicating higher variability, while the remaining parameters show stable and consistent intervals Overall, the estimation procedure performs well for both complete and censored data. The bootstrap intervals are generally narrower than the transformation-based intervals for the scale parameters $\lambda_{21}$ and $\lambda_{22}$, while both methods yield comparable intervals 
for $\alpha$ and $\pi$, suggesting that the parametric bootstrap may offer efficiency gains for the more variable mixture scale parameters.

\begin{table}[!ht]
\centering
\scriptsize
\caption{ EM-based MLEs of model parameters along with confidence intervals and goodness of fit statistics for Simulated and Type-II censored datasets.}
\label{tab:combined}

\begin{tabular}{l|ccc|ccc}
\toprule

& \multicolumn{3}{c|}{Simulated Data(Table \ref{tab:simulated} )}
& \multicolumn{3}{c}{ Censored Data (Table \ref{tab:censored} )} \\

\cmidrule(lr){2-4}
\cmidrule(lr){5-7}

Parameter 
& Estimate 
& Bootstrap CI 
& Transformed CI
& Estimate 
& Bootstrap CI 
& Transformed CI \\

\midrule

$\alpha$
& 1.0115
& (0.7437, 1.7310)
& (0.6503, 1.5733)
& 0.9798
& (0.7299, 1.6762)
& (0.6266, 1.5321) \\

$\lambda_1$
& 0.0359
& (0.0045, 0.0767)
& (0.0104, 0.1243)
& 0.0389
& (0.0048, 0.0806)
& (0.0115, 0.1318) \\

$\lambda_{21}$
& 1.6962
& (0.0949, 5.1820)
& (0.2633, 10.9255)
& 1.8045
& (0.1354, 5.4108)
& (0.2889, 11.2723) \\

$\lambda_{22}$
& 0.1150
& (0.0060, 0.4150)
& (0.0166, 0.7988)
& 0.0981
& (0.0056, 0.6065)
& (0.0137, 0.7034) \\

$\pi$
& 0.4333
& (0.1693, 0.7744)
& (0.2018, 0.7988)
& 0.4671
& (0.1997, 0.7129)
& (0.2307,0.7193)\\

\midrule

Log-likelihood
& -172.5600 & & 
& -250.7872 & & \\

KS Statistic
& 0.2192 & &
& 0.2184 & & \\

p-value
& 0.3670 & &
& 0.4220 & & \\

\bottomrule
\end{tabular}
\end{table}

\noindent Moreover, the adequacy of the proposed model is assessed using both numerical goodness-of-fit measures and graphical comparisons. For the complete dataset, the Kolmogorov–Smirnov (KS) statistic is 0.2192 with a p-value of 0.3670, and the corresponding log-likelihood value is -172.56, indicating an excellent fit of the model to the data. For the Type-II censored dataset, the KS statistic is 0.2184 with a p-value of 0.4220, and the log-likelihood value is -250.7872. The KS value is slightly higher and the log-likelihood is lower than the complete case, which is expected due to the loss of information under censoring, but the model still fits the data quite well.
% \begin{figure}[htbp]
% \centering
% \begin{subfigure}{0.49\textwidth}
% \centering
% \includegraphics[width=\textwidth]{Emperical_vs_fitted_plot.pdf}
% \caption{n=40, r=40}
% \end{subfigure}
% \hfill
% \begin{subfigure}{0.49\textwidth}
% \centering
% \includegraphics[width=\textwidth]{emperical_vs_fitted_censored.pdf}
% \caption{n=40, r=35}
% \end{subfigure}
% \caption{Empirical and fitted CDFs based on the data in Table \ref{tab:simulated} and Table \ref{tab:censored}. }
% \label{fig:emperic vs fit}
% \end{figure}

\begin{figure}[htbp]
\centering
\includegraphics[width=\textwidth]{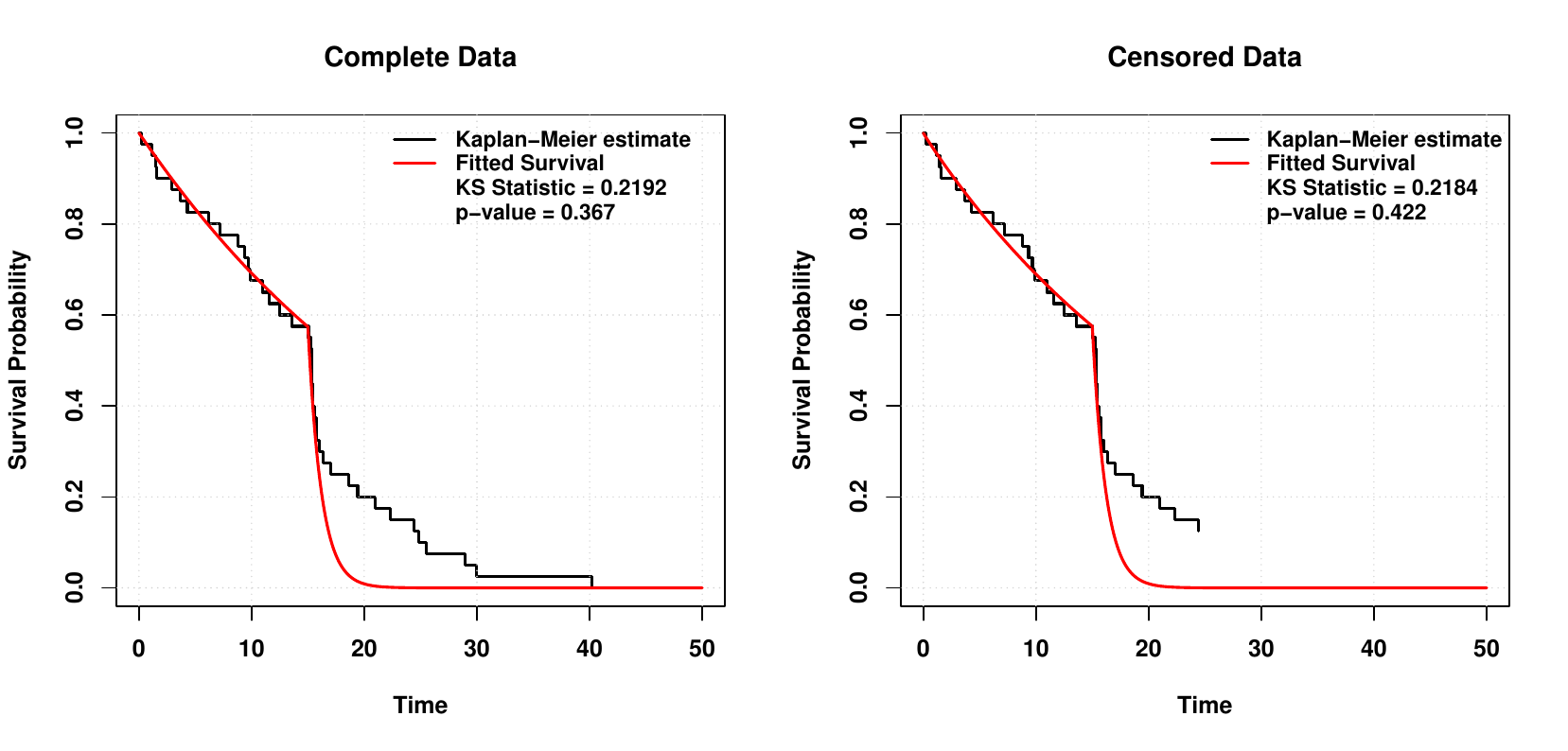}
\caption{Empirical and fitted survival functions based on the data in Table \ref{tab:simulated} and Table \ref{tab:censored}. }
\label{fig:emperic vs fit}
\end{figure}
\noindent The graphical comparison in  Figure~\ref{fig:emperic vs fit} further supports these results. The empirical CDFs, shown as step functions, are plotted alongside the fitted CDFs represented by smooth curves. As seen in Figure~\ref{fig:emperic vs fit}, the fitted CDF closely follows the empirical curve for both complete and censored data, indicating that the proposed model effectively captures the failure behavior across different stress levels, while successfully capturing the underlying heterogeneity in the data.
\FloatBarrier
\subsection{Illustration under Real Data}
\label{subsec:analysis_real}

To illustrate the practical applicability of the proposed $h$-SSALT model, an electronic components dataset originally analyzed in the accelerated life testing literature is considered in this section. Similar electronic component datasets under step-stress accelerated testing schemes have also been studied by \cite{Lee2013} and \cite{Mukhopadhyay2023} for reliability and lifetime performance analysis under accelerated operating conditions.

\begin{table}[htbp]
\scriptsize
\centering
\caption{Transformed Failure Times of the Electronic Component Dataset}
\label{tab:electronicdata}

\begin{tabular}{|c|p{10.5cm}|}
\toprule
\textbf{Stress Level} & \textbf{Transformed Failure Times} \\
\toprule

$s_1$ &
0.32, 0.54, 0.59, 0.86, 1.17, 1.23, 2.13, 2.67, 2.68, 2.73, 2.99, 3.17, 3.21, 3.33, 3.39, 3.86, 4.08, 4.22, 4.35, 4.37, 4.56, 5.18, 5.70, 6.32, 6.66, 6.97, 7.96, 8.54, 8.58, 9.10 \\
\midrule

$s_2$ &
9.26, 9.29, 9.31, 9.46, 9.47, 9.73, 9.80, 9.85, 9.93, 10.05, 10.10, 10.16, 10.20, 10.23, 10.26, 10.45, 10.46, 10.59, 10.82, 10.96 \\
\bottomrule

\end{tabular}
\end{table}

The experiment was conducted under a simple step-stress accelerated life testing scheme with two stress levels. Initially, all test units were operated at the stress level $s_1 = 373\text{ K}$. After a preassigned stress changing time $\tau=910$, the stress was elevated to $s_2 = 423\text{ K}$, while the normal operating condition was assumed to be $s_0 = 298\text{ K}$. 
The experiment was performed under a Type-II censoring mechanism with a censoring number $r=50$ and a total of $n = 100$ electronic components placed on test. At the first stress level, $n_1 = 30$ components failed, whereas at the second stress level, $n_2 = 20$ components failed. The remaining $50$ components were censored under the Type-II censoring scheme.
To improve numerical stability and facilitate efficient implementation of the EM algorithm, the original failure times are transformed by dividing each observation by $100$. Thus, the transformed observations are defined as $y_i=x_i/100, i=1,2,\ldots,50,$ where $x_i$ denotes the original failure time. The transformed failure times corresponding to the two stress levels are presented in Table~\ref{tab:electronicdata}.

\begin{figure}[ht]
\centering
\includegraphics[width=0.95\textwidth]{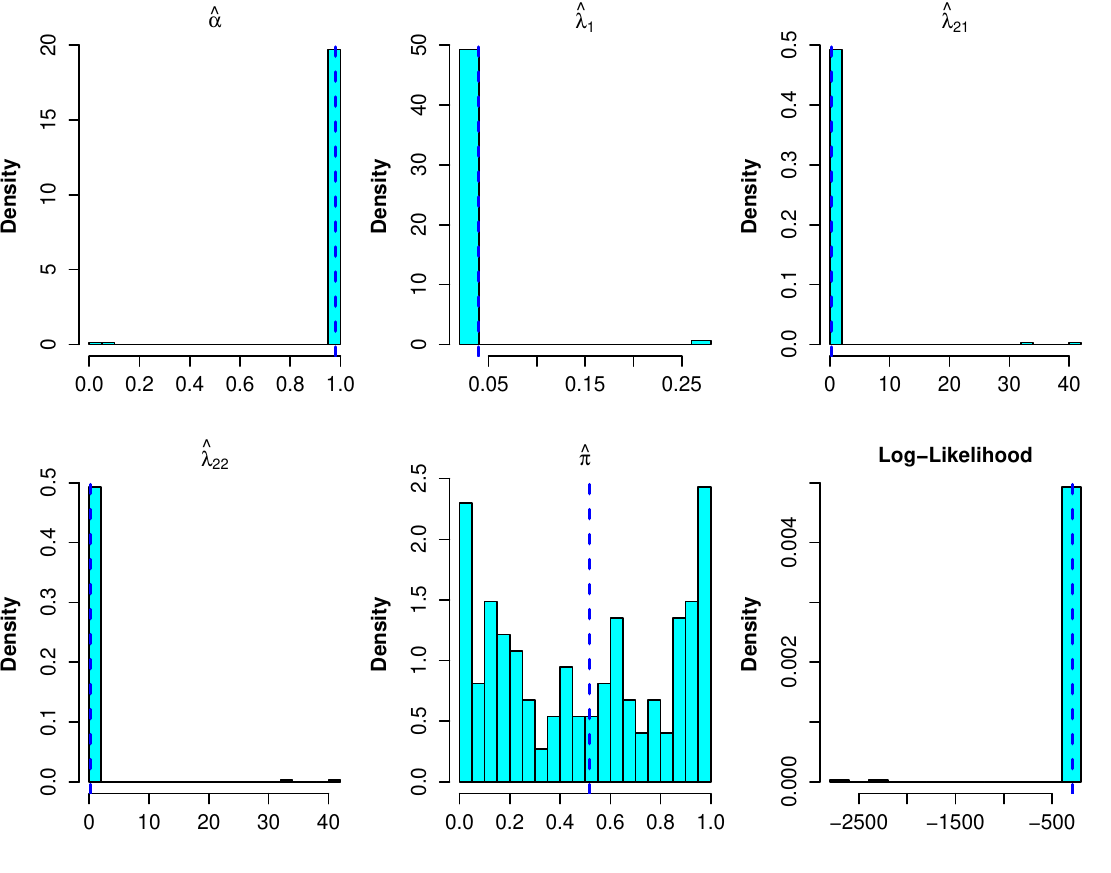}
\caption{Histograms of the EM algorithm for the electronic components dataset under the proposed $h$-SSALT model based on $500$ initialization runs.}
\label{fig:histogram}
\end{figure}

The proposed $h$-SSALT model is then fitted for the dataset using the EM algorithm developed in Section \ref{subsec:EM}. To assess the stability of the estimation procedure and reduce the possibility of convergence toward local maxima, the EM algorithm was implemented using 500 different initialization runs. The initial parameter values are randomly selected from broad uniform distributions to ensure adequate exploration of the parameter space and to examine the convergence behavior of the EM algorithm. Specifically, the initial parameter values are generated independently as 
$\alpha \sim U(0.01,5)$,
$\lambda_1 \sim U(0.001,5)$,
$\lambda_{21} \sim U(0.001,5)$,
$\lambda_{22} \sim U(0.001,5)$,
and
$\pi \sim U(0.001,0.999)$.
Accordingly, the initial parameter vector is taken as
$\boldsymbol{\theta}^{(0)}
=
(\alpha,\lambda_1,\lambda_{21},\lambda_{22},\pi).$

The distributional patterns of the parameter estimates obtained from repeated EM runs are illustrated in Figure~\ref{fig:histogram}, where the histograms corresponding to $\hat{\alpha}$, $\hat{\lambda}_1$, $\hat{\lambda}_{21}$, $\hat{\lambda}_{22}$, $\hat{\pi}$, and the maximized log-likelihood values provide important insight into the stability, robustness, and consistency of the proposed estimation procedure.

% \FloatBarrier
From Figure~\ref{fig:histogram}, it is observed that the estimate of the shape parameter $\hat{\alpha}$ is highly concentrated near 1, with most estimates lying approximately within the interval $0.95$--$1.00$. This indicates that the underlying lifetime behavior is close to the exponential distribution for the considered electronic component dataset.
Similarly, the estimates of $\hat{\lambda}_1$ remain strongly concentrated around small positive values, primarily within the interval $0.02$--$0.05$, indicating stable convergence across different random initializations. Similarly, the estimates of $\hat{\lambda}_{21}$ and $\hat{\lambda}_{22}$ exhibit very similar distributional patterns and are strongly concentrated around small positive values, further supporting the approximately homogeneous behavior. Moreover, the mixing proportion estimate $\hat{\pi}$ exhibits comparatively larger variability, spanning almost the entire interval $(0,1)$, although most values remain concentrated around the middle region near $0.5$. Such instability in $\hat{\pi}$ is commonly observed in finite mixture models when the underlying mixture components overlap substantially and possess similar reliability characteristics, making it difficult for the EM algorithm to clearly distinguish between the latent groups, allowing multiple combinations of the mixture parameters to produce nearly identical likelihood values. Consequently, the likelihood surface becomes relatively flat with respect to the mixing proportion parameter, leading to weaker identifiability and increased instability in the estimation of $\pi$ across different initialization runs. This instability in the estimation of $\hat{\pi}$ across the initial runs indicates a near-homogeneous behavior of the electronic components data.

To further investigate this near-homogeneous behavior suggested by the above heterogeneous analysis, the electronic components dataset was additionally fitted using the homogeneous SSALT model. The resulting maximum likelihood estimates were obtained as
\[
\hat{\alpha}=1.0184, ~
\hat{\lambda}_{1}=0.0380 ~ \text{and} ~
\hat{\lambda}_{2}=0.1690.
\]
The estimate of the shape parameter $\hat{\alpha}$ being close to 1 further supports the approximately exponential lifetime behavior of the dataset. %In addition, the estimates of the other parameters remain fully consistent with the concentration patterns observed under the proposed heterogeneous framework. 
These findings suggest that the electronic components dataset behaves approximately like a homogeneous population under the elevated stress condition. Furthermore, the proposed $h$-SSALT model is sufficiently flexible not only to capture latent heterogeneous structures when meaningful subgroup separation exists, but also to appropriately identify situations where the observed data exhibit near-homogeneous lifetime behavior. This highlights the practical robustness and adaptability of the proposed methodology in modeling accelerated life testing data arising from both homogeneous and heterogeneous populations.

\begin{table}[htbp]
\scriptsize
\centering
\caption{Type-II Censored Heterogeneous Electronic Component Dataset}
\label{tab:censored_electric}
\begin{tabular}{|c|c|p{8cm}|}
\toprule
\textbf{Stress Level} & \textbf{Parameter} & \textbf{Transformed Failure Times} \\
\toprule
$s_1$ & $\lambda_1 = 0.04$ &
0.32, 0.54, 0.59, 0.86, 1.17, 1.23, 2.13, 2.67, 2.68, 2.73, 2.99, 3.17, 3.21, 3.33, 3.39, 3.86, 4.08, 4.22, 4.35, 4.37, 4.56, 5.18, 5.70, 6.32, 6.66, 6.97, 7.96, 8.54, 8.58, 9.10 \\
\midrule
$s_2$ & $\lambda_{21} = 1.4$ &
9.11, 9.36, 9.47, 9.60, 9.70, 13.16, 13.17 \\
      & $\lambda_{22} = 0.16$ &
9.26, 9.29, 9.31, 9.47, 9.73, 9.80, 9.93, 10.20, 10.26, 10.45, 10.46, 10.59, 10.82 \\
\bottomrule
\end{tabular}
\end{table}

To further evaluate the effectiveness of the proposed $h$-SSALT model in identifying latent heterogeneity under elevated stress conditions, a heterogeneous version of the electronic component dataset was artificially constructed. The heterogeneity was incorporated only at the second stress level, while the observations corresponding to the first stress level were retained unchanged. Specifically, among the $n_2=20$ observations observed after the stress elevation, a proportion $(1-\pi) = 0.65$ of the observations was randomly retained from the original dataset, whereas the remaining proportion $\pi = 0.35$ was replaced by observations generated from a Weibull distribution representing a weaker latent subgroup.
The generated subgroup observations were obtained using the parameter $\lambda_{21}=1.4$, while the retained observations approximately represent the relatively stronger subgroup characterized by $\lambda_{22}=0.16$. Consequently, the observations corresponding to the second stress level form a two-component finite mixture structure consisting of two latent subpopulations with different failure characteristics. Accordingly, the resulting heterogeneous dataset contains $n_{21}=7$ observations associated with the weaker subgroup and $n_{22}=13$ observations associated with the relatively stronger subgroup. The transformed heterogeneous failure times are presented in Table~\ref{tab:censored_electric}.
The proposed $h$-SSALT model was fitted to the heterogeneous dataset using the EM algorithm developed in Section~3, and the resulting maximum likelihood estimates together with the associated confidence intervals are reported in Table~\ref{tab:mle_electronic}.
% \begin{table}[!ht]
% \centering
% \caption{Type-II Censored Heterogeneous Electronic Component Dataset}
% \label{tab:censored_electric}
% \begin{tabular}{lll}
% \toprule
% \textbf{Stress level} & \textbf{Parameter} & \textbf{Transformed Failure times} \\
% \midrule
% $s_1$ 
% & $\lambda_1 = 0.04$ 
% & 0.32 \; 0.54 \; 0.59 \; 0.86 \; 1.17 \; 1.23 \; 2.13 \; 2.67 \\
% & &
% 2.68 \; 2.73 \; 2.99 \; 3.17 \; 3.21 \; 3.33 \; 3.39 \; 3.86 \\
% & &
% 4.08 \; 4.22 \; 4.35 \; 4.37 \; 4.56 \; 5.18 \; 5.70 \;6.32\\
% & &
% 6.66 \; 6.97 \; 7.96 \; 8.54 \; 8.58 \; 9.10 \\

% \midrule
% $s_2$ & $\lambda_{21}=1.4$ &
% 9.11 \; 9.36 \; 9.47 \; 9.60 \; 9.70 \; 13.16 \; 13.17 \\
%       &                                     &
%    \\
%       & $\lambda_{22} =0.16$ &
% 9.26 \; 9.29 \; 9.31 \; 9.47 \; 9.73 \; 9.80 \; 9.93 \\
% & &
% 10.20 \; 10.26 \; 10.45 \; 10.46 \; 10.59 \; 10.82 \\
                              
% \bottomrule
% \end{tabular}
% \end{table}

\begin{table}[!ht]
\scriptsize
\centering
\caption{EM-based MLEs and 95\% Confidence Intervals for the Type-II Censored Heterogeneous Electronic Component Dataset}
\label{tab:mle_electronic}
\renewcommand{\arraystretch}{1.15}

\begin{tabular}{l|cccc}
\toprule
\textbf{Parameter} & \textbf{Estimate} & \textbf{Bootstrap CI} & \textbf{Transformed CI} \\
\midrule

$\alpha$ 
& 1.0221
& (0.7461,\;1.4859) 
& (0.7263,\;1.4384) \\

$\lambda_{1}$ 
& 0.0378 
& (0.0121,\;0.0681) 
& (0.0166,\;0.0857) \\

$\lambda_{21}$ 
& 1.3069 
& (0.2207,\;5.3603) 
& (0.3519,\;4.8530) \\

$\lambda_{22}$ 
& 0.0193
& (0.0023,\;0.3421) 
& (0.0028,\;0.1355) \\

$\pi$ 
& 0.2229
& (0.0716,\;0.3693) 
& (0.1212,\;0.3735) \\

\midrule

Log-ikelihood & -433.9026 & & \\
KS Statistic  &  0.334 & & \\
p-value        & 0.563 & & \\

\bottomrule
\end{tabular}

\end{table}
\FloatBarrier
From Table~\ref{tab:mle_electronic}, the estimated Weibull shape parameter is obtained as $\hat{\alpha}=1.0221$, which remains very close to unity, indicating approximately exponential lifetime behavior even under the heterogeneous framework. Similarly, the estimate corresponding to the first stress level parameter is obtained as $\hat{\lambda}_1=0.0378$, which remains stable and consistent with the homogeneous case.

More importantly, the estimated subgroup parameters corresponding to the elevated stress level are obtained as $\hat{\lambda}_{21}=1.3069$ and $\hat{\lambda}_{22}=0.0193$. The substantial separation between these two estimates clearly indicates the presence of distinct latent subpopulations operating under the elevated stress condition. The comparatively larger estimate of $\hat{\lambda}_{21}$ corresponds to the weaker subgroup exhibiting faster failure behavior, whereas the smaller estimate of $\hat{\lambda}_{22}$ corresponds to the relatively stronger subgroup with slower failure characteristics. The estimated mixing proportion, $\hat{\pi}=0.2229$, suggests the existence of a smaller subgroup within the population with relatively different failure characteristics.
The confidence intervals were obtained using the observed Fisher information matrix computed through the Louis' method as discussed in section 4.2, which appropriately accounts for the latent subgroup memberships and Type-II censoring present in the proposed heterogeneous SSALT model. Both the bootstrap and transformation-based confidence intervals indicate satisfactory stability of the parameter estimates.

\begin{figure}[ht]
\centering
\includegraphics[width=0.7\textwidth]{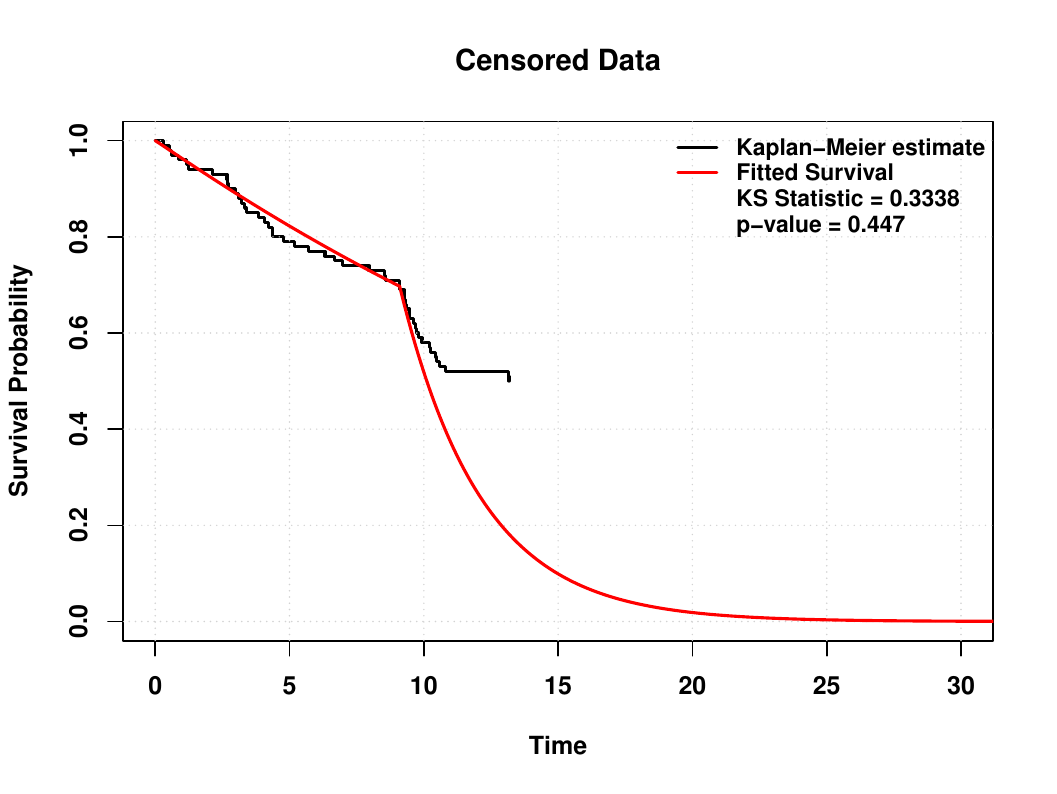}
\caption{Kaplan–Meier estimate of the survival function and fitted survival curve for the Type-II censored heterogeneous electronic component dataset.}
\label{K-M_curve}
\end{figure}

% \FloatBarrier

The Kaplan–Meier curve shown in Figure~\ref{K-M_curve} provides a nonparametric estimate of the survival function for the Type-II censored heterogeneous electronic component dataset. The fitted survival curve obtained from the proposed $h$-SSALT model follows the Kaplan–Meier estimate reasonably well over most of the observed time range, indicating satisfactory agreement between the observed data and the fitted model. In particular, the fitted curve captures the overall survival pattern before and after the stress transition point.

The estimated KS statistic is $0.3338$ with a bootstrap p-value of 0.4470. Since the p-value exceeds the usual significance level of 0.05, the proposed model provides an adequate fit to the data. A moderate deviation between the fitted survival curve and the Kaplan–Meier estimate is observed in the tail region after the stress-change point. This behaviour may be attributed to the relatively small number of observed failures at the second stress level together with the presence of Type-II censoring. Overall, the graphical comparison and goodness-of-fit results indicate that the proposed heterogeneous SSALT model is suitable for analysing the electronic component dataset.
\section{Conclusion}
\label{sec:conclusion}
This paper has proposed a finite mixture failure-rate based $h$-SSALT model for Weibull distributed lifetimes under Type-II censoring. The model extends the exponential heterogeneous SSALT framework of \cite{lu2025} in two important directions: it accommodates non-constant failure-rates through the Weibull shape parameter 
$\alpha$, which is common to both stress levels and governs the aging behaviour of the test units, and it operates under the failure-rate based model of \cite{kateri2015} rather than the cumulative exposure model. The mixture structure at the second stress level captures latent heterogeneity that is invisible at the first stress level and emerges only after the stress is elevated, a pattern commonly observed in industrial reliability testing. Interval estimation is presented based on Louis' missing information identity (\citep{louis1982finding}), and the use of log and logit transformations ensures that the resulting confidence intervals naturally respect the positivity and boundedness constraints on the model parameters.

Parameter estimation is carried out through an EM algorithm that treats the unobserved group membership indicators as latent variables, yielding closed-form updates for $\lambda_1$, $\lambda_{2j}$, and $\pi_j$, with only the shared shape parameter $\alpha$ requiring a one-dimensional numerical solve at each iteration. The simulation study confirms that the EM-based MLEs are consistent and that estimation precision improves steadily with sample size across all parameters. The mixture proportion $\hat{\pi}$, which directly quantifies the fraction of production units belonging to the weaker subgroup, converges reliably already at moderate sample sizes, providing actionable information for burn-in testing and quality screening decisions. The transformed confidence intervals achieve coverage probabilities close to the nominal level throughout, with the intervals for $\lambda_1$ and $\pi$ performing best and those for $\lambda_{22}$ expectedly wider at small censored sample sizes given the high variability of the long-lived subgroup estimates.

The quantile-based comparison with the homogeneous SSALT model reveals a consistent and practically important finding: ignoring population heterogeneity leads to systematic overestimation of upper lifetime quantiles and systematic underestimation of early 
failure quantiles, with neither bias vanishing as the sample size grows. The misspecification is most damaging at the early failure quantiles $q \in \{0.01, 0.05, 0.10\}$, where the SSALT model underestimates the true quantile by approximately 50\% even at $n = 100$. For engineers setting warranty periods or certifying safety-critical components, this represents a serious risk of overestimating product reliability at early times. The $h$-SSALT model eliminates this bias and produces well-centred estimates across all quantile levels and sample sizes considered.

The special case comparison with \cite{lu2025} under the exponential reduction $\alpha = 1$ confirms that both the CEM-based and FRM-based formulations recover the true parameters reliably, with the FRM incurring a modest additional variance for the scale parameters at small samples due to estimating $\alpha$ rather than fixing it. This penalty diminishes with increasing $n$, consistent with asymptotic efficiency, and the mixture proportion $\pi$ is estimated with comparable accuracy by both models regardless of the hazard specification.

The data analysis further illustrates the practical implementation of the proposed model. For the simulated dataset, stable convergence from multiple starting values in EM algorithm and a valid Kolmogorov-Smirnov fit confirm the reliability of the estimation procedure. The real electronic components dataset reveals an equally important feature of the methodology: when no meaningful subgroup structure exists, the instability of $\hat{\pi}$ across distinct initialization runs correctly and precisely signals near-homogeneous behavior rather than forcing a spurious mixture solution.

Several directions remain open for future work. Extension to multiple stress levels with heterogeneity emerging at more than one transition point is a natural generalization. Bayesian estimation with informative priors on the mixture proportion could improve inference when prior knowledge about the fraction of weaker units is available from historical data. Optimal experimental design for the $h$-SSALT model, determining the stress-change time $\tau$ and censoring number $r$ that minimize estimation uncertainty, possibly under cost or resource constraints, is another practically important problem that we leave for our future investigation. Related design frameworks, both 
frequentist under constrained multicriteria criteria \citep{prajapat2026constrained} and Bayesian \citep{prajapat2026exact}, have recently been developed, and extending such approaches to the heterogeneous $h$-SSALT setting represents a natural direction for the future work.

% \section*{Acknowledgements}

\section*{Conflict of Interest}
The author hereby declares that the information provided here is accurate, and there are no apparent conflicts of interest relevant to the content of this article.

\section*{Ethics Statement}
Not applicable.

\section*{Consent}
Not applicable. 

\section*{Data Availability Statement}
The original dataset used to construct the simulated heterogeneous data in Section~\ref{subsec:analysis_simulated} is openly available in \cite{basak2018} at \url{https://doi.org/10.1177/0008068318769506}. The electronic components dataset analysed in Section~\ref{subsec:analysis_real} is openly available in \cite{Lee2013} at \url{https://doi.org/10.1109/TR.2013.2241197}.

\bibliographystyle{chicago}
\bibliography{references}

\appendix 
\section{Likelihood Equations}
Differentiating the log-likelihood equation (8) with respect to the parameters $\alpha,\lambda_{21},\lambda_{22},\pi_j$ yields the following likelihood equations:
\begin{align*}
	&\frac{\partial l}{\partial \pi_j} 
	   = (\lambda_{2j} - \lambda_{2m}) \left[ \frac{\bar{n}_2 - n_1}{\displaystyle\sum_{j=1}^{m-1} \pi_j \lambda_{2j} + \left(1 - \sum_{j=1}^{m-1} \pi_j\right) \lambda_{2m}} \right. \left. \vphantom{\frac{\bar{n}_2 - n_1}{\displaystyle\sum_{j=1}^{m-1} \pi_j \lambda_{2j} + \left(1 - \sum_{j=1}^{m-1} \pi_j\right) \lambda_{2m}}}- \left( \sum_{i=n_1+1}^{\bar{n}_2} t_{i:n}^\alpha + (n - \bar{n}_2)t_{r:n}^\alpha - (n - n_1)\tau^\alpha \right)\right] = 0,&\\	
	&\frac{\partial l}{\partial \lambda_{2j}} 
	= \pi_j \left[ \frac{\bar{n}_2 - n_1}{\displaystyle\sum_{j=1}^{m-1} \pi_j \lambda_{2j} + \left(1 - \sum_{j=1}^{m-1} \pi_j\right) \lambda_{2m}} \right.  \left. \vphantom{\frac{\bar{n}_2 - n_1}{\displaystyle\sum_{j=1}^{m-1} \pi_j \lambda_{2j} + \left(1 - \sum_{j=1}^{m-1} \pi_j\right) \lambda_{2m}}} - \left( \sum_{i=n_1+1}^{\bar{n}_2} t_{i:n}^\alpha + (n - \bar{n}_2)t_{r:n}^\alpha - (n - n_1)\tau^\alpha \right) \right] = 0 \\	
	&\text{and } ~\frac{\partial l}{\partial \alpha} 
	= \sum_{i=1}^{n_1} \ln t_{i:n} - \lambda_1 \left( \sum_{i=1}^{n_1} t_{i:n}^\alpha \ln t_{i:n} + (n - n_1)\tau^\alpha \ln \tau \right) \\ 
	& \ \ \qquad \qquad+ \frac{\bar{n}_2}{\alpha} + \sum_{i=n_{1}+1}^{\bar{n}_{2}}\ln t_{i:n} - \left\{ \sum_{j=1}^{m-1} \pi_j \lambda_{2j} + \left(1 - \sum_{j=1}^{m-1} \pi_j\right) \lambda_{2m} \right\}\\
	&\ \ \qquad \qquad\times \left( \sum_{i=n_1+1}^{\bar{n}_2} t_{i:n}^\alpha \ln t_{i:n} + (n - \bar{n}_2)t_{r:n}^\alpha \ln t_{r:n} - (n - n_1)\tau^\alpha \ln \tau \right) = 0.
\end{align*}
%\[
%\begin{aligned}
%\quad
%\frac{\partial l}{\partial \pi_j} &= (\lambda_{2j} - \lambda_{2m}) \left\Bigg[\frac{\bar{n}_2 - n_1}{\displaystyle\sum_{j=1}^{m-1} \pi_j \lambda_{2j} + \left(1 - \sum_{j=1}^{m-1} \pi_j\right) \lambda_{2m}} \right. \\
%&\quad\qquad- \left. \left( \sum_{i=n_1+1}^{\bar{n}_2} t_{i:n}^\alpha + (n - \bar{n}_2)t_{r:n}^\alpha - (n - n_1)\tau^\alpha \right)\right] = 0 \\[1em]
%%
%\quad
%\frac{\partial l}{\partial \lambda_{2j}} &= \pi_j \left[ \frac{\bar{n}_2 - n_1}{\displaystyle\sum_{j=1}^{m-1} \pi_j \lambda_{2j} + \left(1 - \sum_{j=1}^{m-1} \pi_j\right) \lambda_{2m}} \right\\
%&\quad\quad\quad - \left. \left( \sum_{i=n_1+1}^{\bar{n}_2} t_{i:n}^\alpha + (n - \bar{n}_2)t_{r:n}^\alpha - (n - n_1)\tau^\alpha \right) \right] = 0 \\[1em]
%%
%\quad
%\frac{\partial l}{\partial \alpha} &= \sum_{i=1}^{n_1} \ln t_{i:n}- \lambda_1 \left( \sum_{i=1}^{n_1} t_{i:n}^\alpha \ln t_{i:n} + (n - n_1)\tau^\alpha \ln \tau \right) \\ 
%&\quad\quad+\frac{\bar{n}_2}{\alpha}+\sum_{i=n_{1}+1}^{\bar{n}_{2}}\ln t_{i:n}- \left\{ \sum_{j=1}^{m-1} \pi_j \lambda_{2j} + \left(1 - \sum_{j=1}^{m-1} \pi_j\right) \lambda_{2m} \right\}\\
%&\quad\quad\times \left( \sum_{i=n_1+1}^{\bar{n}_2} t_{i:n}^\alpha \ln t_{i:n} + (n - \bar{n}_2)t_{r:n}^\alpha \ln t_{r:n} - (n - n_1)\tau^\alpha \ln \tau \right) = 0
%\end{aligned}
%\]

\section{Elements of the Observed Information 
Matrix using the Louis' Method}
\label{app:louis}

Let $\hat{{\boldsymbol\theta}} = (\hat{\alpha}, \hat{\lambda}_1, 
\hat{\lambda}_{21}, \hat{\lambda}_{22}, \hat{\pi})$ 
denote the EM-based MLE. The elements of the 
complete data information matrix 
$I_{\text{com}}(\hat{{\boldsymbol\theta}})$ are as follows:
\begin{align*}
&I^{\text{com}}_{\alpha\alpha}
=
\frac{n_1 +\displaystyle\sum_{i=n_1+1}^{n} \delta_i}{\alpha^2}
+ \lambda_1
\left[
\sum_{i=1}^{n_1} (t_i)^\alpha (\ln t_i)^2
+
(n - n_1)\,\tau^\alpha (\ln \tau)^2
\right]&
\end{align*}
\begin{align*}
&\qquad\qquad
+ \sum_{i=n_1+1}^{n}
\sum_{j=1}^{2}
\eta_{ij}\lambda_{2j}
\left[
(t_i^*)^\alpha (\ln t_i^*)^2
-
\tau^\alpha (\ln \tau)^2
\right],&
\\
&I^{\text{com}}_{\lambda_1\lambda_1}
=
\frac{n_1}{\lambda_1^2}, &
\\
&I^{\text{com}}_{\lambda_{2j}\lambda_{2j}}
=
\frac{
\displaystyle\sum_{i=n_1+1}^{n} \eta_{ij}\delta_i
}{\lambda_{2j}^2},
\quad j=1,2, &\\
&I^{\text{com}}_{\pi\pi}
=\sum_{i=n_1+1}^{n} \frac{\eta_{i1}}{\pi^2}
+
\sum_{i=n_1+1}^{n} \frac{\eta_{i2}}{(1-\pi)^2},&
\\
&I^{\text{com}}_{\alpha\lambda_1}
=
\sum_{i=1}^{n_1} (t_i)^\alpha \ln(t_i)
+
(n - n_1)\,\tau^\alpha \ln(\tau),&\\	
&I^{\text{com}}_{\alpha\lambda_{2j}}
=
\sum_{i=n_1+1}^{n}
\eta_{ij}
\left[
(t_i^*)^\alpha \ln(t_i^*)
-
\tau^\alpha \ln(\tau)
\right],
\quad j=1,2.&
\end{align*}
All other off-diagonal elements of $I_{\text{com}} (\hat{{\boldsymbol\theta}})$ are zero. The elements of the missing information matrix $I_{\text{mis}}(\hat{{\boldsymbol\theta}})$ are as follows:
\begin{align*}
&I^{\text{mis}}_{\alpha\alpha}
=
(\lambda_{22}-\lambda_{21})^2
\sum_{i=n_1+1}^{n}
\eta_{i1}\eta_{i2}
\left[
(t_i^*)^\alpha \ln(t_i^*)
-
\tau^\alpha \ln(\tau)
\right]^2,&
\\
&I^{\text{mis}}_{\alpha\lambda_{21}}
=
(\lambda_{22}-\lambda_{21})
\sum_{i=n_1+1}^{n}
\eta_{i1}\eta_{i2}
\left[
(t_i^*)^\alpha \ln(t_i^*)
-
\tau^\alpha \ln(\tau)
\right] \times
\left[
\frac{\delta_i}{\lambda_{21}}
-
\left((t_i^*)^\alpha - \tau^\alpha\right)
\right],&
\\
&I^{\text{mis}}_{\alpha\lambda_{22}}
=
-(\lambda_{22}-\lambda_{21})
\sum_{i=n_1+1}^{n}
\eta_{i1}\eta_{i2}
\left[
(t_i^*)^\alpha \ln(t_i^*)
-
\tau^\alpha \ln(\tau)
\right] \times
\left[
\frac{\delta_i}{\lambda_{22}}
-
\left((t_i^*)^\alpha - \tau^\alpha\right)
\right],&\\
&I^{\text{mis}}_{\alpha\pi}
=
\frac{\lambda_{22}-\lambda_{21}}{\pi(1-\pi)}
\sum_{i=n_1+1}^{n}
\eta_{i1}\eta_{i2}
\left[
(t_i^*)^\alpha \ln(t_i^*)
-
\tau^\alpha \ln(\tau)
\right],&\\
&I^{\text{mis}}_{\lambda_{2j}\lambda_{2j}}
=
\sum_{i=n_1+1}^{n}
\eta_{i1}\eta_{i2}
\left[
\frac{\delta_i}{\lambda_{2j}}
-
\left((t_i^*)^\alpha - \tau^\alpha\right)
\right]^2,
\quad j=1,2,&
\\
&I^{\text{mis}}_{\lambda_{21}\lambda_{22}}
=
-\sum_{i=n_1+1}^{n}
\eta_{i1}\eta_{i2}
\left[
\frac{\delta_i}{\lambda_{21}}
-
\left((t_i^*)^\alpha - \tau^\alpha\right)
\right] \times
\left[
\frac{\delta_i}{\lambda_{22}}
-
\left((t_i^*)^\alpha - \tau^\alpha\right)
\right],&
\\
&I^{\text{mis}}_{\lambda_{2j}\pi}
=
\frac{1}{\pi(1-\pi)}
\sum_{i=n_1+1}^{n}
\eta_{i1}\eta_{i2}
\left[
\frac{\delta_i}{\lambda_{2j}}
-
\left((t_i^*)^\alpha - \tau^\alpha\right)
\right],
\quad j=1,2, \text{ and }
I^{\text{mis}}_{\pi\pi}
=
\sum_{i=n_1+1}^{n}
\frac{\eta_{i1}\eta_{i2}}{\left(\pi(1-\pi)\right)^2}.&
\end{align*}
All remaining entries of $I_{\text{mis}}(\hat{{\boldsymbol\theta}})$ 
are zero. The observed information matrix is then 
obtained as $I_{\text{obs}}(\hat{{\boldsymbol\theta}}) = 
I_{\text{com}}(\hat{{\boldsymbol\theta}}) - 
I_{\text{mis}}(\hat{{\boldsymbol\theta}})$, and asymptotic 
confidence intervals are constructed using the 
diagonal elements of 
$I_{\text{obs}}^{-1}(\hat{{\boldsymbol\theta}})$.
\end{document}